\documentclass[11pt]{article}
\usepackage[dvips]{epsfig} % this package is for manipulating postscript graphics
\usepackage{amssymb,amsmath,amsthm,mathrsfs} %this package contains all the special ams fonts etc.
\usepackage{graphicx}

\usepackage[authoryear,sort&compress]{natbib}
\usepackage{url}
\usepackage{enumerate}
\usepackage{colordvi,color,pspicture}
\usepackage{psfrag}

\usepackage{setspace}
\graphicspath{{Figures/}}

\newcommand{\Y}{\mathcal{Y}}
\newcommand{\X}{\mathcal{X}}
\newcommand{\NY}{N_{\Y}}
\newcommand{\G}{\Gamma}
\newcommand{\N}{\mathcal{N}}
\newcommand{\NCSt}{N_{CS}^{\tau}}
\newcommand{\TY}{T(\mathcal Y)}
\newcommand{\V}{\mathcal{V}}
\newcommand{\A}{\mathcal{A}}
\newcommand{\R}{\mathbb R}
\newcommand{\I}{\mathbf{I}}
\newcommand{\y}{\mathsf{y}}
\newcommand{\ve}{\varepsilon}
\newcommand{\lam}{\lambda}
\newcommand{\U}{\mathcal{U}}
\newcommand{\E}[1]{\mathbf{E}\,#1}
\newcommand{\Var}[1]{\mathbf{Var}\,#1}
\newcommand{\Cov}[1]{\mathbf{Cov}\,#1}

\newcommand{\argsup}{\mbox{argsup}}
\newcommand{\arginf}{\mbox{arginf}}
\newcommand{\PAE}{\mbox{PAE}}

\begin{document}

\title{A New Family of Random Graphs for Testing Spatial Segregation}

\author{Elvan Ceyhan \thanks{Department of
Mathematics, Ko\c{c} University, Sar{\i}yer, 34450 Istanbul, Turkey
(elceyhan@ku.edu.tr)},
    Carey E. Priebe\thanks{Department of Applied Mathematics and Statistics, The Johns Hopkins University,
Baltimore, MD. 21218 (cep@jhu.edu)},
 \& David J. Marchette\thanks{Department of
Applied Mathematics and Statistics, The Johns Hopkins University,
Baltimore, MD. 21218 (marchettedj@nswc.navy.mil)}
%\thanks[KU]{Department of Mathematics, Ko\c{c} University, Sar{\i}yer, 34450, Istanbul, Turkey}
}

\date{\today}

\maketitle

%\baselineskip0.25in
%\newpage

{\small

\begin{singlespace}
\begin{abstract}
We discuss a graph-based approach for testing spatial point
patterns. This approach falls under the category of data-random
graphs, which have been introduced and used for statistical pattern
recognition in recent years. Our goal is to test complete spatial
randomness against segregation and association between two or more
classes of points. To attain this goal, we use a particular type of
parametrized random digraph called proximity catch digraph (PCD)
which is based based on relative positions of the data points from
various classes. The statistic we employ is the relative density of
the PCD. When scaled properly, the relative density of the PCD is a
$U$-statistic. We derive the asymptotic distribution of the relative
density, using the standard central limit theory of $U$-statistics.
The finite sample performance of the test statistic is evaluated by
Monte Carlo simulations, and the asymptotic performance is assessed
via Pitman's asymptotic efficiency, thereby yielding the optimal
parameters for testing. Furthermore, the methodology discussed in
this article is also valid for data in multiple dimensions.
\end{abstract}


%\vspace{.25 in}

\indent $^\star$This work was partially supported by Office of Naval
Research Grant and Defense Advanced
Research Projects Agency Grant.\\
\indent
$^*$Corresponding author.\\
\indent {\it e-mail:} elceyhan@ku.edu.tr (E.~Ceyhan)

\end{singlespace}
}

\newpage

\section{Introduction}
\label{sec:intro}
 In this article, we discuss a graph-based approach for testing
spatial point patterns. In statistical literature, the analysis of
spatial point patterns in natural populations has been extensively
studied and have important implications in epidemiology, population
biology, and ecology. We investigate the patterns of one class with
respect to other classes, rather than the pattern of one-class with
respect to the ground. The spatial relationships among two or more
groups have important implications especially for plant species.
See, for example, \cite{pielou:1961}, \cite{dixon:1994}, and
\cite{dixon:2002b}.

Our goal is to test the spatial pattern of complete spatial
randomness against spatial segregation or association. Complete
spatial randomness (CSR) is roughly defined as the lack of spatial
interaction between the points in a given study area. Segregation is
the pattern in which points of one class tend to cluster together,
i.e., form one-class clumps. In association, the points of one class
tend to occur more frequently around points from the other class.
%For example, considering two classes of points, $\X$ and $\Y$, a
%realization of segregation, CSR, and association are presented in
%this order in Figure \ref{fig:deldata}.
For convenience and
generality, we call the different types of points as ``classes", but
the class can be replaced by any characteristic of an observation at
a particular location. For example, the pattern of spatial
segregation has been investigated for species (\cite{diggle:2003}),
age classes of plants (\cite{hamill:1986}) and sexes of dioecious
plants (\cite{nanami:1999}).

\begin{figure}[ht]
\centering
 \rotatebox{-0}{ \resizebox{1.8 in}{!}{
\includegraphics{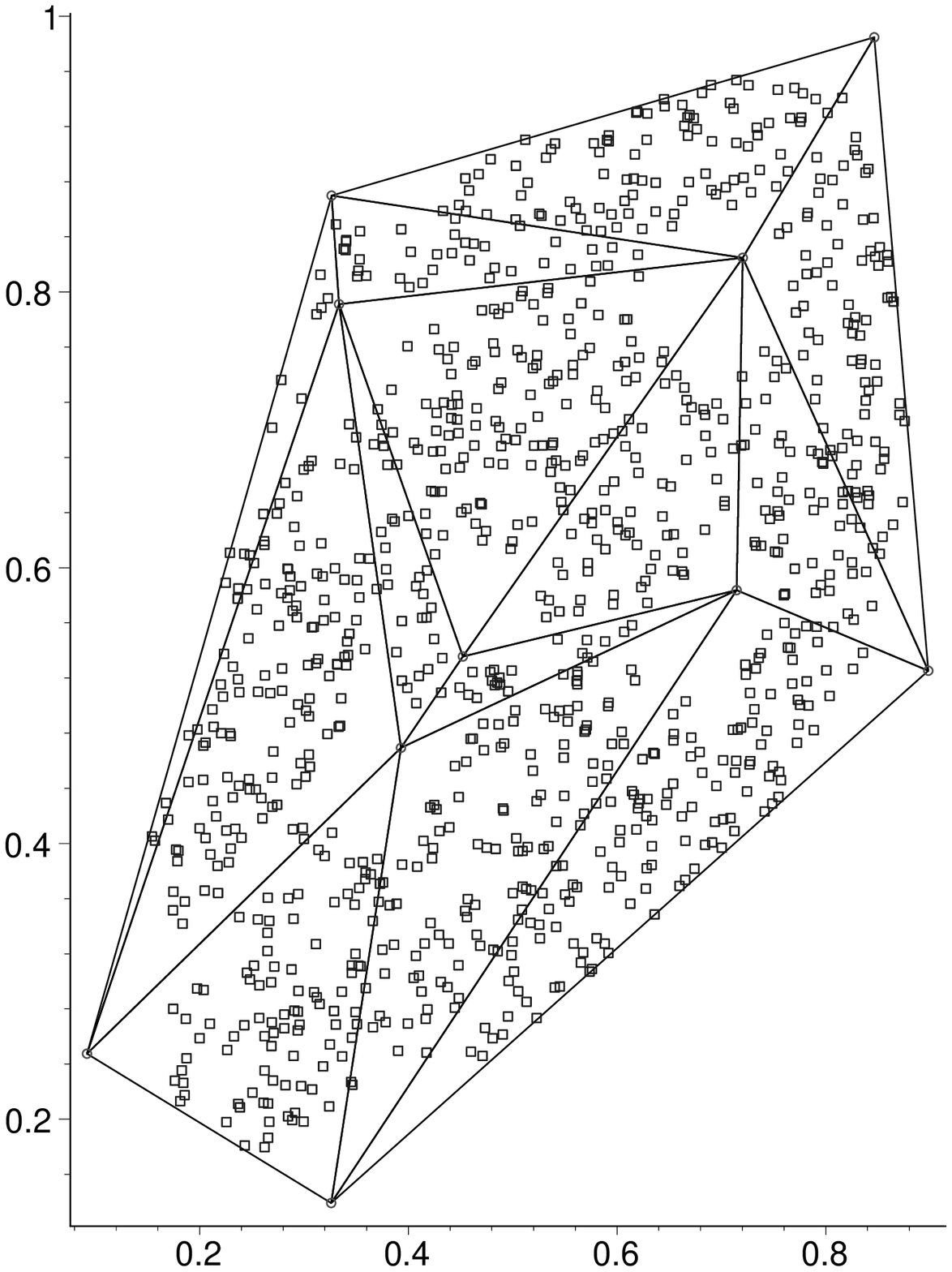} } }
 \rotatebox{-0}{ \resizebox{1.8 in}{!}{ \includegraphics{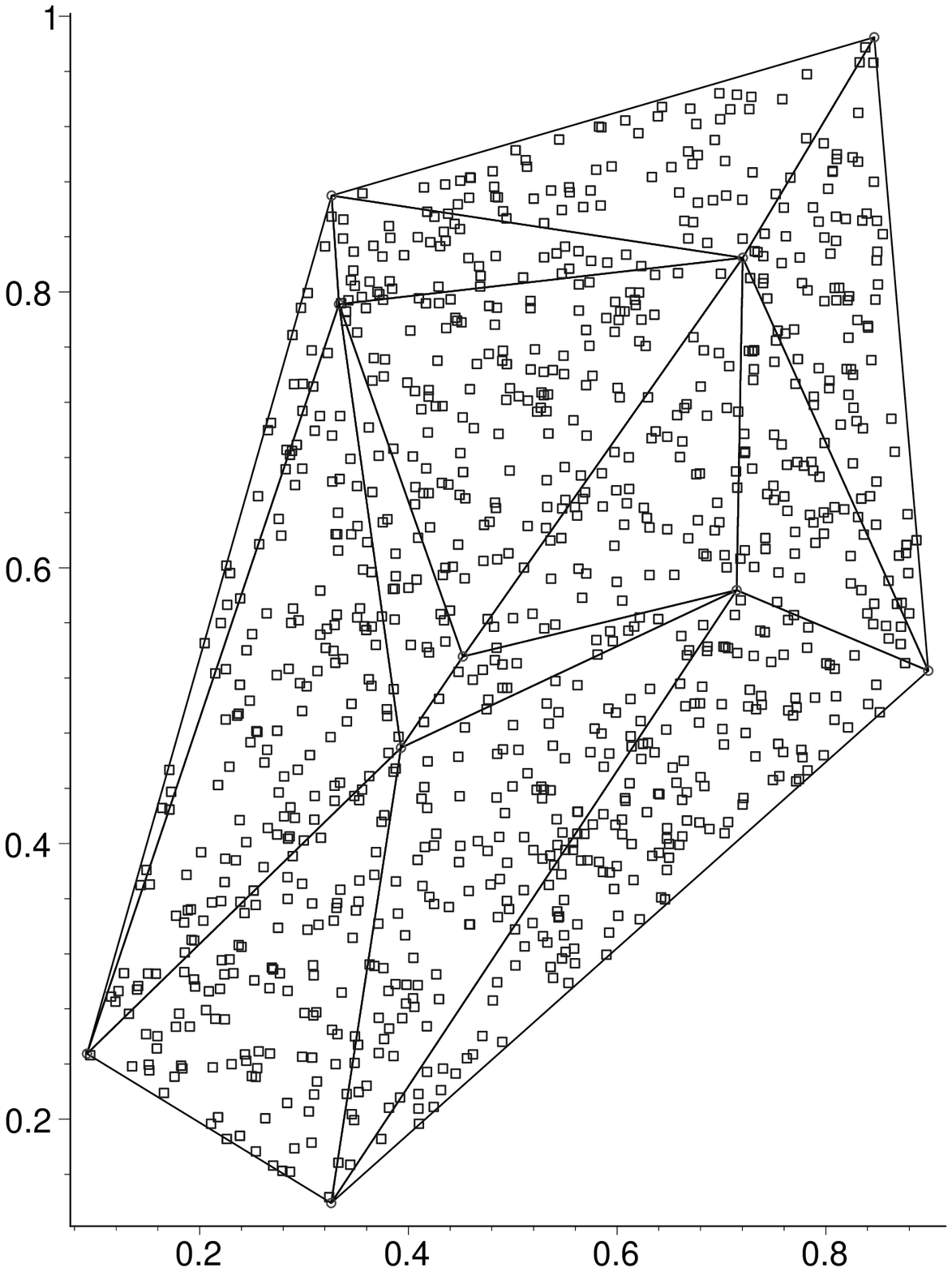} } }
 \rotatebox{-0}{ \resizebox{1.8 in}{!}{ \includegraphics{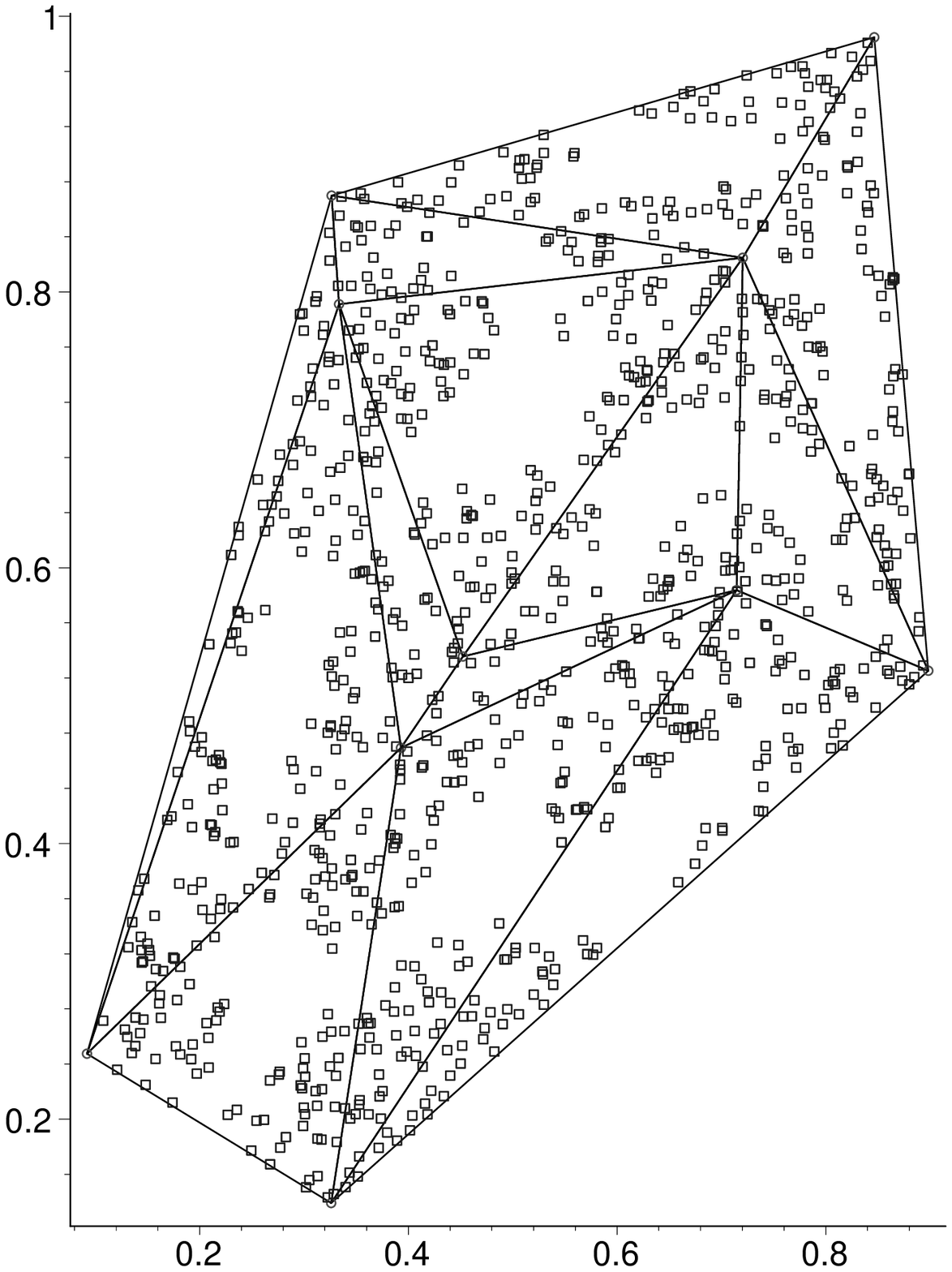} } }
 \caption{
\label{fig:deldata} Realizations of segregation (left), CSR
(middle), and association (right) for $|\Y|=10$ and $|\X|=1000$. $Y$
points are at the vertices of the triangles, and $X$ points are
squares. }
\end{figure}

We use special graphs called proximity catch digraphs (PCDs) for
testing CSR against segregation or association. In recent years,
\cite{priebe:2001} introduced a random digraph related to PCDs
(called class cover catch digraphs) in $\R$ and extended it to
multiple dimensions. \cite{devinney:2002a}, \cite{marchette:2003},
\cite{priebe:2003a}, and \cite{priebe:2003b} demonstrated relatively
good performance of it in classification. In this article, we define
a new class of random digraphs (called PCDs) and apply it in testing
against segregation or association. A PCD is comprised by a set of
vertices and a set of (directed) edges. For example, in the two
class case, with classes $\X$ and $\Y$, the $\X$ points are the
vertices and there is an arc (directed edge) from $x_1 \in \X$ to
$x_2\in \X$, based on a binary relation which measures the relative
allocation of $x_1$ and $x_2$ with respect to $\Y$ points. By
construction, in our PCDs, $\X$ points further away from $\Y$ points
will be more likely to have more arcs directed to other $\X$ points,
compared to the $\X$ points closer to the $\Y$ points. Thus, the
relative density (number of arcs divided by the total number of
possible arcs) is a reasonable statistic to apply to this problem.
To illustrate our methods, we provide three artificial data sets,
one for each pattern. These data sets are plotted in Figure
\ref{fig:deldata}, where $\Y$ points are at the vertices of the
triangles, and $\X$ points are depicted as squares. Observe that we
only consider the $\X$ points in the convex hull of $\Y$ points;
since in the current form, our proposed methodology only works for
such points. Hence we avoid using a real life example, but use these
artificial pattern realizations for illustrative purposes. Under
segregation (left) the relative density of our PCD will be larger
compared to the CSR case (middle), while under association (right)
the relative density will be smaller compared to the CSR case.

The statistical tool we utilize is the asymptotic theory of
$U$-statistics. Properly scaled, we demonstrate that the relative
density of our PCDs is a $U$-statistic, which have asymptotic
normality by the general central limit theory of $U$-statistics. The
digraphs introduced by \cite{priebe:2001}, whose relative density is
also of the $U$-statistic form, the asymptotic mean and variance of
the relative density is not analytically tractable, due to geometric
difficulties encountered. However, the PCD we introduce here is a
parametrized family of random digraphs, whose relative density has
tractable asymptotic mean and variance.

Ceyhan and Priebe introduced an (unparametrized) version of this PCD
and another parametrized family of PCDs in \cite{ceyhan:CS-JSM-2003} and
\cite{ceyhan:dom-num-NPE}, respectively. \cite{ceyhan:dom-num-NPE} used the
domination number (which is another statistic based on the number of
arcs from the vertices) of the second parametrized family for
testing segregation and association. The domination number approach
is appropriate when both classes are comparably large.
\cite{ceyhan:NPE-rel-dens} used the relative density of the same PCD for
testing the spatial patterns. The new parametrized family of PCDs we
introduce has more geometric appeal, simpler in distributional
parameters in the asymptotics, and the range of the parameters is
bounded.

Using the Delaunay triangulation of the $\Y$ observations, we will
define the parametrized version of the proximity maps of
\cite{ceyhan:CS-JSM-2003} in Section \ref{sec:tau-factor-PCD} for which
the calculations ---regarding the distribution of the relative
density--- are tractable. We then can use the relative density of
the digraph to construct a test of complete spatial randomness
against the alternatives of segregation or association which are
defined explicitly in Sections \ref{sec:spat-pattern} and
\ref{sec:null-and-alt}. We will calculate the asymptotic
distribution of the relative density for these digraphs, under both
the null distribution and the alternatives in Sections
\ref{sec:asy-norm-null} and \ref{sec:asy-norm-alt}, respectively.
This procedure results in a consistent test, as will be shown in
Section \ref{sec:consistency}. The finite sample behaviour (in terms
of power) is analyzed using Monte Carlo simulation in Section
\ref{sec:monte-carlo}. The Pitman asymptotic efficiency is analyzed
in Section \ref{sec:Pitman}. The multiple-triangle case is presented
in Section \ref{sec:mult-tri} and the extension to higher dimensions
is presented in Section \ref{sec:NCS-higher-D}. All proofs are
provided in the Appendix.

\section{Spatial Point Patterns}
\label{sec:spat-pattern}

 For simplicity, we describe the spatial point patterns for two-class
populations. The null hypothesis for spatial patterns have been a
controversial topic in ecology from the early days
(\cite{gotelli:1996}). But in general, the null hypothesis consists
of two random pattern types: complete spatial randomness or random
labeling.

Under {\em complete spatial randomness} (CSR) for a spatial point
pattern $\{X(D), A(D)  \in \R^2 \}$ where $A(\cdot)$ denotes the
area functional, we have
\begin{itemize}
\item[(i)] given $n$ events in domain $D$, the events are an independent
random sample from a uniform distribution on $D$;
\item[(ii)] there is no spatial interaction.
\end{itemize}
Furthermore, the number of events in any planar region with area
$A(D)$ follows a Poisson distribution with mean $\lam \cdot A(D)$,
whose probability mass function is given by
$$f_{X(D)}(x)=\frac{e^{-\lam\cdot A(D)}(\lam\cdot A(D))^x}{x!},\;\; x \in \{0,1,2,\ldots\}$$
where $\lam$ is the intensity of the Poisson distribution.

Under {\em random labeling}, class labels are assigned to a fixed
set of points randomly so that the labels are independent of the
locations. Thus, random labeling is less restrictive than CSR. But
conditional on a set of points from CSR, both processes are
equivalent. We only consider a special case of CSR as our null
hypothesis in this article. That is, only $X$ points are assumed to
be uniformly distributed over the convex hull of $Y$ points.

The alternative patterns fall under two major categories called
\emph{association} and \emph{segregation}. {\em Association} occurs
if the points from the two classes together form clumps or clusters.
That is,  association occurs when members of one class have a
tendency to attract members of the other class, as in symbiotic
species, so that the $X_i$ will tend to cluster around the elements
of $\Y$. For example, in plant biology, $\X$ points might be
parasitic plants exploiting $\Y$ points. As another example, $\X$
and $\Y$ points might represent mutualistic plant species, so they
depend on each other to survive. In epidemiology, $\Y$ points might
be contaminant sources, such as a nuclear reactor, or a factory
emitting toxic gases, and $\X$ points might be the residence of
cases (incidences) of certain diseases caused by the contaminant,
e.g., some type of cancer. {\em Segregation} occurs if the members
of the same class tend to be clumped or clustered (see, e.g.,
\cite{pielou:1961}). Many different forms of segregation are
possible. Our methods will be useful only for the segregation
patterns in which the two classes more or less share the same
support (habitat), and members of one class have a tendency to repel
members of the other class.  For instance, it may be the case that
one type of plant does not grow well in the vicinity of another type
of plant, and vice versa. This implies, in our notation, that $X_i$
are unlikely to be located near any elements of $\Y$. See, for
instance, (\cite{dixon:1994, coomes:1999}). In plant biology, $\Y$
points might represent a tree species with a large canopy, so that,
other plants ($\X$ points) that need light cannot grow around these
trees. As another interesting but contrived example, consider the
arsonist who wishes to start fires with maximum duration time (hence
maximum damage), so that he starts the fires at the furthest points
possible from fire houses in a city. Then $\Y$ points could be the
fire houses, while $\X$ points will be the locations of arson cases.

We consider \emph{completely mapped data}, i.e., the locations of
all events in a defined space are observed rather than sparsely
sampled data (only a random subset of locations are observed).

\section{Data-Random Proximity Catch Digraphs}
\label{sec:prox-map}
 In general, in a random digraph, there is an arc between two
 vertices, with a fixed probability, independent of other arcs and
 vertex pairs. However, in our approach, arcs with a shared vertex
 will be dependent. Hence the name \emph{data-random digraphs}.

Let $(\Omega,\mathcal{M})$ be a measurable space and consider a
function $N:\Omega \times 2^\Omega \rightarrow 2^\Omega$, where
$2^\Omega$ represents the power set of $\Omega$. Then given $\Y
\subseteq \Omega$, the {\em proximity map} $\NY(\cdot) =
N(\cdot,\Y): \Omega \rightarrow 2^\Omega$ associates a {\em
proximity region} $\NY(x) \subseteq \Omega$ with each point $x \in
\Omega$.  The region $\NY(x)$ is defined in terms of the distance
between $x$ and $\Y$.

If $\X_n:=\{X_1,X_2,\cdots,X_n\}$ is a set of $\Omega$-valued random
variables, then the $\NY(X_i),\; i=1,\cdots,n$, are random sets. If
the $X_i$ are independent and identically distributed, then so are
the random sets, $\NY(X_i)$.

Define the data-random proximity catch digraph $D$ with vertex set
$\V=\{X_1,\cdots,X_n\}$ and arc set $\A$ by $(X_i,X_j) \in \A \iff
X_j \in \NY(X_i)$ where point $X_i$ ``catches" point $X_j$. The
random digraph $D$ depends on the (joint) distribution of the $X_i$
and on the map $\NY$. The adjective {\em proximity}
--- for the catch digraph $D$ and for the map $\NY$ ---
comes from thinking of the region $\NY(x)$ as representing those
points in $\Omega$ ``close'' to $x$ (\cite{toussaint:1980} and
\cite{jaromczyk:1992}).

The \emph{relative density} of a digraph $D=(\V,\A)$ of order $|\V|
= n$ (i.e., number of vertices is $n$), denoted $\rho(D)$, is
defined as
$$
\rho(D) = \frac{|\A|}{n(n-1)}
$$
where $|\cdot|$ denotes the set cardinality functional (\cite{janson:2000}).

Thus $\rho(D)$ represents the ratio of the number of arcs in the
digraph $D$ to the number of arcs in the complete symmetric digraph
of order $n$, namely $n(n-1)$.
%For brevity we use \emph{relative density} rather than relative arc density henceforth.

If $X_1,\cdots,X_n \stackrel{iid}{\sim} F$, then
the relative density
of the associated data-random proximity catch digraph $D$,
denoted $\rho(\X_n;h,\NY)$, is a U-statistic,
\begin{eqnarray}
\rho(\X_n;h,\NY) =
  \frac{1}{n(n-1)}
    \sum\hspace*{-0.1 in}\sum_{i < j \hspace*{0.25 in}}
      \hspace*{-0.1 in}h(X_i,X_j;\NY)
\end{eqnarray}
where
\begin{eqnarray}
h(X_i,X_j;\NY)&=& \I \{(X_i,X_j) \in \A\}+ \I \{(X_j,X_i) \in \A\} \nonumber \\
       &=& \I \{X_j \in \NY(X_i)\}+ \I \{X_i \in \NY(X_j)\}
\end{eqnarray}
with $\I(\cdot)$ being the indicator function. We denote
$h(X_i,X_j;\NY)$ as $h_{ij}$ henceforth for brevity of notation.
Although the digraph is not symmetric (since $(x,y) \in \A$ does not
necessarily imply $(y,x) \in \A$), $h_{ij}$ is defined as the number
of arcs in $D$ between vertices $X_i$ and $X_j$, in order to produce
a symmetric kernel with finite variance (\cite{lehmann:1988}).

The random variable $\rho_n := \rho(\X_n;h,\NY)$ depends on $n$ and $\NY$ explicitly
and on $F$ implicitly.
The expectation $\E[\rho_n]$, however, is independent of $n$
and depends on only $F$ and $\NY$:
\begin{eqnarray}
0 \leq \E[\rho_n] = \frac{1}{2}\E[h_{12}] \leq 1 \mbox{ for all $n\ge 2$}.
\end{eqnarray}
The variance $\Var[\rho_n]$ simplifies to
\begin{eqnarray}
\label{eq:var-rho}
0 \leq
  \Var[\rho_n] =
     \frac{1}{2n(n-1)} \Var[h_{12}] +
     \frac{n-2}{n(n-1)} \Cov[h_{12},h_{13}]
  \leq 1/4.
\end{eqnarray}
A central limit theorem for $U$-statistics (\cite{lehmann:1988}) yields
\begin{eqnarray}
\sqrt{n}(\rho_n-\E[\rho_n]) \stackrel{\mathcal{L}}{\longrightarrow} \N(0,\Cov[h_{12},h_{13}])
\end{eqnarray}
provided that $\Cov[h_{12},h_{13}] > 0$.
The asymptotic variance of $\rho_n$, $\Cov[h_{12},h_{13}]$,
depends on only $F$ and $\NY$.
Thus, we need determine only
$\E[h_{12}]$
and
$\Cov[h_{12},h_{13}]$
in order to obtain the normal approximation
\begin{eqnarray}
\rho_n \stackrel{\mbox{approx}}{\sim}
\N\left(\E[\rho_n],\Var[\rho_n]\right) =
\N\left(\frac{\E[h_{12}]}{2},\frac{\Cov[h_{12},h_{13}]}{n}\right) \mbox{ for large $n$}.
\end{eqnarray}

\subsection{The $\tau$-Factor Central Similarity Proximity Catch Digraphs}
\label{sec:tau-factor-PCD}
\begin{figure} []
    \centering
   \scalebox{.4}{\input{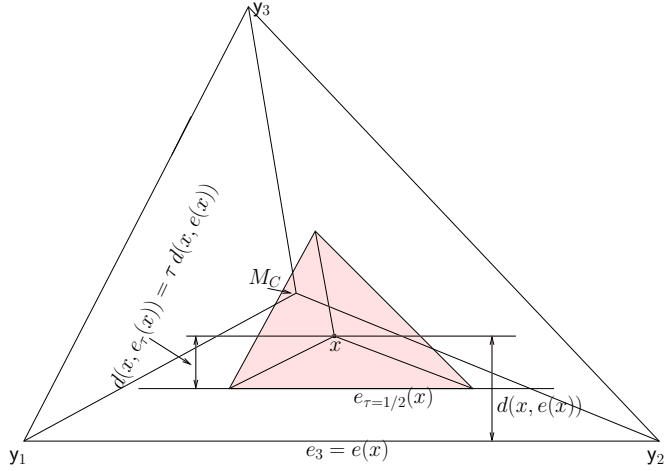}}
    \caption{Construction of $\tau$-factor central similarity proximity region,
    $N_{CS}^{1/2}(x)$ (shaded region).}
\label{fig:ProxMapDef}
    \end{figure}

 We define the $\tau$-factor central
similarity proximity map briefly. Let $\Omega = \R^2$ and let $\Y =
\{\y_1,\y_2,\y_3\} \subset \R^2$ be three non-collinear points.
Denote the triangle --- including the interior --- formed by the
points in $\Y$ as $\TY$. For $\tau \in [0,1]$, define $\NY^{\tau}$
to be the {\em $\tau$-factor central similarity proximity map} as
follows; see also Figure \ref{fig:ProxMapDef}.  Let $e_j$ be the
edge opposite vertex $\y_j$ for $j=1,2,3$, and let ``edge regions''
$R(e_1)$, $R(e_2)$, $R(e_3)$ partition $\TY$ using segments from the
center of mass of $\TY$ to the vertices. For $x \in \TY \setminus
\Y$, let $e(x)$ be the edge in whose region $x$ falls; $x \in
R(e(x))$. If $x$ falls on the boundary of two edge regions we assign
$e(x)$ arbitrarily. For ${\tau} \in (0,1]$, the {\em $\tau$-factor}
central similarity proximity region $\NCSt(x)=\NY^{\tau}(x)$ is
defined to be the triangle $T_{\tau}(x)$ with the following
properties:
\begin{itemize}
\item[(i)] $T_{\tau}(x)$ has an edge $e_{\tau}(x)$ parallel to $e(x)$ such that
$d(x,e_{\tau}(x))=\tau\, d(x,e(x))$ and $d(e_{\tau}(x),e(x)) \le d(x,e(x))$ where
$d(x,e(x))$ is the Euclidean (perpendicular) distance from $x$ to $e(x)$,
\item[(ii)] $T_{\tau}(x)$ has the same orientation as and is similar to $\TY$,
\item[(iii)] $x$ is at the center of mass of $T_{\tau}(x)$.
\end{itemize}
 Note that (i) implies the ``$\tau$-factor", (ii) implies ``similarity", and
(iii) implies ``central" in the name, {\em $\tau$-factor central
similarity proximity map}. Notice that $\tau>0$ implies that $x \in
\NCSt(x)$ and $\tau \le 1$ implies that $\NCSt(x)\subseteq \TY$ for
all $x \in \TY$. For $x \in
\partial(\TY)$ and $\tau \in [0,1]$, we define
$\NCSt(x)=\{x\}$; for $\tau=0$ and $x \in \TY$ we also define
$\NCSt(x)=\{x\}$. Let $\TY^o$ be the interior of the triangle $\TY$.
Then for all $ x\in \TY^o$ the edges $e_{\tau}(x)$ and $e(x)$ are
coincident iff $\tau=1$. Observe that the central similarity
proximity map in (\cite{ceyhan:CS-JSM-2003}) is $\NCSt(\cdot)$ with
$\tau=1$. Hence by definition, $(x,y)$ is an arc of the
$\tau$-factor central similarity PCD iff $y \in \NCSt(x)$.

Notice that $X_i \stackrel{iid}{\sim} F$, with the additional
assumption that the non-degenerate two-dimensional probability
density function $f$ exists with support in $\TY$, implies that the
special case in the construction of $\NCSt$ --- $X$ falls on the
boundary of two edge regions --- occurs with probability zero.

\begin{figure}[]
\centering
 \rotatebox{-90}{ \resizebox{2.0 in}{!}{ \includegraphics{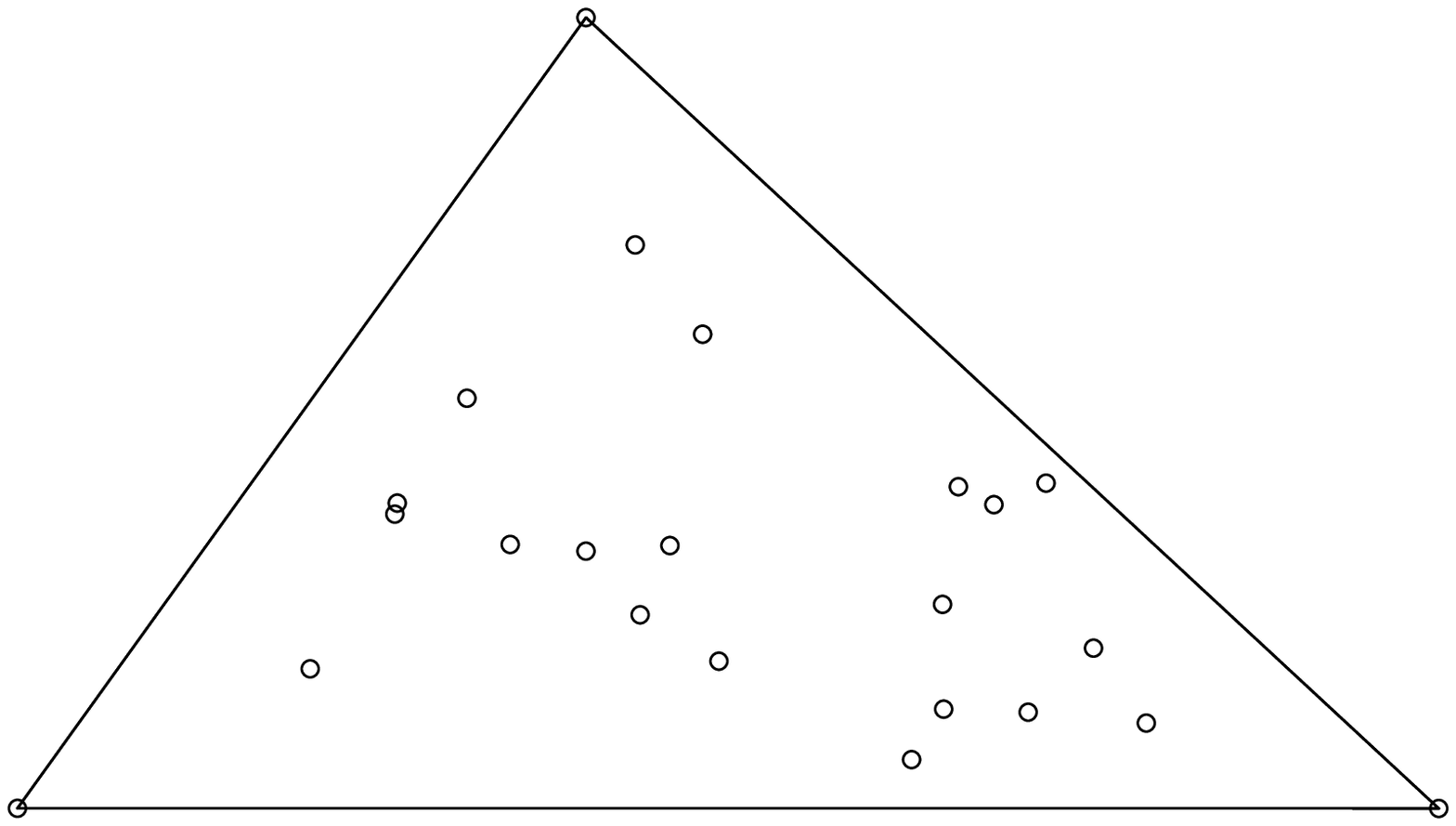} } }
 \rotatebox{-90}{ \resizebox{2.0 in}{!}{ \includegraphics{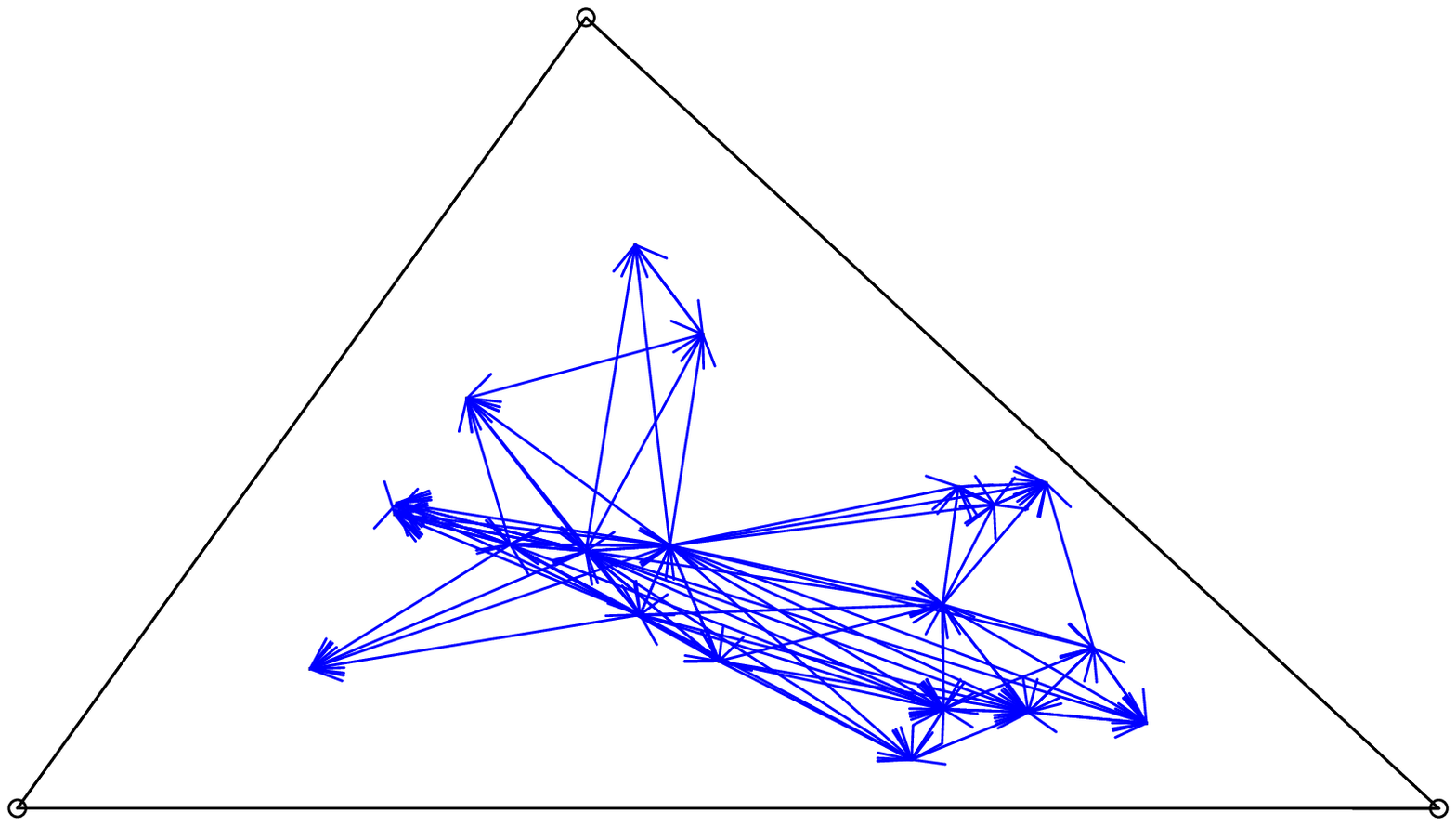} } }
 \rotatebox{-90}{ \resizebox{2.0 in}{!}{ \includegraphics{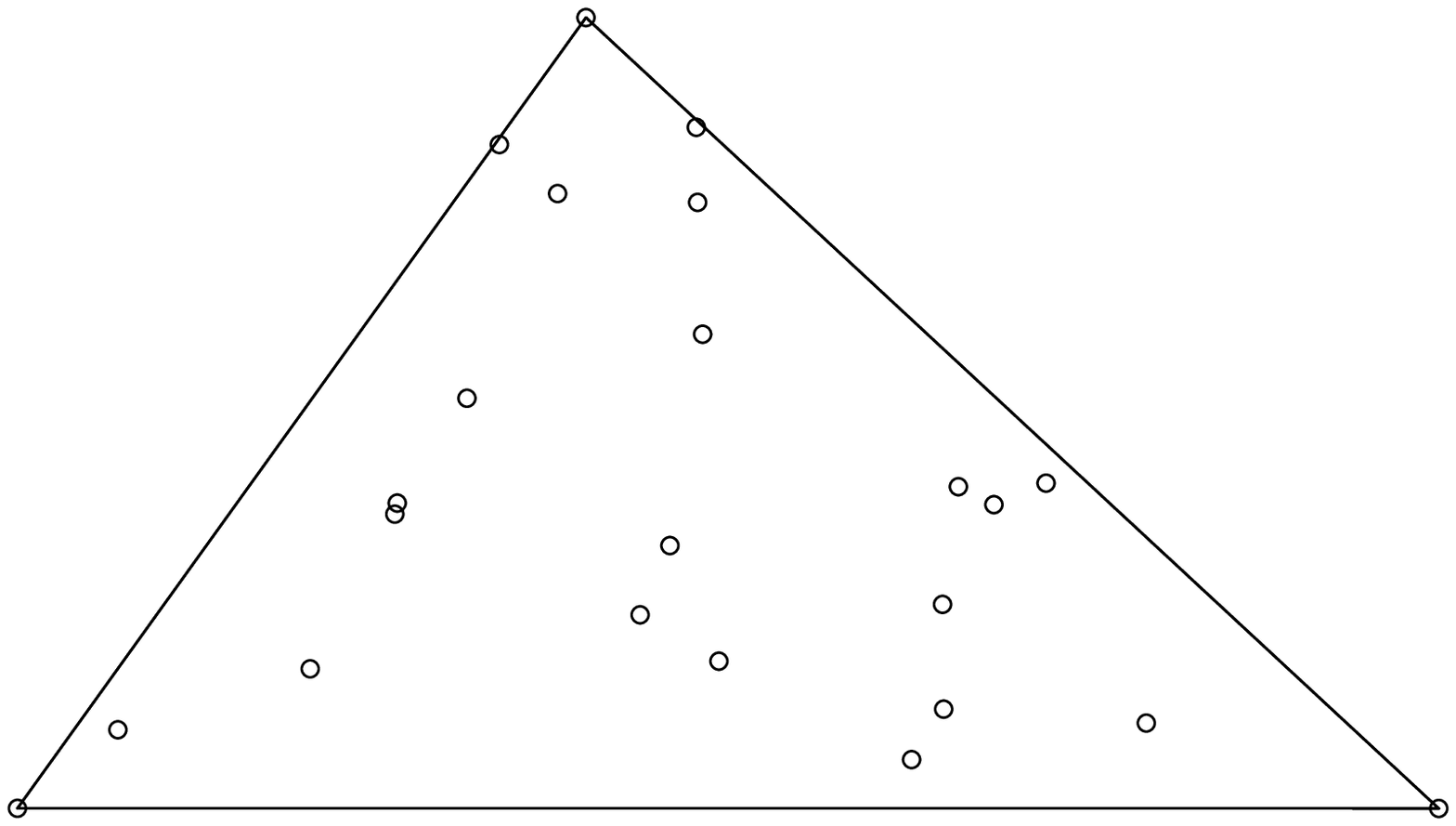} } }
 \rotatebox{-90}{ \resizebox{2.0 in}{!}{ \includegraphics{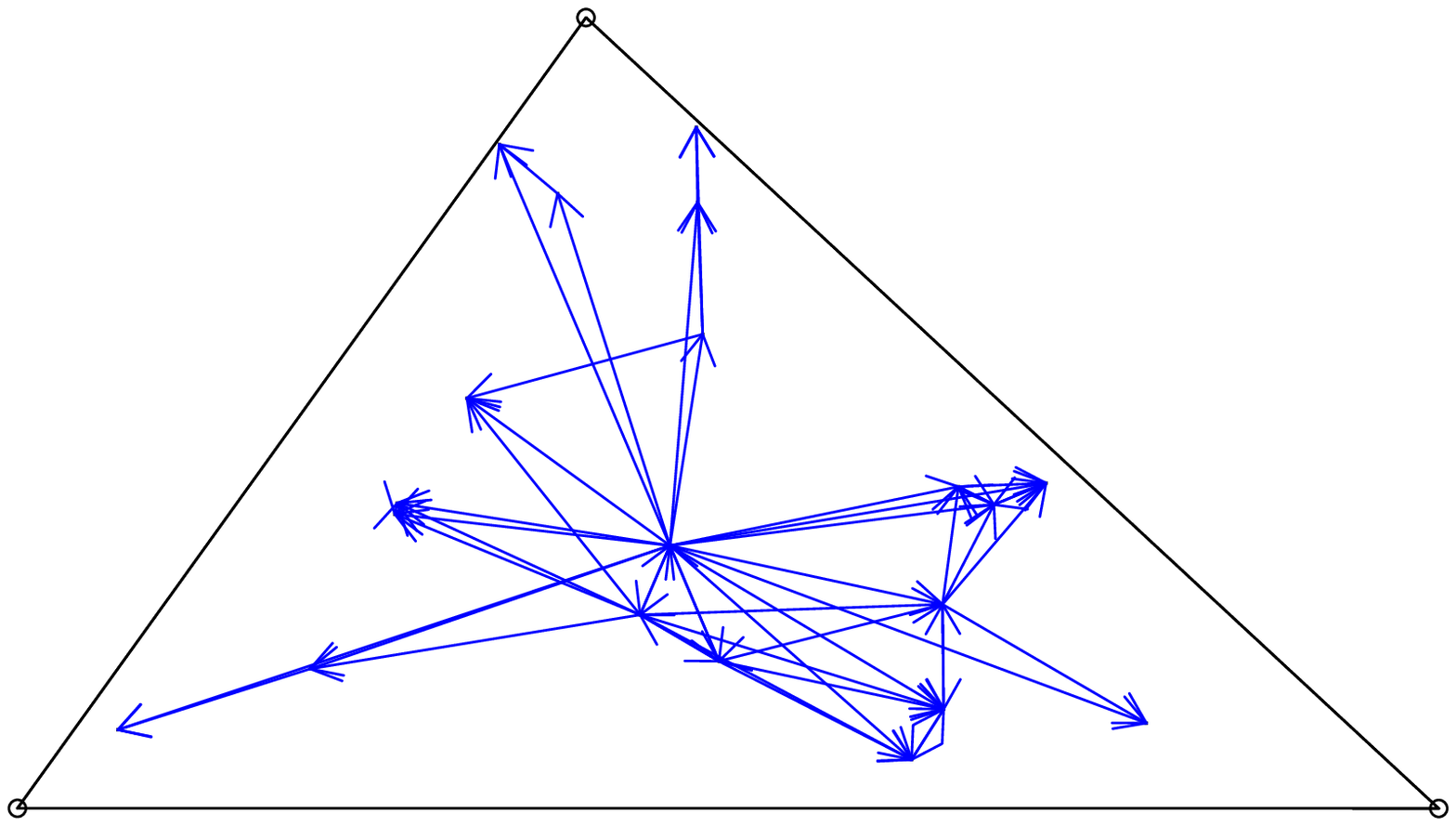} } }
 \rotatebox{-90}{ \resizebox{2.0 in}{!}{ \includegraphics{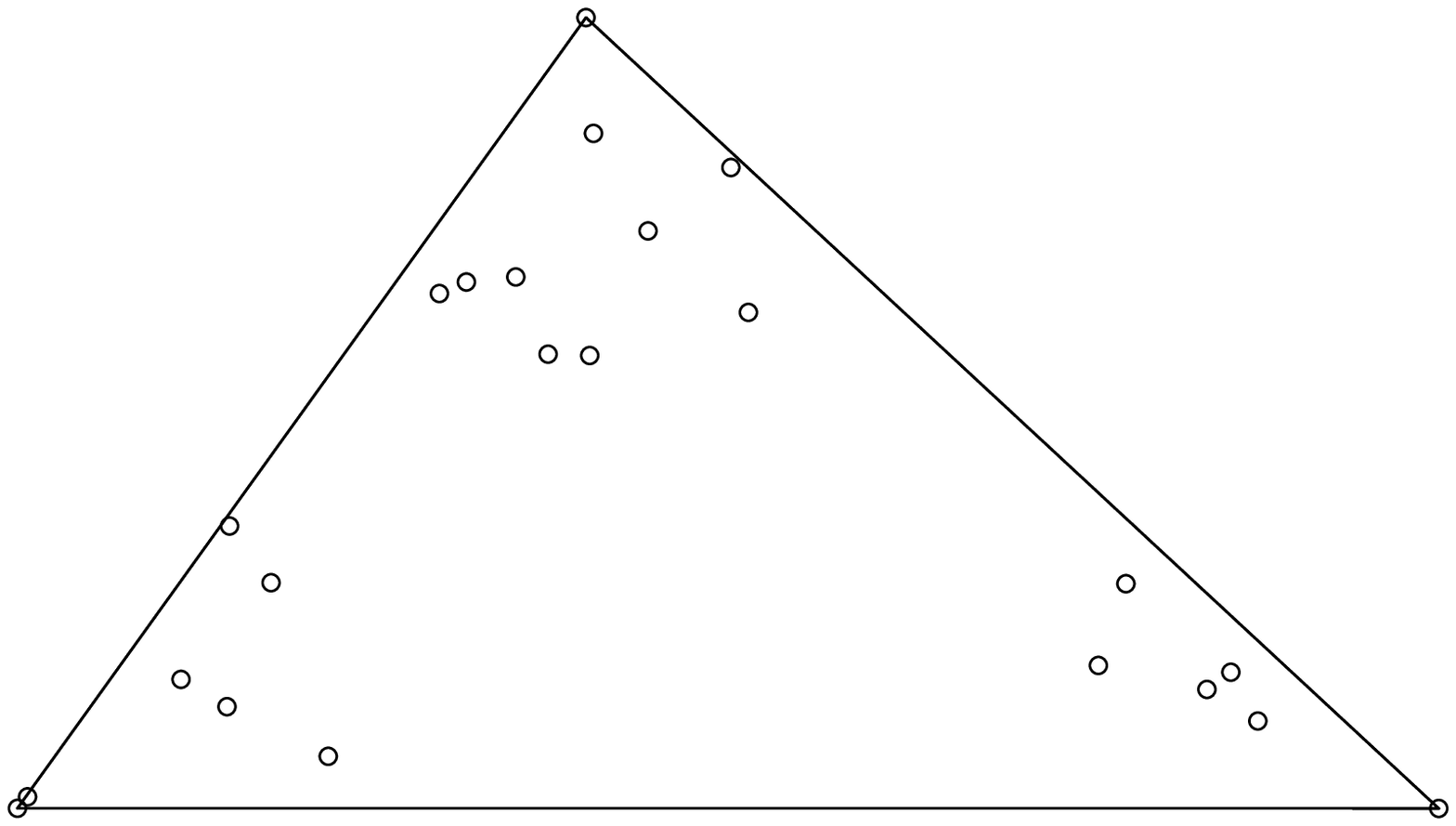} } }
 \rotatebox{-90}{ \resizebox{2.0 in}{!}{ \includegraphics{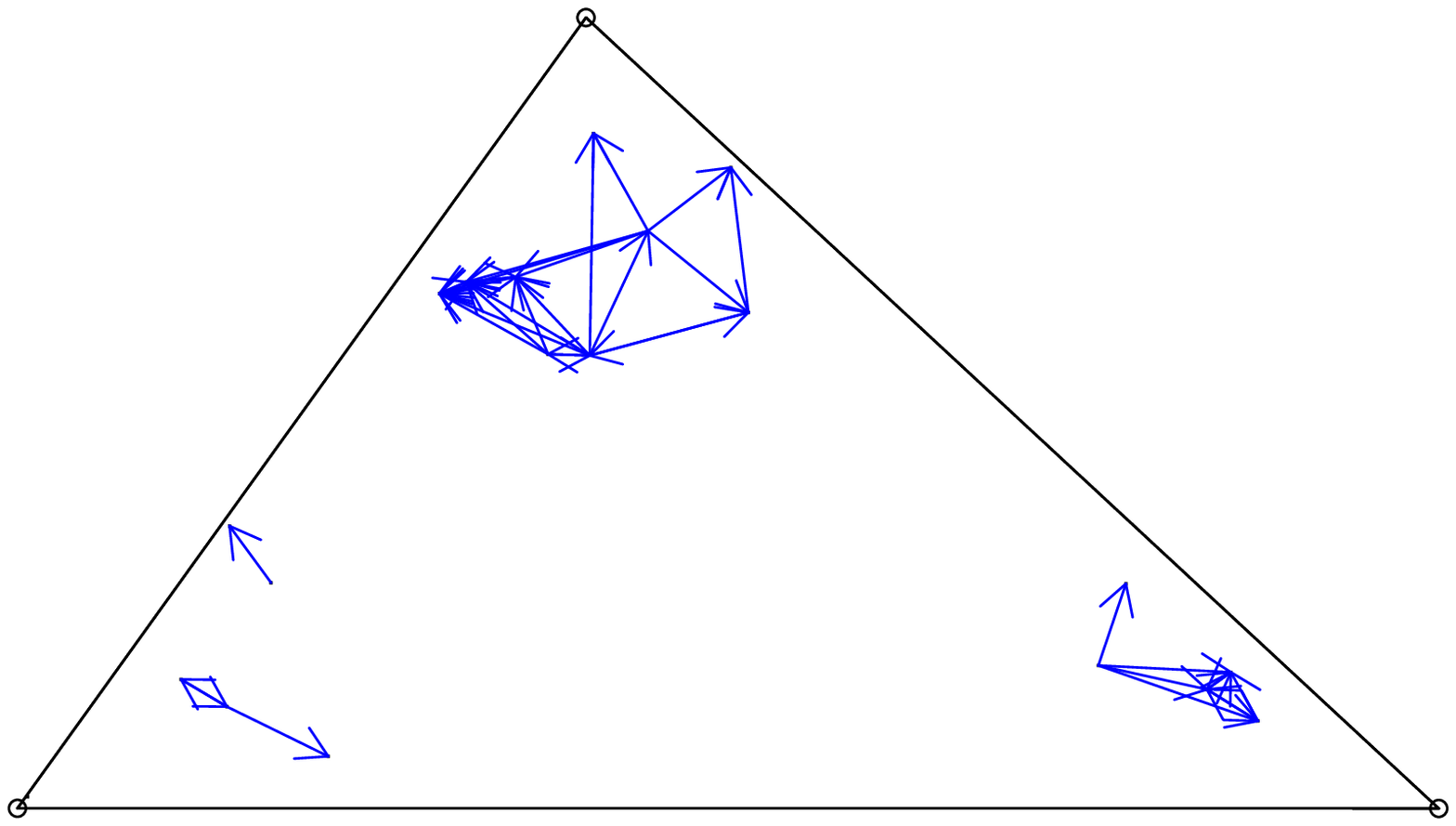} } }
 \caption{
\label{fig:deldata-J=1} Realizations of segregation (left), CSR
(middle), and association (right) for $|\Y|=3$ and $|\X|=20$. $Y$
points are at the vertices of the triangle, and $X$ points are
squares. The number of arcs with $\tau=1$ are 98, 53, and 37,
respectively. So, relative density values are .258, .139, and .097,
respectively.}
\end{figure}

For a fixed $\tau \in (0,1]$, $\NCSt(x)$ gets larger (in area) as
$x$ gets further away from the edges (or equivalently gets closer to
the center of mass, $C_M$) in the sense that as $d(x,e(x))$
increases (or equivalently $d(C_M,e_{\tau}(x))$ decreases. Hence for
points in $\TY$, the further the points away from the vertices $\Y$
(or closer the points to $C_M$ in the above sense), the larger the
area of $\NCSt(x)$. Hence, it is more likely for such points to
catch other points, i.e., have more arcs directed to other points.
Therefore, if more $X$ points are clustered around the center of
mass, then the digraph is more likely to have more arcs, hence
larger relative density. So, under segregation, relative density is
expected to be larger than that in CSR or association. On the other
hand, in the case of association, i.e., when $X$ points are
clustered around $Y$ points, the regions $\NCSt(x)$ tend to be
smaller in area, hence, catch less points, thereby resulting in a
small number of arcs, or a smaller relative density compared to CSR
or segregation. See, for example, Figure \ref{fig:deldata-J=1} with
3 $Y$ points, and 20 $X$ points for segregation (top left), CSR
(middle left) and association (bottom right). The corresponding arcs
in the $\tau$-factor central similarity PCD with $\tau=1$ are
plotted in the right in Figure \ref{fig:deldata-J=1}. The
corresponding relative density values (for $\tau=1$) are .1395,
.2579, and .0974, respectively.
%For a range of $\tau$ values ($\tau
%\in \{.1,.2,\ldots,.9,1.0 \}$) the relative density values are
%plotted in Figure \ref{fig:relden-n=20} (left). Observe that at any
%fixed $\tau$, the relative density values are smallest for
%association and largest for segregation.

Furthermore, for a fixed $x \in \TY^o$, $\NCSt(x)$ gets larger (in
area) as $\tau$ increases. So, as $\tau$ increases, it is more
likely to have more arcs, hence larger relative density for a given
realization of $\X$ points in $\TY$.
%See Figure \ref{fig:relden-n=20} (left).
% where for the realizations considered
%for the three patterns in Figure \ref{fig:deldata-J=1}, relative
%density values increase as $\tau$ increases.

\section{Asymptotic Distribution of the Relative Density}
\label{sec:asy-dist-rho}
 We first describe the null and alternative patterns we consider
 briefly, and then provide the asymptotic distribution of the
 relative density for these patterns.

 There are two major types of asymptotic structures for spatial data (\cite{lahiri:1996}).
In the first, any two observations are required to be at least a
fixed distance apart, hence as the number of observations increase,
the region on which the process is observed eventually becomes
unbounded. This type of sampling structure is called ``increasing
domain asymptotics". In the second type, the region of interest is a
fixed bounded region and more and more points are observed in this
region. Hence the minimum distance between data points tends to zero
as the sample size tends to infinity. This type of structure is
called ``infill asymptotics", due to \cite{cressie:1991}. The
sampling structure for our asymptotic analysis is infill, as only
the size of the type $X$ process tends to infinity, while the
support, the convex hull of a given set of points from type $Y$
process, $C_H(\Y)$ is a fixed bounded region.

\subsection{Null and Alternative Patterns}
\label{sec:null-and-alt}
 For statistical testing for segregation and
association, the null hypothesis is generally some form of {\em
complete spatial randomness}; thus we consider
$$H_o: X_i \stackrel{iid}{\sim} \mathcal{U}(\TY).$$
If it is desired to have the sample size be a random variable, we
may consider a spatial Poisson point process on $\TY$ as our null
hypothesis.

\subsection*{Geometry Invariance Property}
\label{sec:geo-inv} We first present a ``geometry invariance" result
that will simplify our calculations by allowing us to consider the
special case of the equilateral triangle.

{\bf Theorem 1:} Let $\Y = \{\y_1,\y_2,\y_3\} \subset \R^2$ be three
non-collinear points. For $i=1,\cdots,n$ let $X_i
\stackrel{iid}{\sim} F = \mathcal{U}(\TY)$, the uniform distribution
on the triangle $\TY$. Then for any $\tau \in [0,1]$ the
distribution of $\rho_n(\tau):=\rho(\X_n;h,\NCSt)$ is independent of
$\Y$, hence the geometry of $\TY$.

Based on Theorem 1 and our uniform null hypothesis, we may assume
that $\TY$ is the standard equilateral triangle with $\Y =
\bigl\{(0,0),(1,0),\bigl(1/2,\sqrt{3}/2\bigr) \bigr\}$, henceforth.
For our $\tau$-factor central similarity proximity map and uniform
null hypothesis, the asymptotic null distribution of $\rho_n(\tau) =
\rho(\X_n;h,\NCSt)$ as a function of $\tau$ can be derived.  Let
$\mu(\tau):=\E[\rho_n]$, then $\mu(\tau)=\E[h_{12}]/2=P(X_2 \in
\NCSt(X_1))$ is the probability of an arc occurring between any two
vertices and let $\nu(\tau):=\Cov[h_{12},h_{13}]$.

\begin{figure} []
\centering
 \scalebox{.4}{\input{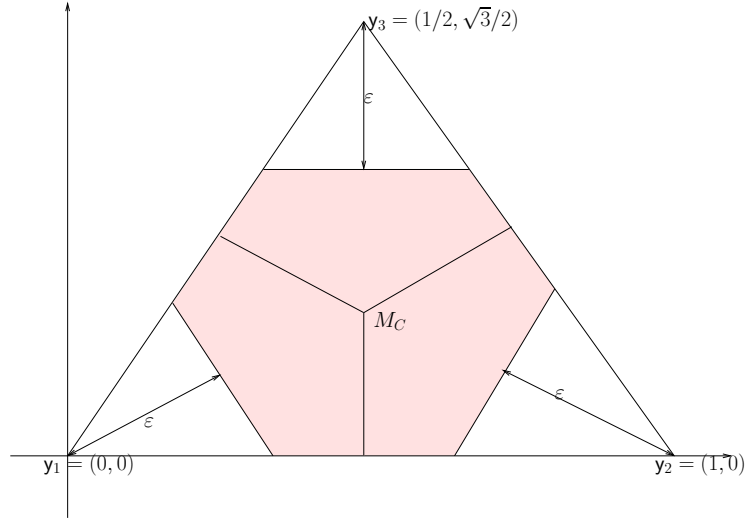}}
 \caption{An example for the segregation alternative for a particular $\ve$ (shaded region),
 and its complement is for the association alternative (unshaded region) on the standard
 equilateral triangle.}
 \label{fig:seg-alt}
\end{figure}

We define two simple classes of alternatives, $H^S_{\ve}$ and
$H^A_{\ve}$ with $\ve \in \bigl( 0,\sqrt{3}/3 \bigr)$, for
segregation and association, respectively. See also Figure
\ref{fig:seg-alt}. For $\y \in \Y$, let $e(\y)$ denote the edge of
$\TY$ opposite vertex $\y$, and for $x \in \TY$ let $\ell_{\y}(x)$
denote the line parallel to $e(\y)$ through $x$. Then define
$T(\y,\ve) = \bigl\{x \in \TY: d(\y,\ell_{\y}(x)) \le \ve \bigr\}$.
Let $H^S_{\ve}$ be the model under which $X_i \stackrel{iid}{\sim}
\mathcal{U} \Bigl(\TY \setminus \cup_{\y \in \Y} T(\y,\ve)\Bigr)$
and $H^A_{\ve}$ be the model under which $X_i \stackrel{iid}{\sim}
\mathcal{U}\Bigl(\cup_{\y \in \Y} T\bigl(\y,\sqrt{3}/3 - \ve
\bigr)\Bigr)$. The shaded region in Figure \ref{fig:seg-alt} is the
support for segregation for a particular $\ve$ value; and its
complement is the support for the association alternative with
$\sqrt{3}/3-\ve$. Thus the segregation model excludes the
possibility of any $X_i$ occurring near a $\y_j$, and the
association model requires that all $X_i$ occur near a $\y_j$. The
$\sqrt{3}/3 - \ve$ in the definition of the association alternative
is so that $\ve=0$ yields $H_o$ under both classes of alternatives.
We consider these types of alternatives among many other
possibilities, since relative density is geometry invariant for
these alternatives as the alternatives are defined with parallel
lines to the edges.

{\bf Remark:}
These definitions of the alternatives
are given for the standard equilateral triangle.
The geometry invariance result of Theorem 1 from Section \ref{sec:geo-inv} still holds
under the alternatives, in the following sense.
If, in an arbitrary triangle,
a small percentage $\delta \cdot 100\%$ where $\delta \in (0,4/9)$ of the area is carved
away as forbidden from each vertex using line segments parallel
to the opposite edge, then
under the transformation to the standard equilateral triangle
this will result in the alternative $H^S_{\sqrt{3 \delta / 4}}$.
This argument is for
segregation with $\delta < 1/4$;
a similar construction is available for the other cases.

\subsection{Asymptotic Normality Under the Null Hypothesis}
\label{sec:asy-norm-null}

By detailed geometric  probability calculations provided in the
Appendix, the mean and the asymptotic variance of the relative
density of the $\tau$-factor proximity catch digraph can be
calculated explicitly. The central limit theorem for $U$-statistics
then establishes the asymptotic normality under the uniform null
hypothesis. These results are summarized in the following theorem.

{\bf Theorem 2:} For $\tau \in (0,1]$, the relative density of the
$\tau$-factor central similarity proximity digraph converges in law
to the normal distribution, i.e., as $n \rightarrow \infty$,
\begin{eqnarray}
 \frac{\sqrt{n}(\rho_n(\tau)-\mu(\tau))}{\sqrt{\nu(\tau)} }
 \stackrel{\mathcal{L}}{\longrightarrow}
 \N(0,1)
\end{eqnarray}
where
\begin{eqnarray}
\label{eq:CSAsymean} \mu(\tau) = \tau^2/6
\end{eqnarray}
and
\begin{eqnarray}
\label{eq:CSAsyvar} \nu(\tau) =\frac
{\tau^4(6\,\tau^5-3\,\tau^4-25\,\tau^3+\tau^2+49\,\tau+14)}{45\,(\tau+1)(2\,\tau+1)(\tau+2)}
\end{eqnarray}
For $\tau=0$, $\rho_n(\tau)$ is degenerate for all $n>1$.

See the Appendix for the derivation.

\begin{figure}[ht]
\centering
 \psfrag{t}{ {\large $\tau$}} \psfrag{m}{ {\large $\mu(\tau)$}}
\epsfig{figure=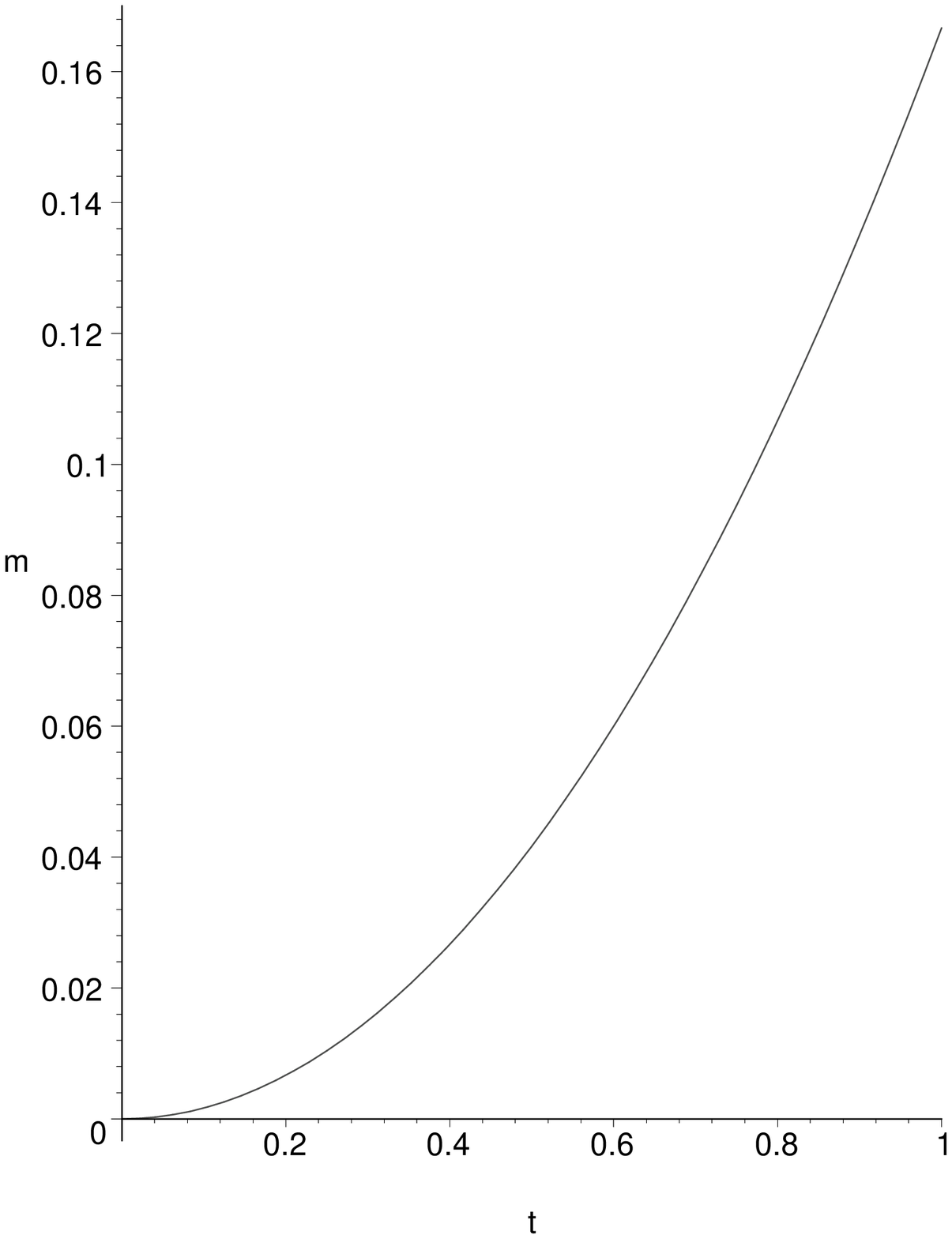, height=150pt, width=200pt}
%\rotatebox{-0}{ \resizebox{2.25 in}{!}{ \includegraphics{muCS.eps} } }
\psfrag{v}{ {\large $\nu(\tau)$}}
 \epsfig{figure=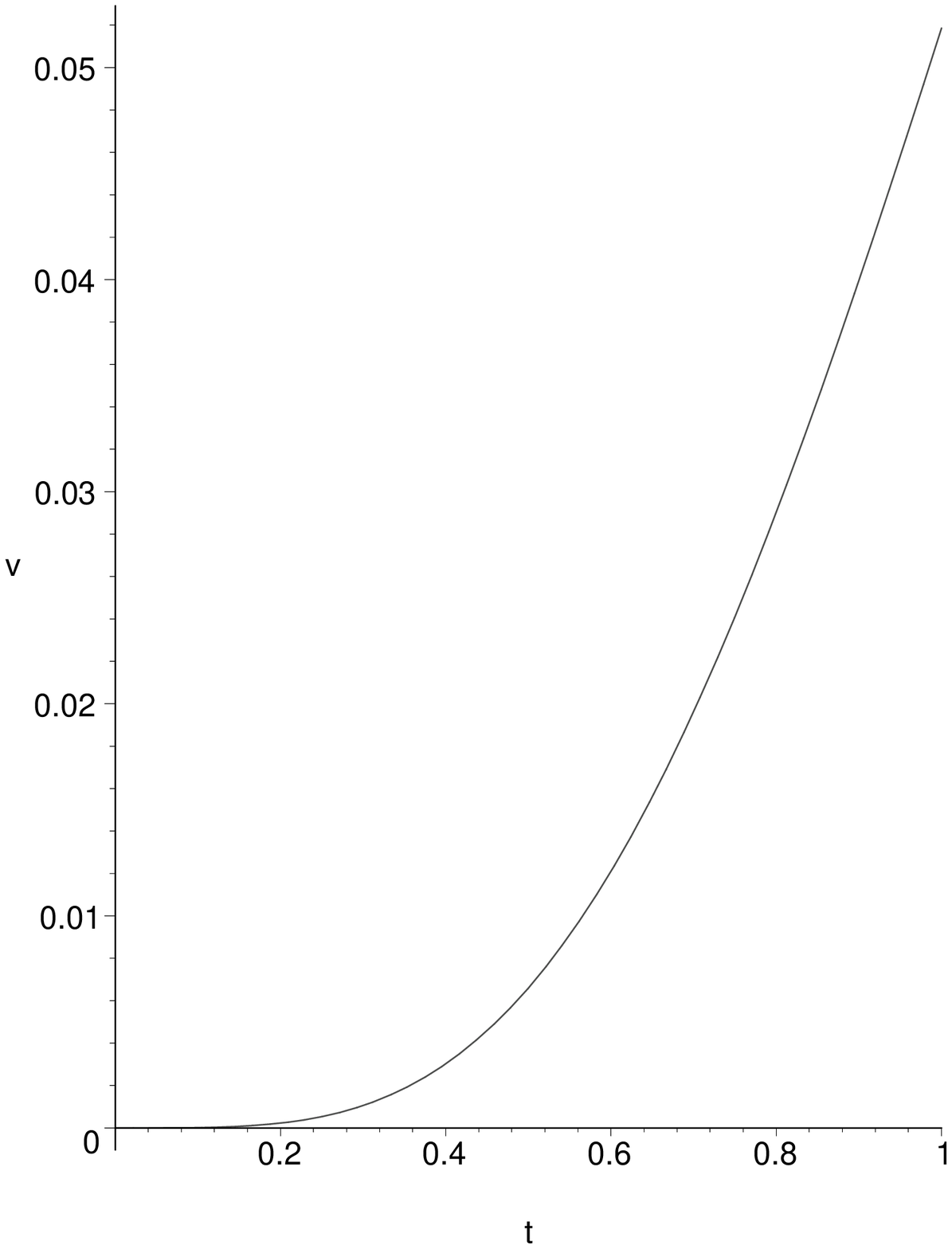, height=150pt, width=200pt}
%\rotatebox{-0}{ \resizebox{2.25 in}{!}{\includegraphics{varCS.eps}}}
 \caption{
 \label{fig:AsyNormCurves}
 Result of Theorem 2:
asymptotic null mean $\mu(\tau)=\mu(\tau)$ (left) and variance
$\nu(\tau)=\nu(\tau)$ (right), from Equations (\ref{eq:CSAsymean})
and (\ref{eq:CSAsyvar}), respectively. }
\end{figure}

Consider the form of the mean and the variance functions, which are
depicted in Figure \ref{fig:AsyNormCurves}.  Note that $\mu(\tau)$
is monotonically increasing in $\tau$, since $\NCSt(x)$ increases
with $\tau$ for all $x \in \TY^o$.  Note also that $\mu(\tau)$ is
continuous in $\tau$ with $\mu(\tau=1)=1/6$ and $\mu(\tau=0)=0$.

Regarding the asymptotic variance, note that $\nu(\tau)$ is
continuous in $\tau$ and $\nu(\tau=1)=7/135$ and $\nu(\tau=0)=0$
---there are no arcs when $\tau=0$ a.s.--- which explains why
$\rho_n(\tau=0)$ is degenerate.

%\vspace*{0.1 in}

As an example of the limiting distribution, $\tau=1/2$ yields
$$
\frac{\sqrt{n} \bigl(\rho_n(1/2) - \mu(1/2)\bigr)}{\sqrt{\nu(1/2)}}
=
\sqrt{\frac{2880\,n}{19}} \bigl(\rho_n(1/2) - 1/24\bigr)\stackrel{\mathcal{L}}{\longrightarrow} \mathcal{N}(0,1)$$
or equivalently,
$$
\rho_n(1/2) \stackrel{\mbox{\tiny{approx}}}{\sim}
\N\left(\frac{1}{24},\frac{19}{2880\,n}\right) .$$

The finite sample variance and skewness may be derived analytically
in much the same way as was $\Cov[h_{12},h_{13}]$ for the asymptotic
variance. In fact, the exact distribution of $\rho_n(\tau)$ is, in
principle, available by successively conditioning on the values of
the $X_i$. Alas, while the joint distribution of $h_{12},h_{13}$ is
available, the joint distribution of $\{h_{ij}\}_{1 \leq i < j \leq
n}$, and hence the calculation for the exact distribution of
$\rho_n(\tau)$, is extraordinarily tedious and lengthy for even
small values of $n$.

\begin{figure}[ht]
\centering
 \psfrag{Density}{ \Huge{\bf{density}}}
 \rotatebox{-90}{\resizebox{1.8 in}{!}{ \includegraphics{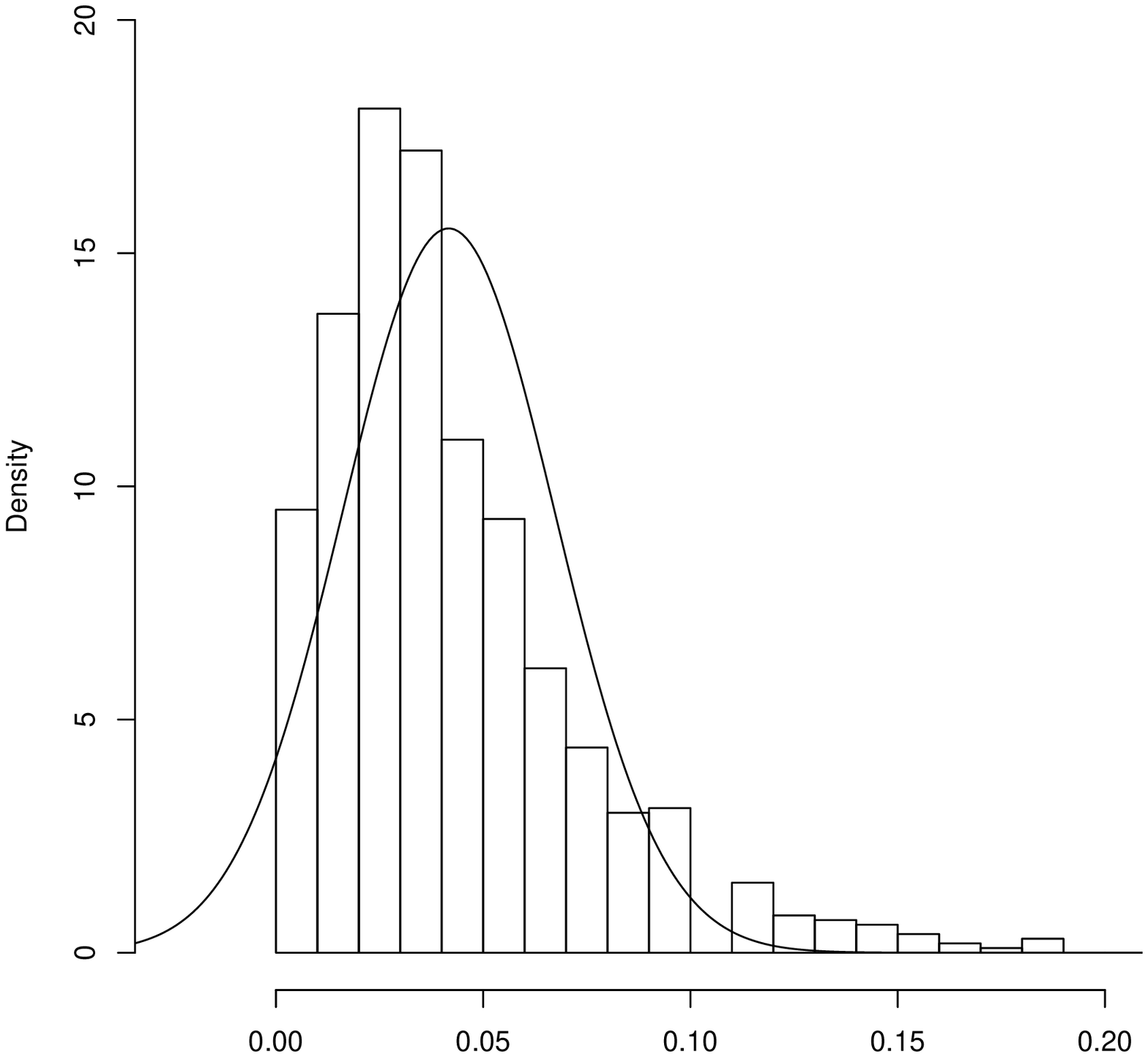} } }
 \rotatebox{-90}{ \resizebox{1.8 in}{!}{ \includegraphics{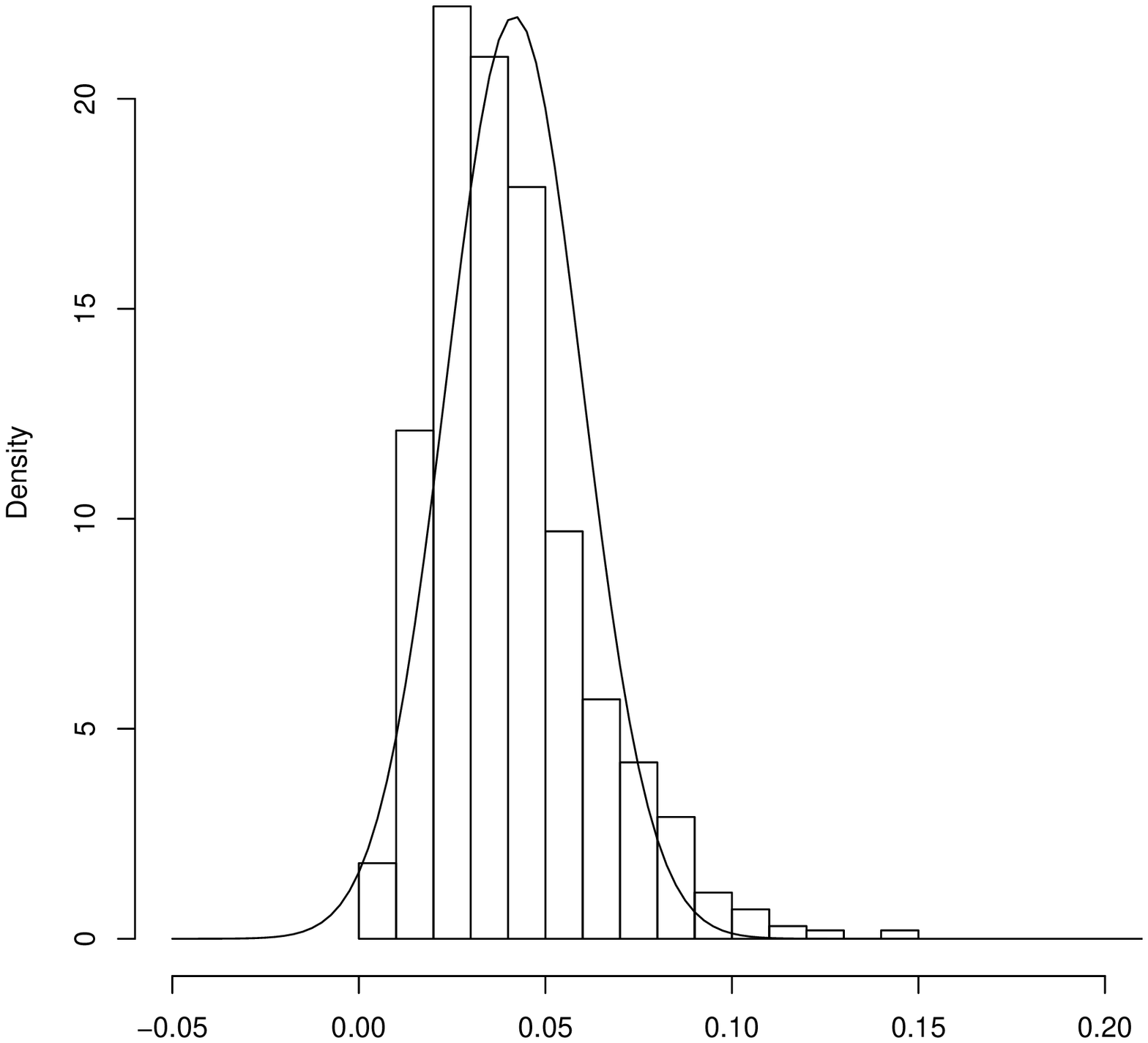} } }
 \rotatebox{-90}{ \resizebox{1.8 in}{!}{ \includegraphics{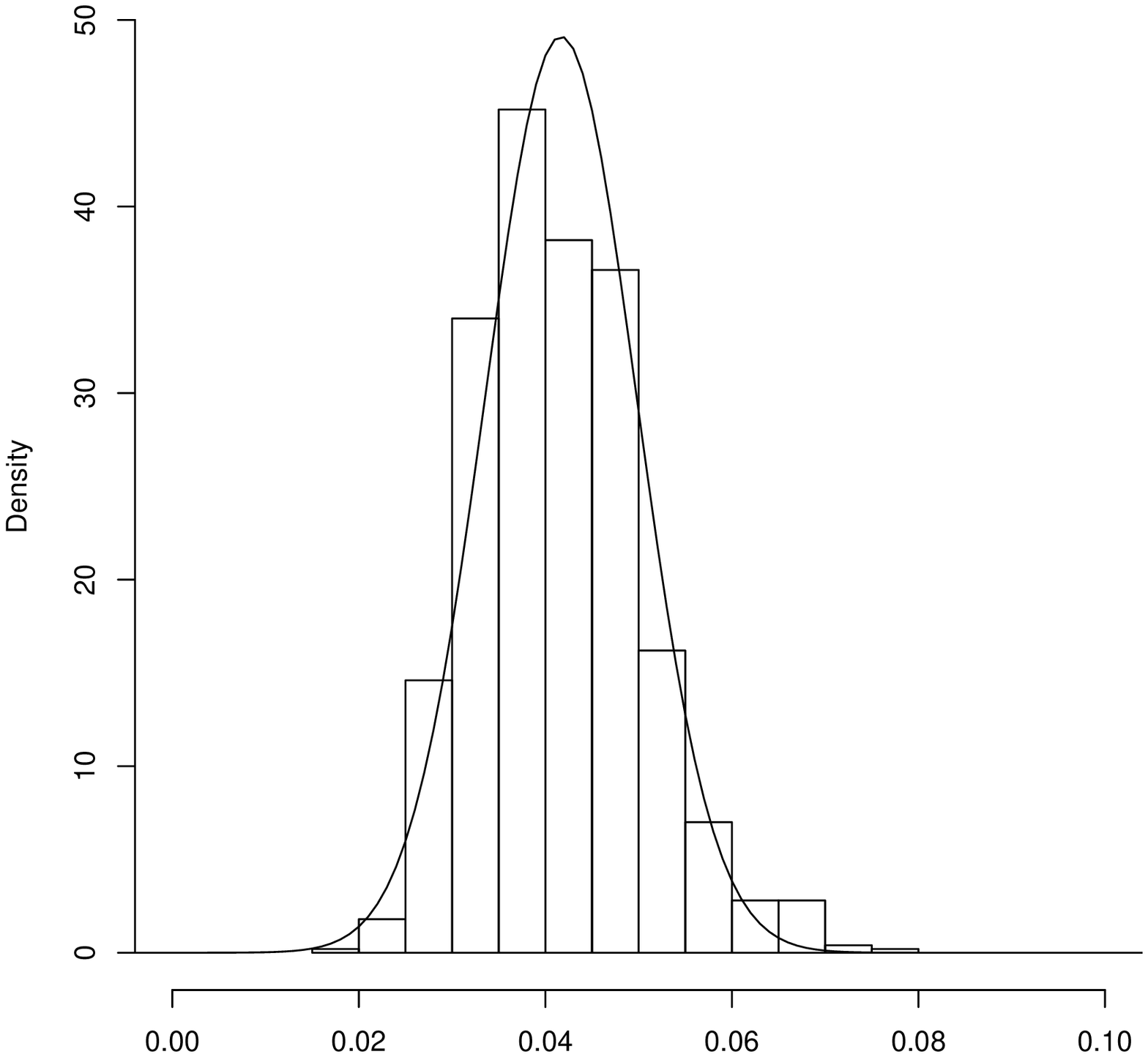} } }
 \caption{\label{fig:NormSkewCS} Depicted are $\rho_n(1/2)
\stackrel{\mbox{approx}}{\sim}
\mathcal{N}\left(\frac{1}{24},\frac{19}{2880\,n}\right)$ for
$n=10,\,20,\,100$ (left to right).  Histograms are based on 1000
Monte Carlo replicates. Solid curves represent the approximating
normal densities given in Theorem 2. Note that the vertical axes are
differently scaled. }
\end{figure}

\begin{figure}[ht]
\centering
 \psfrag{Density}{ \Huge{\bf{density}}}
 \rotatebox{-90}{\resizebox{1.8 in}{!}{ \includegraphics{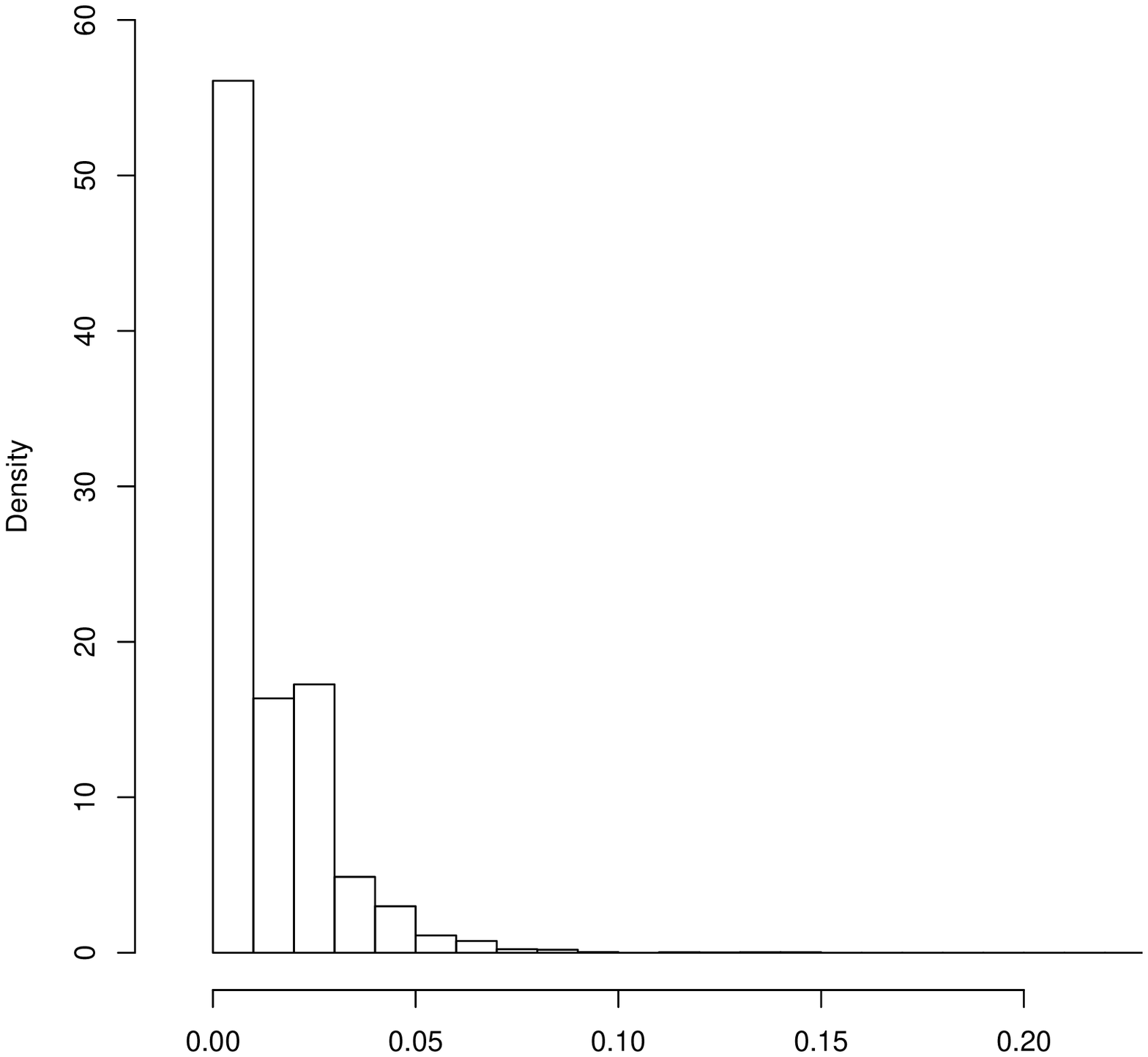} } }
 \rotatebox{-90}{ \resizebox{1.8 in}{!}{\includegraphics{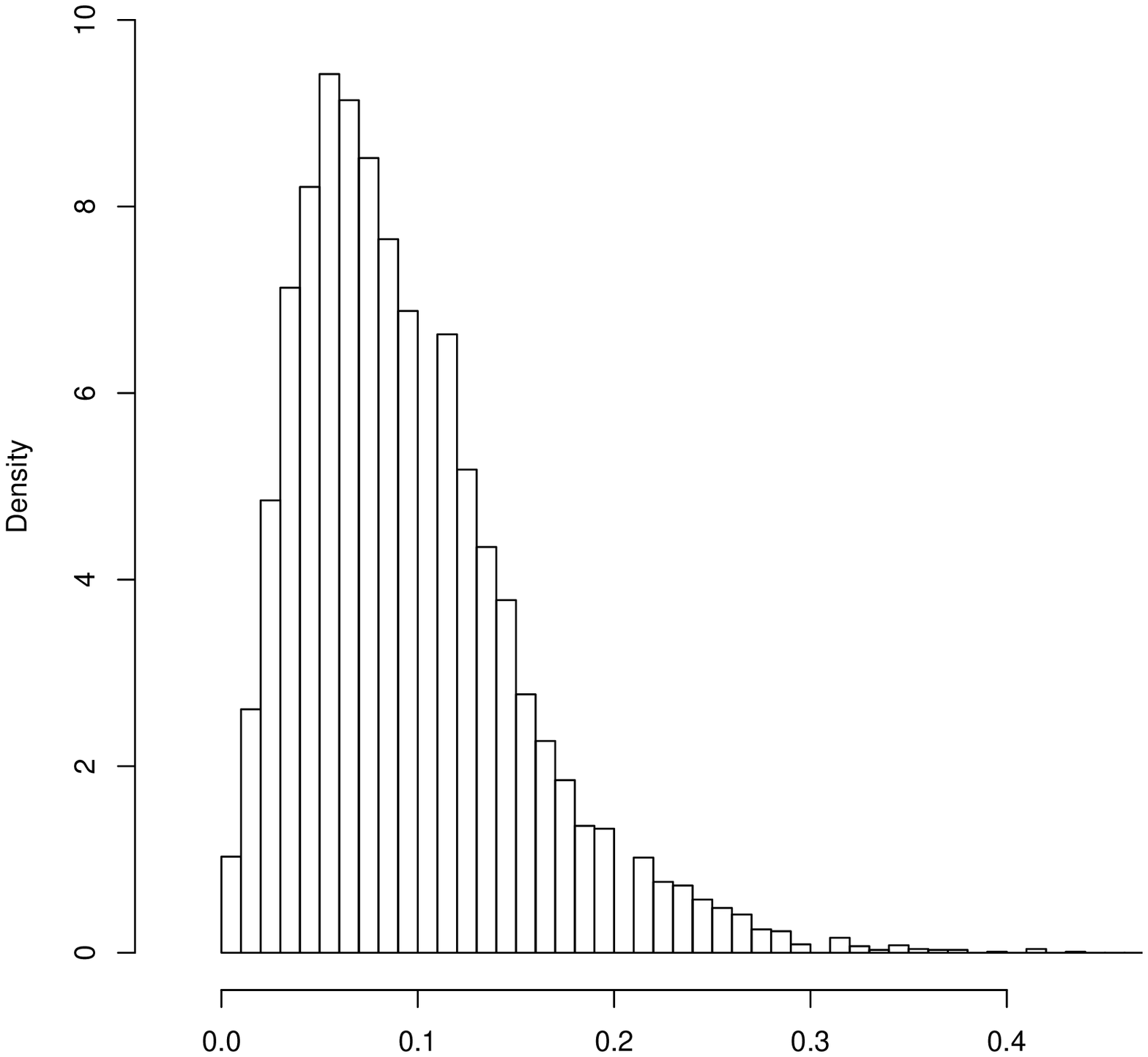} } }
 \rotatebox{-90}{ \resizebox{1.8 in}{!}{ \includegraphics{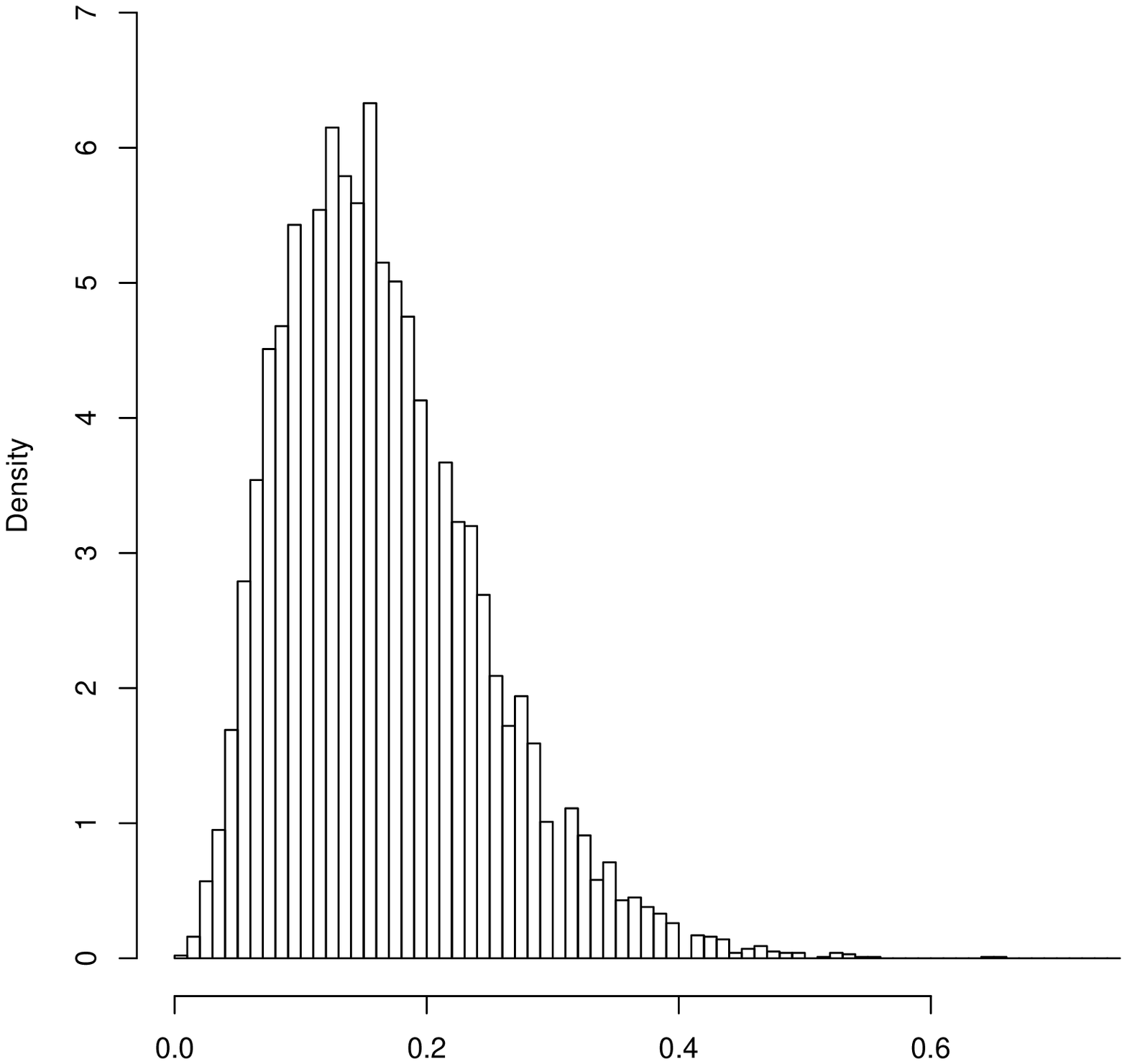} } }
\caption{ \label{fig:CSNormSkew1}
 Depicted are the histograms for
10000 Monte Carlo replicates of $\rho_{10}(1/4)$ (left),
$\rho_{10}(3/4)$ (middle), and $\rho_{10}(1)$ (right) indicating
severe small sample skewness for small values of $\tau$. }
\end{figure}

Figure \ref{fig:NormSkewCS}
indicates that, for $\tau=1/2$,
the normal approximation is accurate even for small $n$
(although kurtosis and skewness may be indicated for $n=10,\,20$).
Figure \ref{fig:CSNormSkew1} demonstrates,
however, that the smaller the value of $\tau$ the more severe the skewness of the probability density.

\subsection{Asymptotic Normality Under the Alternatives}
\label{sec:asy-norm-alt}
 Asymptotic normality of the relative
density of the proximity catch digraph can be established under the
alternative hypotheses of segregation and association by the same
method as under the null hypothesis. Let $\E_{\ve}[\cdot]$ be the
expectation with respect to the uniform distribution under the
segregation and association alternatives with $\ve \in \bigl(
0,\sqrt{3}/3 \bigr)$.

{\bf Theorem 3:} Let $\mu_S(\tau,\ve)$ ( $\mu_A(\tau,\ve)$ ) be the
mean and $\nu_S(\tau,\ve)$ ( $\nu_A(\tau,\ve)$ ) be the covariance,
$\Cov[h_{12},h_{13}]$ for $\tau \in (0,1]$ and $\ve \in \bigl(
0,\sqrt{3}/3 \bigr)$ under segregation ( association ). Then under
$H^S_{\ve}$, $\sqrt{n}\bigl(\rho_n(\tau)-\mu_S(\tau,\ve)\bigr)
\stackrel {\mathcal L}{\longrightarrow}
\N\bigl(0,\nu_S(\tau,\ve)\bigr)$ for the values of the pair
$(\tau,\ve)$ for which $\nu_S(\tau,\ve)>0$. $\rho_n(\tau)$ is
degenerate when $\nu_S(\tau,\ve)=0$. Likewise, under $H^A_{\ve}$,
$\sqrt{n}\bigl(\rho_n(\tau)-\mu_A(\tau,\ve)\bigr) \stackrel
{\mathcal L}{\longrightarrow} \N\bigl(0,\nu_A(\tau,\ve)\bigr)$ for
the values of the pair $(\tau,\ve)$ for which $\nu_A(\tau,\ve)>0$.
$\rho_n(\tau)$ is degenerate when $\nu_A(\tau,\ve)=0$.

Notice that under the association alternatives any $\tau \in (0,1]$
yields asymptotic normality for all $\ve \in \bigl( 0,\sqrt{3}/3
\bigr)$, while under the segregation alternatives only $\tau=1$
yields this universal asymptotic normality.

\section{The Test and Analysis}
\label{sec:test-analiz}
 The relative density of the central
similarity proximity catch digraph is a test statistic for the
segregation/association alternative; rejecting for extreme values of
$\rho_n(\tau)$ is appropriate since under segregation, we expect
$\rho_n(\tau)$ to be large; while under association, we expect
$\rho_n(\tau)$ to be small. Using the test statistic
\begin{eqnarray}
R(\tau) = \frac{\sqrt{n} \bigl(\rho_n(\tau) -
\mu(\tau)\bigr)}{\sqrt{\nu(\tau)}},
\end{eqnarray}
which is the normalized relative density, the asymptotic critical
value for the one-sided level $\alpha$ test against segregation is
given by
\begin{eqnarray*}
z_{\alpha} = \Phi^{-1}(1-\alpha).
\end{eqnarray*}
%where $\Phi(\cdot)$ is the standard normal distribution function.
Against segregation, the test rejects for $R(\tau)>z_{\alpha}$ and
against association, the test rejects for $R(\tau) < z_{1-\alpha}$.
For the example patterns in Figure \ref{fig:deldata-J=1},
$R(\tau=1)=1.792,-.534$, and -1.361, respectively.
%For the patterns
%in Figure \ref{fig:deldata-J=1}, we also calculate $R(\tau)$ values
%for $\tau \in \{.1,.2,\ldots,.9,1.0\}$.
%These values are plotted in
%Figure \ref{fig:relden-n=20} (right) where the horizontal lines are
%at values -1.96, 0, and 1.96. Observe that, as expected $R(\tau)$
%values for CSR are around zero (i.e., do not fall outside
%$[-1.96,1.96]$. Although, $R(\tau)$ values are smaller for the
%association case than those for CSR case, still no $R(\tau)$ value
%falls below -1.96, i.e., for none of the $\tau$ values considered,
%the association is deemed to be significant. On the other hand, for
%the segregation case, $R(\tau)$ values are all larger than those for
%CSR case, and are above 1.96 for $\tau=.6,.7,.8$ values (i.e.,
%segregation is found to be significant at these $\tau$ values).

%\begin{figure}[ht]
%\centering
% \psfrag{kernel density estimate}{ \Huge{\bf{kernel density estimate}}}
% \rotatebox{-90}{ \resizebox{1.8 in}{!} { \includegraphics[width=6cm, height=10cm]{Relden_CS1Tn20.ps}}}
% \rotatebox{-90}{ \resizebox{1.8 in}{!} { \includegraphics[width=6cm, height=10cm]{ReldenStd_CS1Tn20.ps}}}
% \caption{
%\label{fig:relden-n=20} Relative density (left) and $R(\tau)$
%(standardized relative density) (right) values for the realizations
%provided in Figure \ref{fig:deldata-J=1}. Squares are for CSR,
%triangles are for association, and plus signs are for segregation
%alternatives. Horizontal lines in the left plot are for
%$R(\tau)=-1.96,0$, and 1.96.}
%\end{figure}

\subsection{Consistency of the Tests Under the Alternatives}
\label{sec:consistency} {\bf Theorem 4:} The test against
$H^S_{\ve}$ which rejects for $R(\tau) >z_{\alpha}$ and the test
against $H^A_{\ve}$ which rejects for $R(\tau) <z_{1-\alpha}$ are
consistent for $\tau \in (0,1]$ and $\ve \in \bigl( 0,\sqrt{3}/3
\bigr)$.

In fact, the analysis of the means under the alternatives reveals
more than what is required for consistency.  Under segregation, the
analysis indicates that $\mu_S(\tau,\ve_1) < \mu_S(\tau,\ve_2)$ for
$\ve_1<\ve_2$. On the other hand, under association, the analysis
indicates that $\mu_A(\tau,\ve_1) > \mu_A(\tau,\ve_2)$ for
$\ve_1<\ve_2$.

\subsection{Monte Carlo Power Analysis}
\label{sec:monte-carlo}
 In this section, we asses the finite sample
behaviour of the relative density using Monte Carlo simulations for
testing CSR against segregation or association. We provide the
kernel density estimates, empirical significance levels, and
empirical power estimates under the null case and various
segregation and association alternatives.

\subsubsection{Monte Carlo Power Analysis for Segregation Alternatives}
\begin{figure}[ht]
\centering
\psfrag{kernel density estimate}{ \Huge{\bf{kernel density estimate}}}
\psfrag{relative density}{ \Huge{\bf{relative density}}}
\rotatebox{-90}{ \resizebox{1.8 in}{!}{ \includegraphics{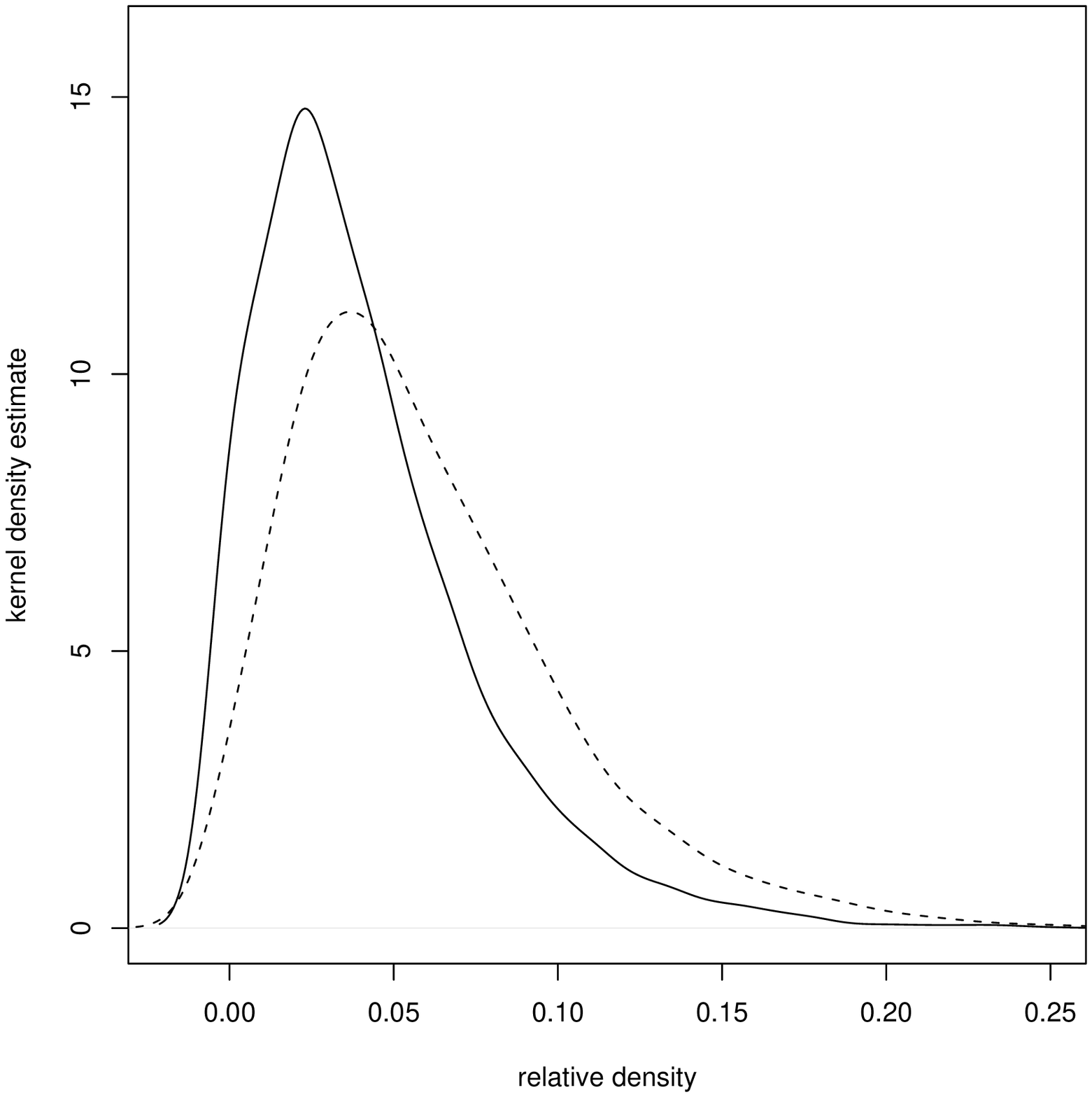}}}
\rotatebox{-90}{ \resizebox{1.8 in}{!}{ \includegraphics{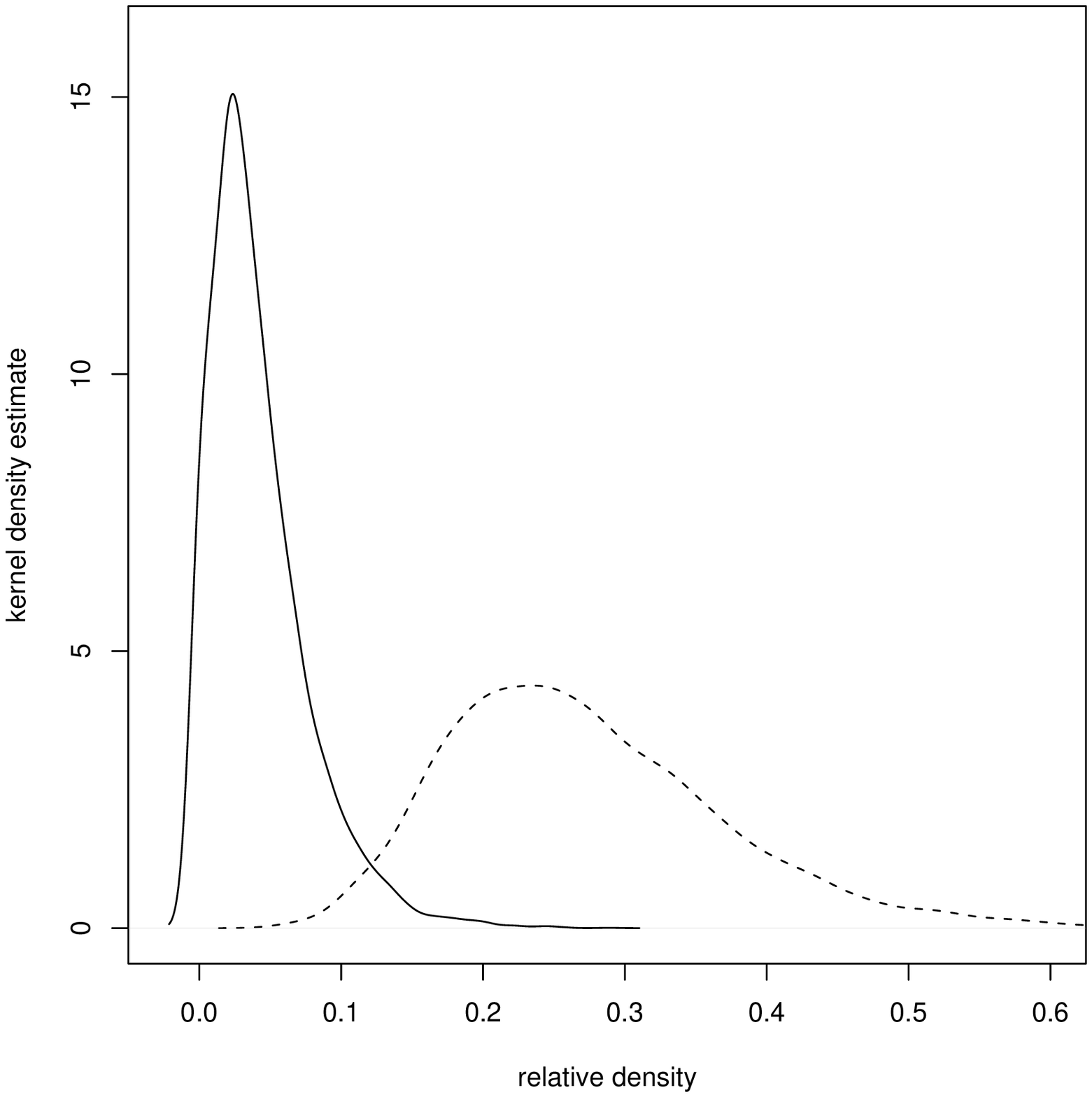}}}
\rotatebox{-90}{ \resizebox{1.8 in}{!}{ \includegraphics{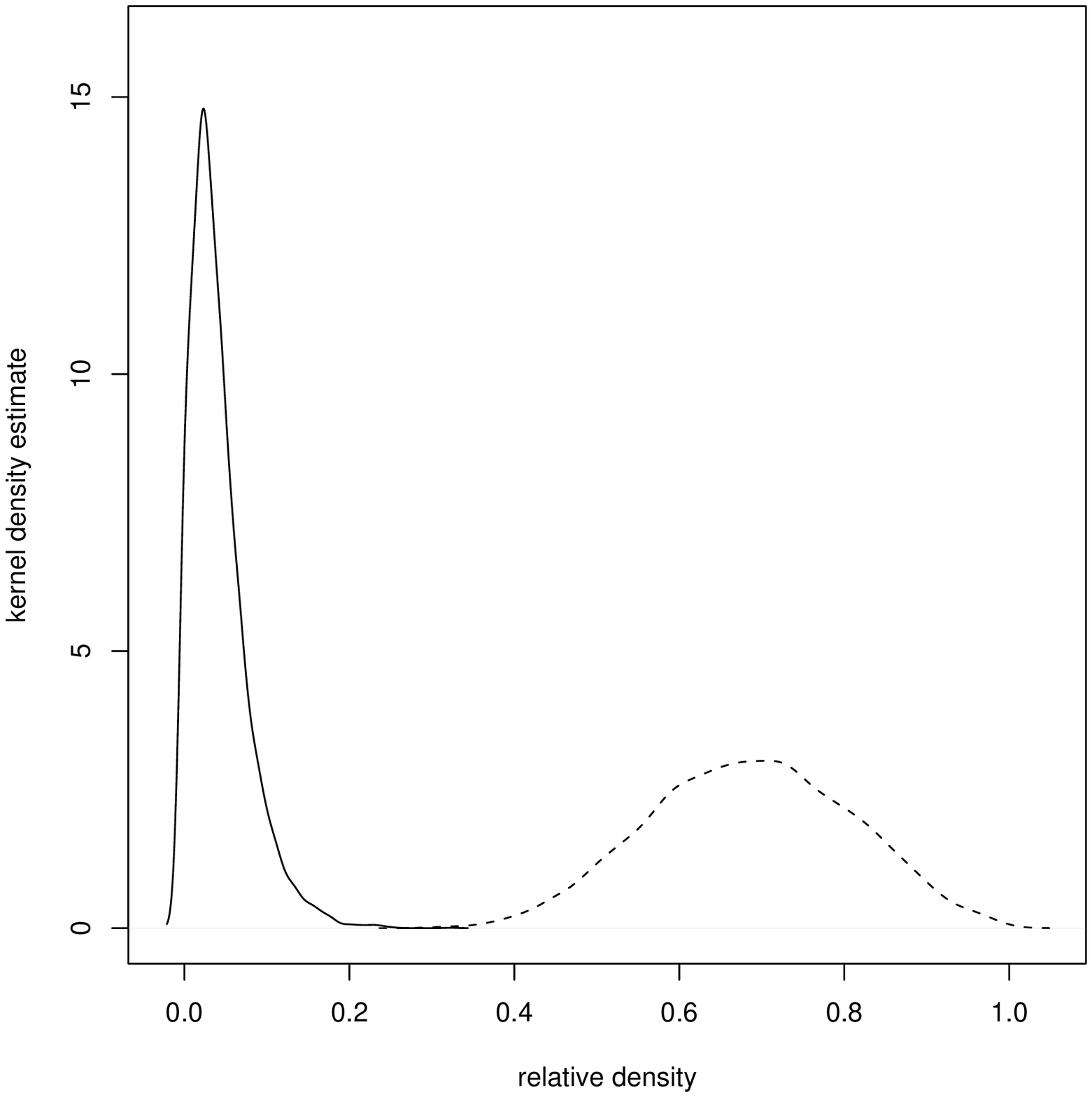}}}
\caption{
\label{fig:CSsegsim5}
Kernel density estimates for the null (solid) and the segregation alternative
$H^S_{\ve}$ (dashed) with $\tau=1/2$, $n=10$, $N=10000$, and
$\ve=\sqrt{3}/8$ (left), $\ve=\sqrt{3}/4$ (middle), and $\ve=2\,\sqrt{3}/7$ (right).
}
\end{figure}

\begin{figure}[ht]
\centering
\psfrag{kernel density estimate}{ \Huge{\bf{kernel density estimate}}}
\psfrag{relative density}{ \Huge{\bf{relative density}}}
%\rotatebox{-90}{ \resizebox{1.8 in}{!}{ \includegraphics{CSsegsim5.ps}}}
\rotatebox{-90}{ \resizebox{1.8 in}{!}{ \includegraphics{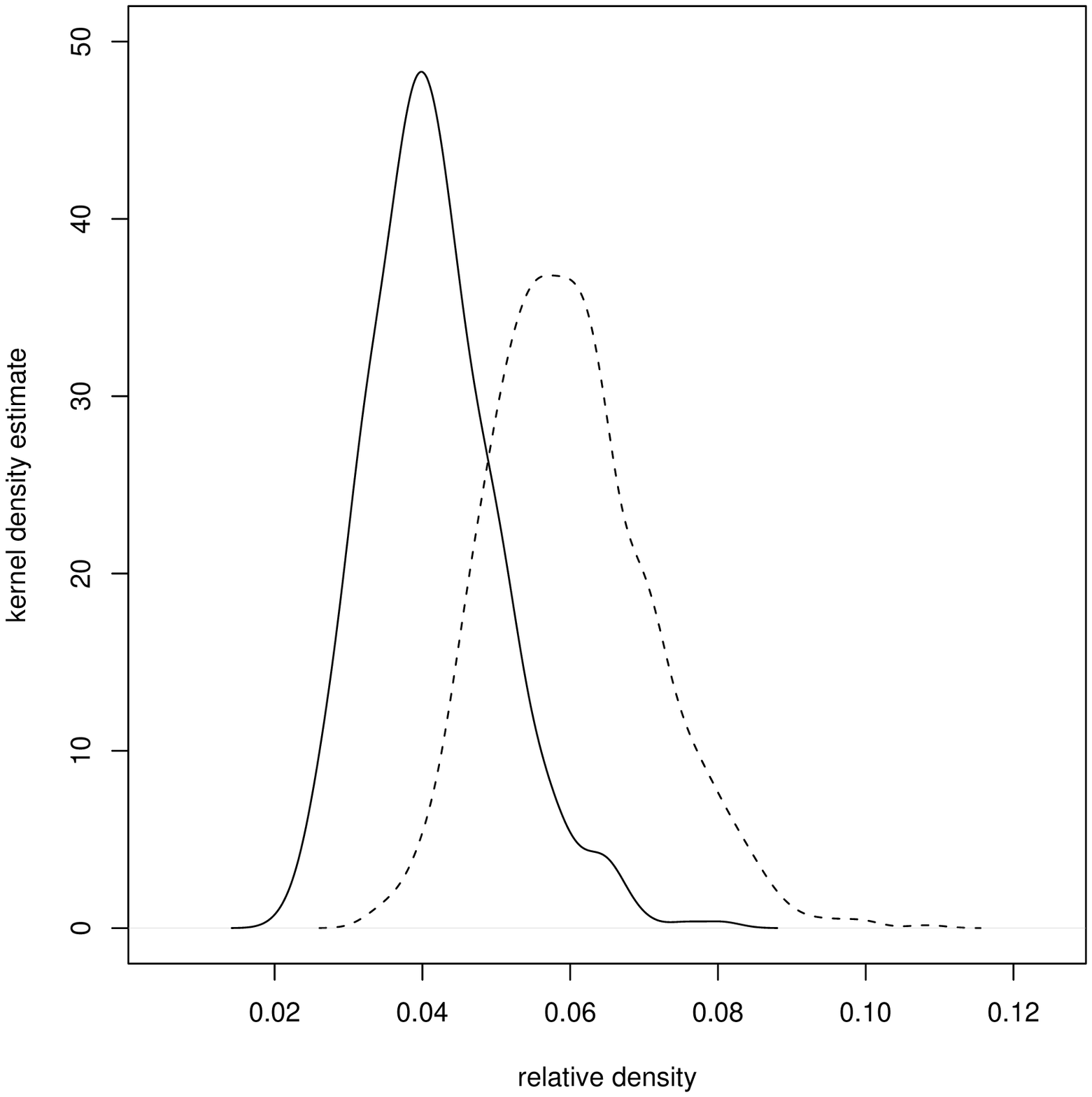}}}
%\rotatebox{-90}{ \resizebox{1.8 in}{!}{ \includegraphics{CSseg2sim5.ps}}}
\rotatebox{-90}{ \resizebox{1.8 in}{!}{ \includegraphics{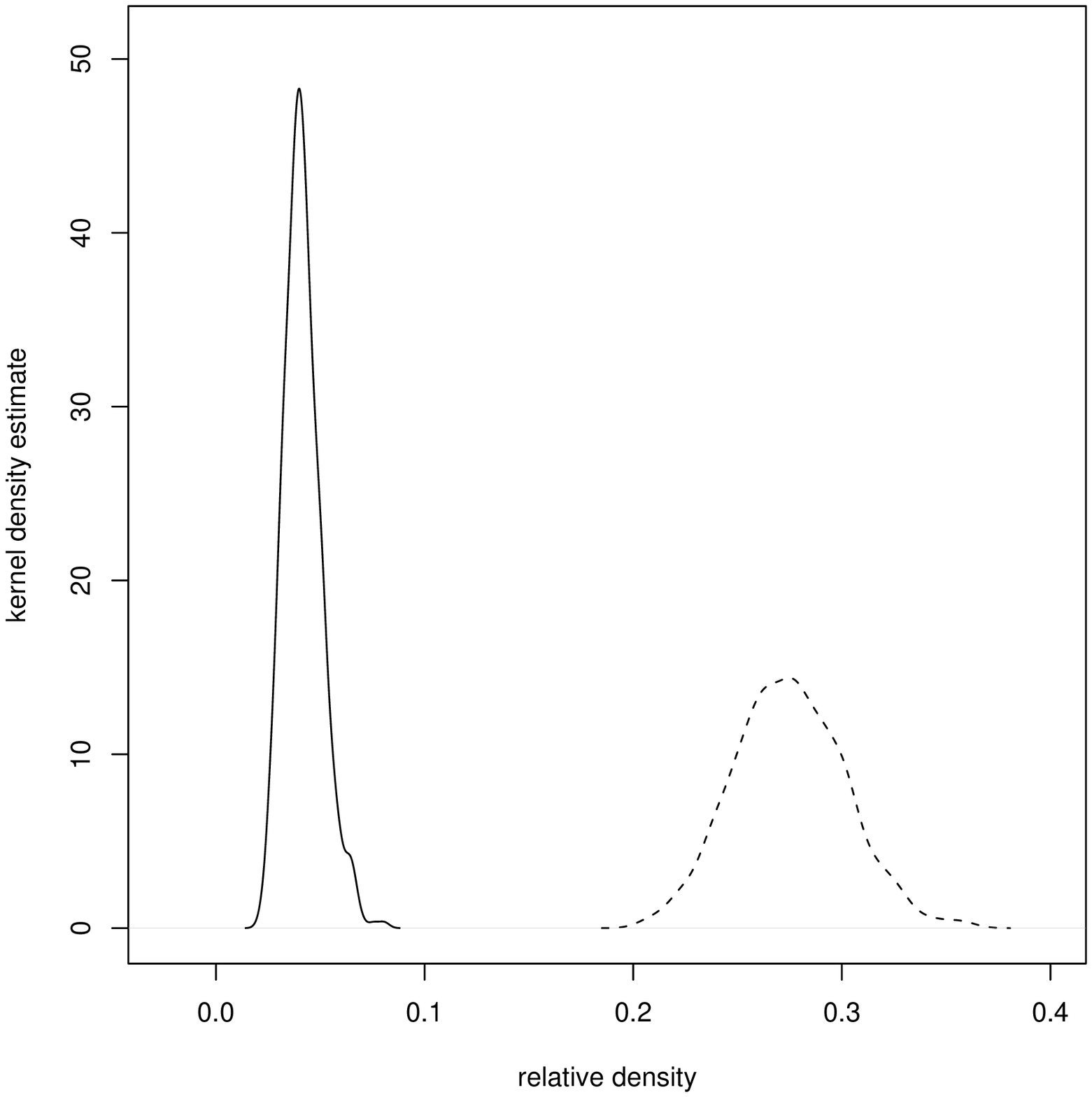}}}
\caption{
\label{fig:CSsegsim5-n100}
Kernel density estimates for the null (solid) and the segregation alternative
$H^S_{\sqrt{3}/4}$ (dashed) for $\tau=1/2$ with $n=10$ and $N=10000$ (left) and $n=100$ and $N=1000$ (right).
}
\end{figure}

In Figures \ref{fig:CSsegsim5} and \ref{fig:CSsegsim5-n100}, we
present the kernel density estimates under $H_o$ and $H^S_{\ve}$
with $\ve=\sqrt{3}/8,\,\sqrt{3}/4,\,2\,\sqrt{3}/7$.  Observe that
with $n=10$, and $\ve=\sqrt{3}/8$, the density estimates are very
similar implying small power; and as $\ve$ gets larger, the
separation between the null and alternative curves gets larger,
hence the power gets larger. With $n=10$, 10000 Monte Carlo
replicates yield power estimates
$\widehat{\beta}^S_{mc}(\ve)=.0994,\,.9777,\,1.000$, respectively.
With $n=100$, there is more separation between the null and
alternative curves at each $\ve$, which implies that power increases
as $\ve$ increases. With $n=100$, 1000 Monte Carlo replicates yield
$\widehat{\beta}^S_{mc}(\ve)=.5444,\,1.000,\,1.000$.

\begin{figure}[ht]
\centering
\psfrag{power}{ \Huge{\bf{power}}}
%\psfrag{t}{ \Huge{$\tau$}}
 \rotatebox{-90}{ \resizebox{2.2 in}{!}{ \includegraphics{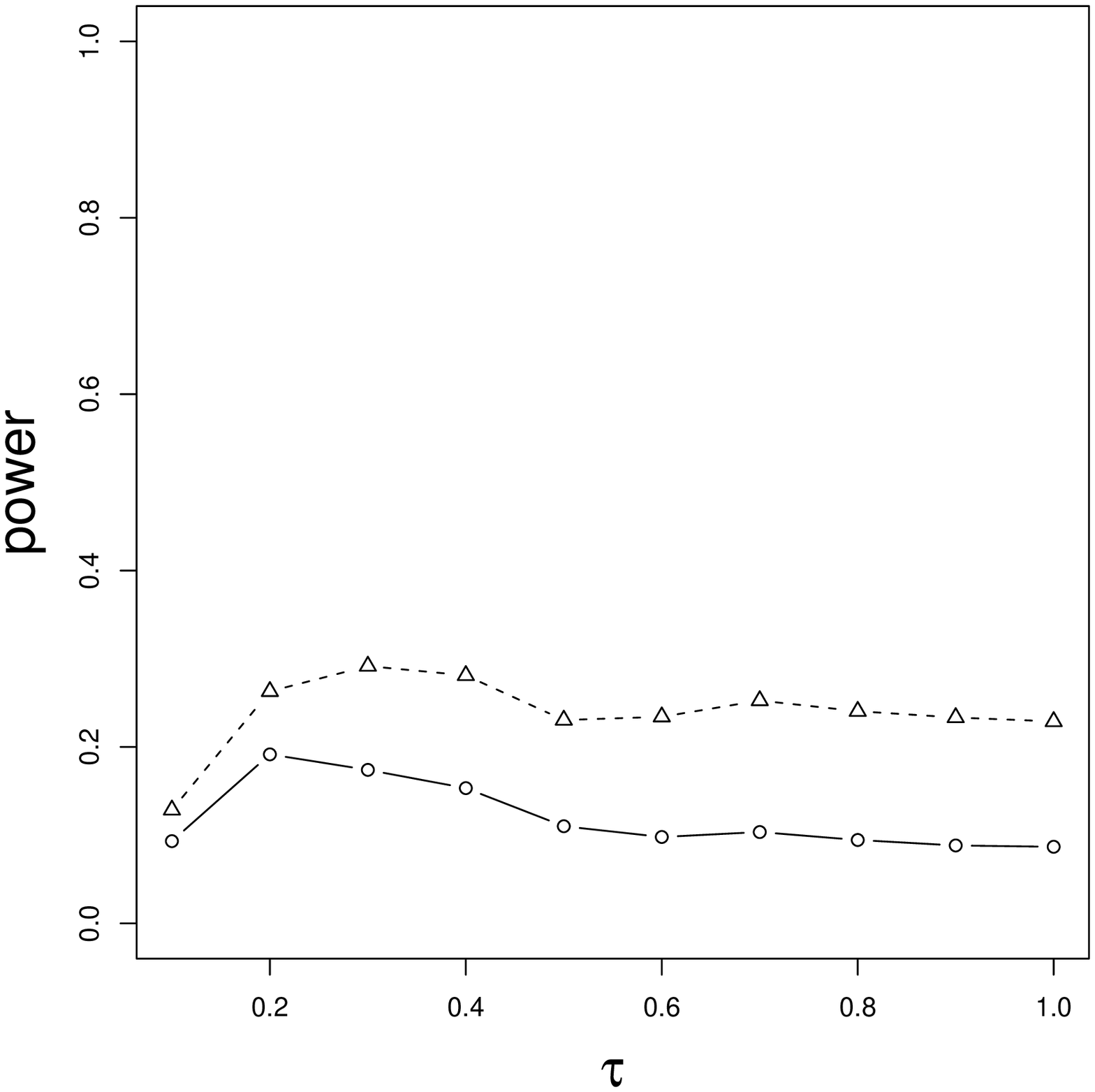}}}
 \rotatebox{-90}{ \resizebox{2.2 in}{!}{ \includegraphics{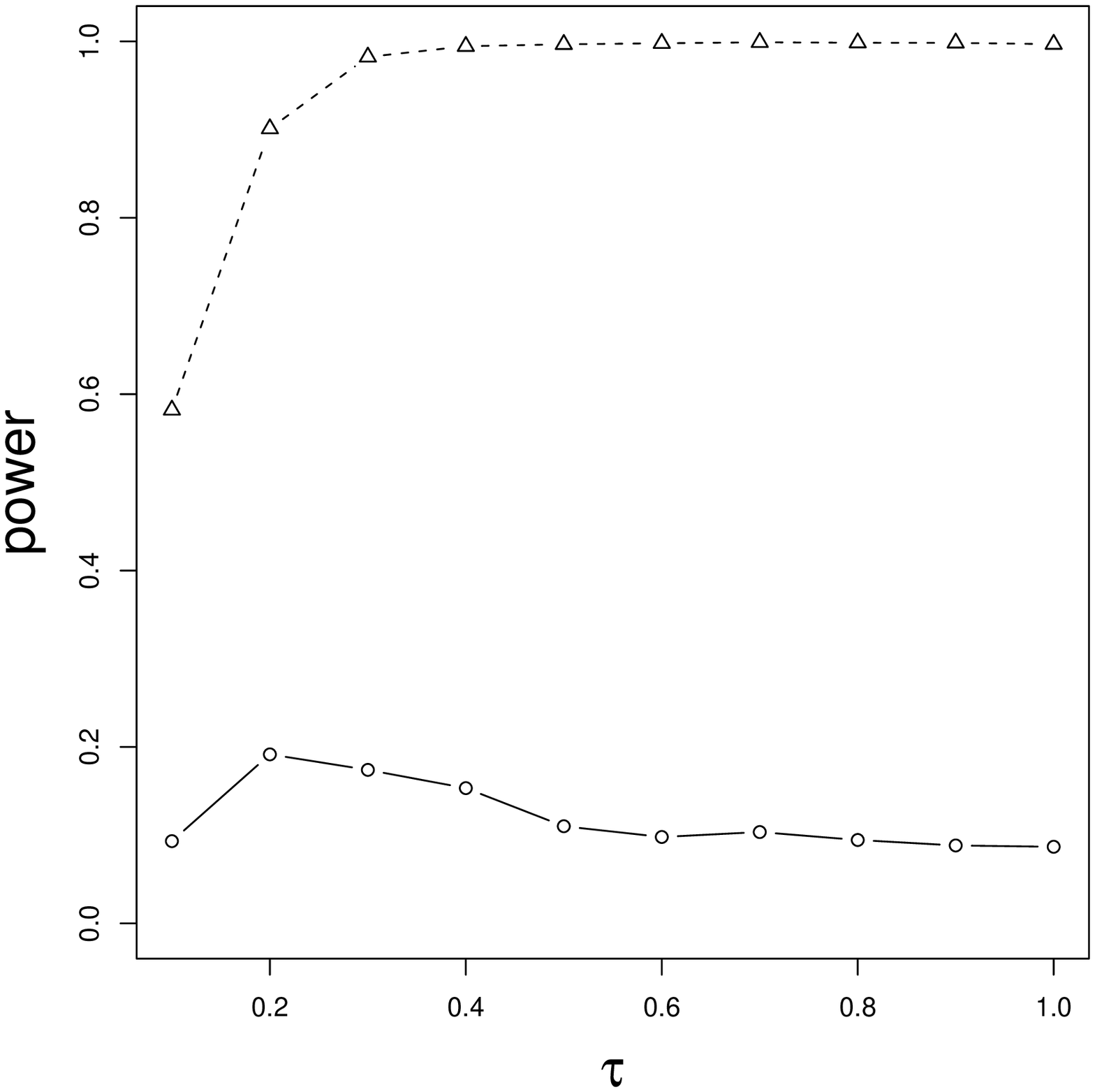}}}
%\rotatebox{-90}{ \resizebox{1.8 in}{!}{ \includegraphics{CSsegtilde3.ps}}}
\caption{ \label{fig:CSSegSimPowerCurve} Monte Carlo power using the
asymptotic critical value against segregation alternatives
$H^S_{\sqrt{3}/8}$ (left), $H^S_{\sqrt{3}/4}$ (right),
%  and $H^S_{2\,\sqrt{3}/7}$ (right)
as a function of $\tau$, for $n=10$ and $N=10000$.  The circles
represent the empirical significance levels while triangles
represent the empirical power values. }
\end{figure}

For a given alternative and sample size, we may consider analyzing
the power of the test --- using the asymptotic critical value (i.e.,
the normal approximation)--- as a function of $\tau$. Figure
\ref{fig:CSSegSimPowerCurve} presents a Monte Carlo investigation of
power against $H^S_{\sqrt{3}/8}$, $H^S_{\sqrt{3}/4}$
%, and $H^S_{2\,\sqrt{3}/7}$
as a function of $\tau$ for $n=10$. The corresponding empirical
significance levels and power estimates are presented in Table
\ref{tab:CS-asy-emp-val-S}. The empirical significance levels,
$\widehat{\alpha}_{n=10}$, are all greater than $.05$ with smallest
being $.0868$ at $\tau=1.0$ which have the empirical power
$\widehat{\beta}_{10}(\sqrt{3}/8) = .2289$,
$\widehat{\beta}_{10}(\sqrt{3}/4) = .9969$.
%, and $\widehat{\beta}_{10}(2\,\sqrt{3}/7) = 1.000$.
 However, the
empirical significance levels imply that $n=10$ is not large enough
for normal approximation. Notice that as $n$ gets larger, the
empirical significance levels gets closer to $.05$ (except for
$\tau=0.1$), but still are all greater than $.05$, which indicates
that for $n \le 100$, the test is liberal in rejecting $H_o$ against
segregation. Furthermore, as $n$ increases, for fixed $\ve$ the
empirical power estimates increase, the empirical significance
levels get closer to $.05$; and for fixed $n$ as $\tau$ increases
power estimates get larger. Therefore, for segregation, we recommend
the use of large $\tau$ values ($\tau \lesssim 1.0$).

\begin{table}[]
\centering
\begin{tabular}{|c|c|c|c|c|c|c|c|c|c|c|}
\hline
$\tau$  & .1 & .2 & .3  & .4 & .5 & .6 & .7 & .8 & .9 & 1.0\\
\hline
\multicolumn{11}{c}{$n=10$, $N=10000$} \\
\hline
$\widehat{\alpha}_S(n)$ & .0932 & .1916 & .1740 & .1533 & .1101 & .0979 & .1035 & .0945 & .0883 & .0868 \\
\hline
$\widehat{\beta}^S_{n}(\tau,\sqrt{3}/8)$ & .1286 & .2630 & .2917 & .2811 & .2305 & .2342 & .2526 & .2405 & .2334 & .2289 \\
\hline
$\widehat{\beta}^S_{n}(\tau,\sqrt{3}/4)$ & .5821 & .9011 & .9824 & .9945 & .9967 & .9979 & .9990 & .9985 & .9983 & .9969 \\
\hline
%$\widehat{\beta}^S_{n}(\tau,2\,\sqrt{3}/7)$ & .9834 & 1.000 & 1.000 & 1.000 & 1.000 & 1.000 & 1.000 & 1.000 & 1.000 & 1.000 \\
%\hline
\multicolumn{11}{c}{$n=20$, $N=10000$} \\
\hline
$\widehat{\alpha}_S(n)$ & .2018 & .1707 & .1151 & .1099 & .0898 & .0864 & .0866 & .0800 & .0786 & .0763 \\
\hline
$\widehat{\beta}^S_{n}(\tau,\sqrt{3}/8)$ & .2931 & .3245 & .2744 & .3021 & .2844 & .2926 & .3117 & .3113 & .3119 & .3038 \\
\hline
\multicolumn{11}{c}{$n=100$, $N=1000$} \\
\hline
$\widehat{\alpha}_S(n)$ & .155 & .101 & .080 & .077 & .075 & .066 & .065 & .063 & .066 & .069 \\
\hline
$\widehat{\beta}^S_{n}(\tau,\sqrt{3}/8)$ & .574 & .574 & .612 & .655 & .709 & .742 & .774 & .786 & .793 & .793 \\
\hline
\end{tabular}
\caption{ \label{tab:CS-asy-emp-val-S} The empirical significance
level and empirical power values under $H^S_{\ve}$ for
$\ve=\sqrt{3}/8,\,\sqrt{3}/4$ at $\alpha=.05$.}
\end{table}

\subsubsection{Monte Carlo Power Analysis for Association Alternatives}

\begin{figure}[ht]
\centering
\psfrag{kernel density estimate}{ \Huge{\bf{kernel density estimate}}}
\psfrag{relative density}{ \Huge{\bf{relative density}}}
\rotatebox{-90}{ \resizebox{1.8 in}{!}{ \includegraphics{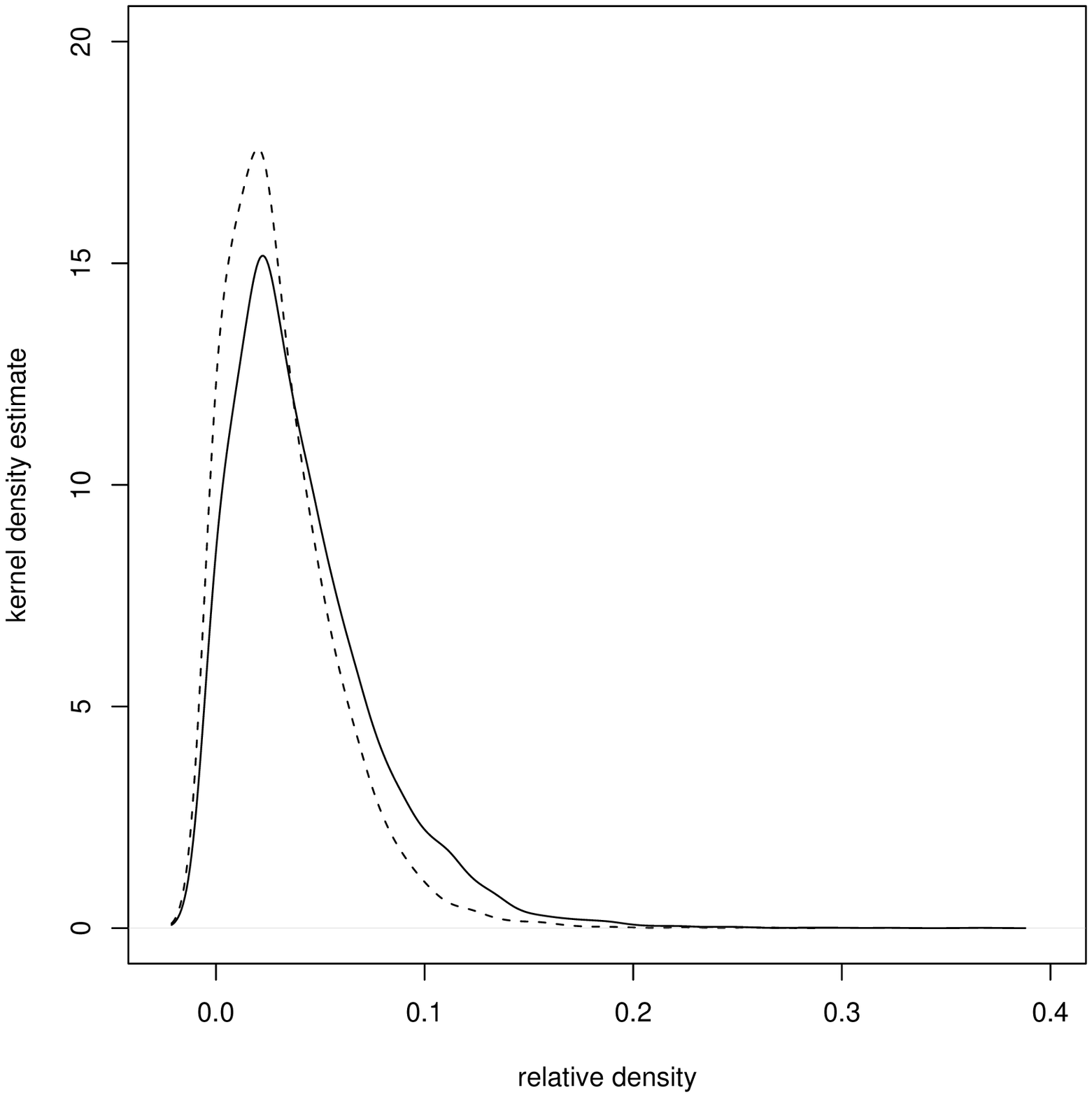}}}
\rotatebox{-90}{ \resizebox{1.8 in}{!}{ \includegraphics{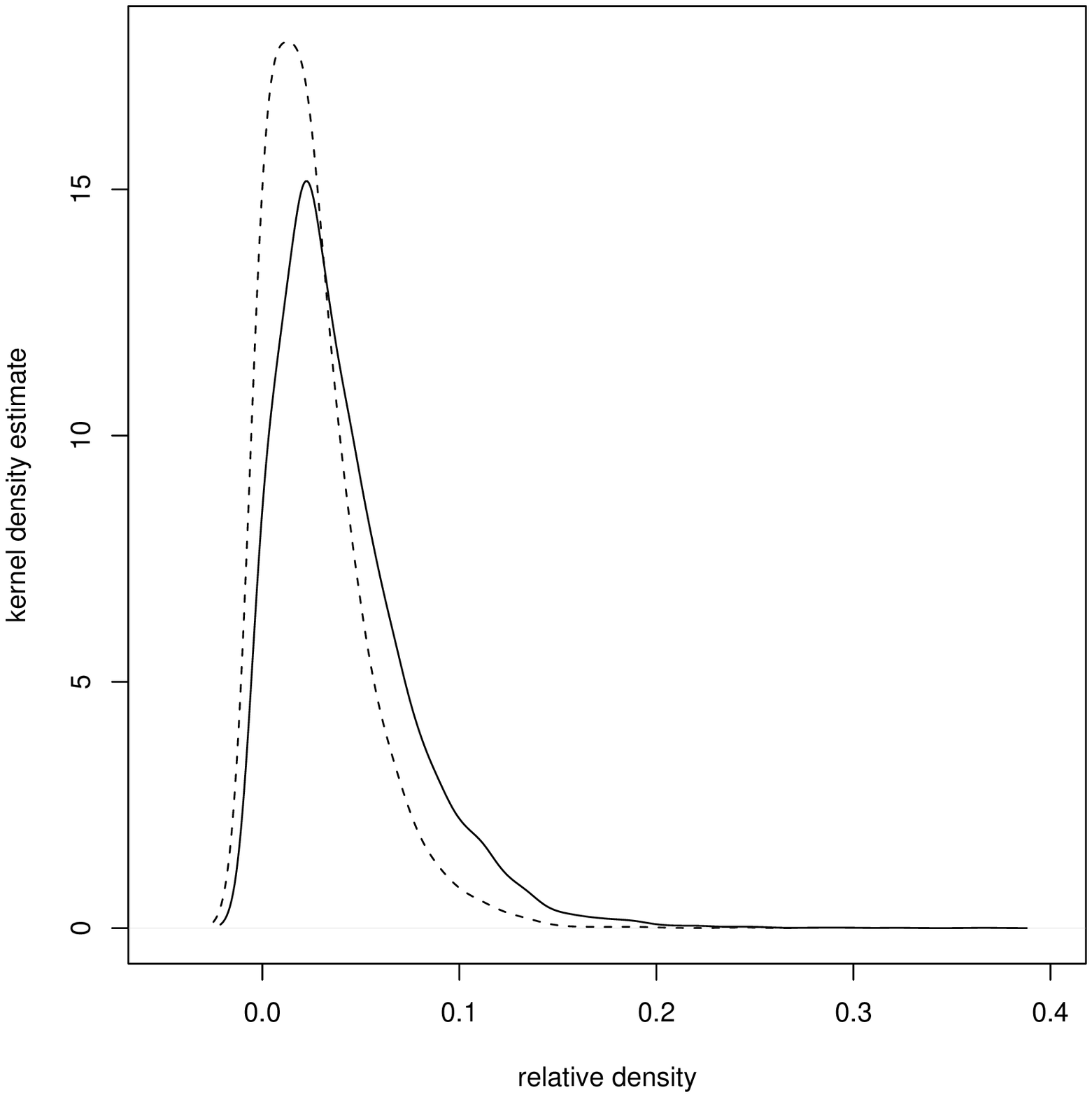}}}
\rotatebox{-90}{ \resizebox{1.8 in}{!}{ \includegraphics{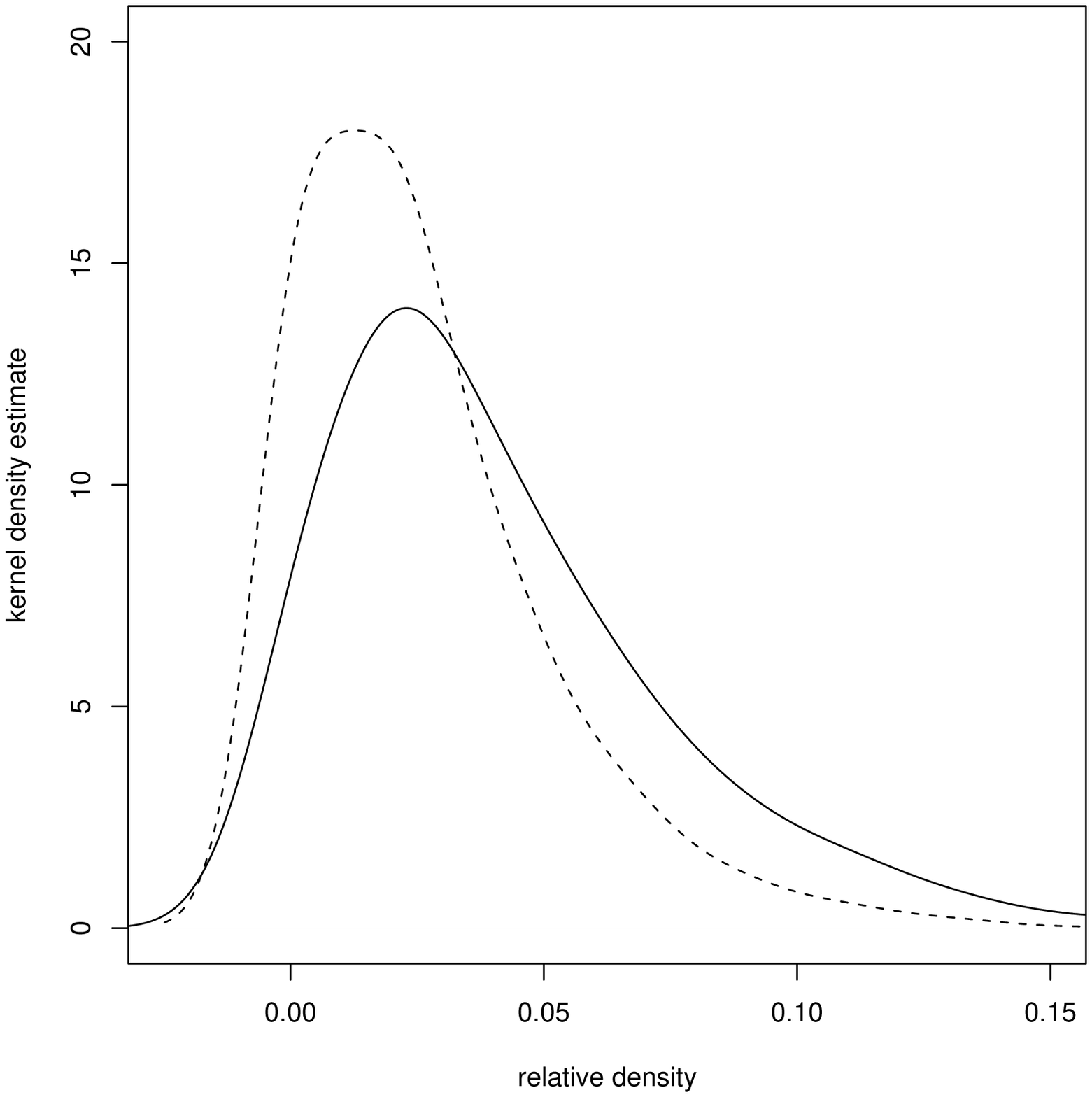}}}
\caption{
\label{fig:CSaggsim5}
Kernel density estimates for the null (solid) and the association alternative $H^A_{\ve}$ (dashed) for $\tau=1/2$ with $n=10$, $N=10000$ and $\ve=\sqrt{3}/21$ (left), $\ve=\sqrt{3}/12$ (middle), $\ve=5\,\sqrt{3}/24$ (right).
}
\end{figure}

\begin{figure}[ht]
\centering
\psfrag{kernel density estimate}{ \Huge{\bf{kernel density estimate}}}
\psfrag{relative density}{ \Huge{\bf{relative density}}}
%\rotatebox{-90}{ \resizebox{1.8 in}{!}{ \includegraphics{CSagg3sim5.ps}}}
\rotatebox{-90}{ \resizebox{1.8 in}{!}{ \includegraphics{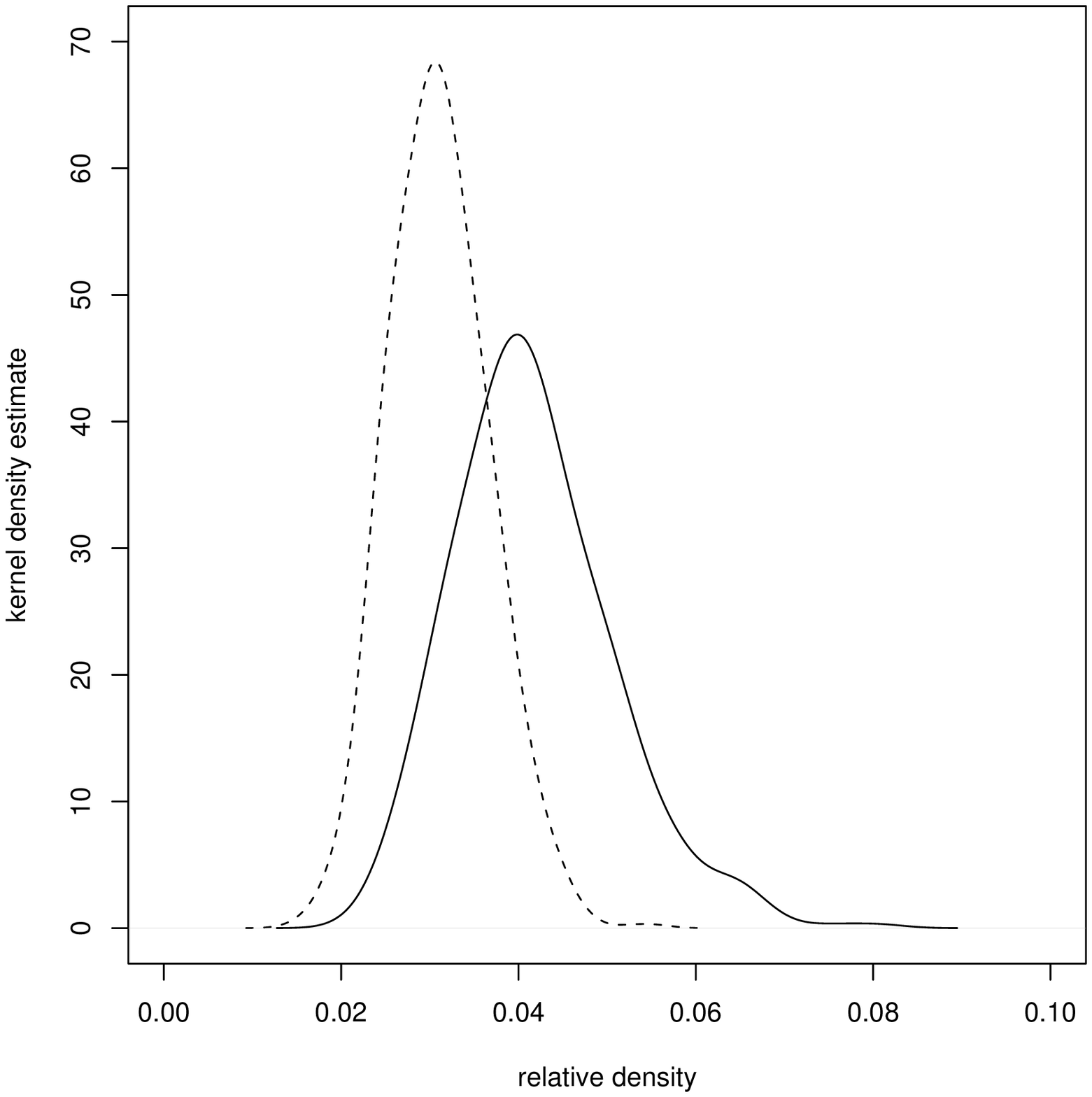}}}
%\rotatebox{-90}{ \resizebox{1.8 in}{!}{ \includegraphics{CSagg2sim5.ps}}}
\rotatebox{-90}{ \resizebox{1.8 in}{!}{ \includegraphics{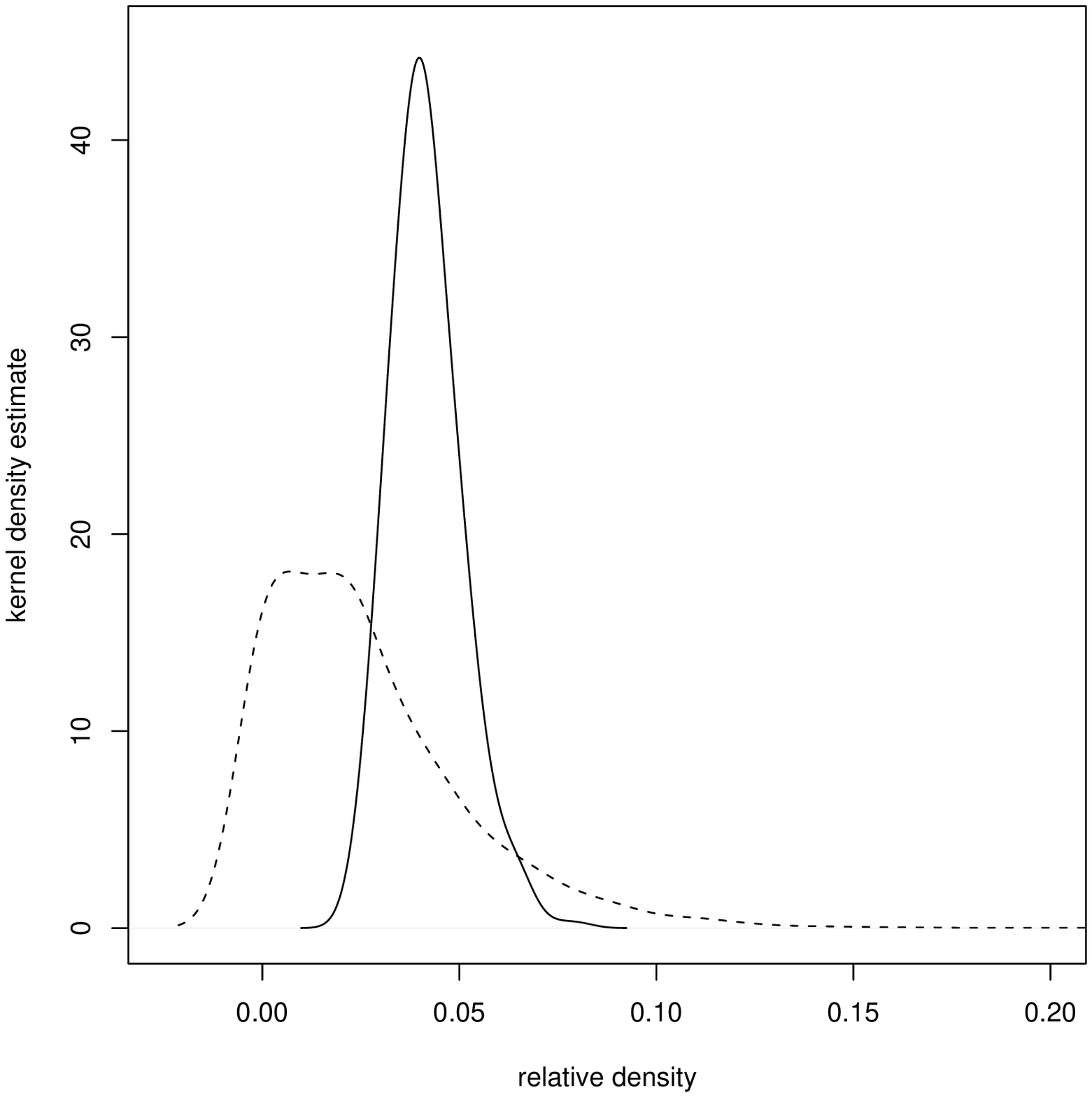}}}
\caption{
\label{fig:CSaggsim5-n100}
Kernel density estimates for the null (solid) and the association alternative $H^A_{\ve}$ (dashed) for $\tau=1/2$ with $n=100$, $N=1000$ and $\ve=\sqrt{3}/21$ (left), $\ve=\sqrt{3}/12$ (right).
}
\end{figure}

In Figures \ref{fig:CSaggsim5} and \ref{fig:CSaggsim5-n100}, we
present the kernel density estimates under $H_o$ and $H^A_{\ve}$
with $\ve=\sqrt{3}/21,\,\sqrt{3}/12,\,5\,\sqrt{3}/24$.  Observe that
with $n=10$, the density estimates are very similar for all $\ve$
values (with slightly more separation for larger $\ve$) implying
small power.  10000 Monte Carlo replicates yield power estimates
$\widehat{\beta}^A_{mc}\approx 0$. With $n=100$, there is more
separation between the null and alternative curves at each $\ve$,
which implies that power increases as $\ve$ increases. 1000 Monte
Carlo replicates yield $\widehat{\beta}^A_{mc}=.324,\,.634,\,.634$,
respectively.

For a given alternative and sample size, we may consider analyzing
the power of the test --- using the asymptotic critical value--- as
a function of $\tau$.

%\begin{figure}[ht]
%\centering
%\psfrag{power}{ \Huge{\bf{power}}}
%\rotatebox{-90}{ \resizebox{1.8 in}{!}{ \includegraphics{CSaggtilde3.ps}}}
%\rotatebox{-90}{ \resizebox{1.8 in}{!}{ \includegraphics{CSaggtilde2.ps}}}
%\rotatebox{-90}{ \resizebox{1.8 in}{!}{ \includegraphics{CSaggtilde1.ps}}}
%\caption{ \label{fig:CSAggSimPowerCurve}
%Monte Carlo power using the asymptotic critical value against association alternatives
%$H^A_{\sqrt{3}/21}$ (left),
%$H^A_{\sqrt{3}/12}$ (middle),
%and
%$H^A_{5\,\sqrt{3}/24}$ (right)
%as a function of $\tau$, for $n=10$.  The circles represent the empirical significance levels while triangles represent the empirical power values.}
%\end{figure}

%Figure \ref{fig:CSAggSimPowerCurve} presents a Monte Carlo
%investigation of power against $H^A_{\ve}$, with
%$\ve=\sqrt{3}/21,\,\sqrt{3}/12,\,5\,\sqrt{3}/24$ as a function of
%$\tau$ for $n=10$.
The empirical significance levels and power estimates against
$H^A_{\ve}$, with $\ve=\sqrt{3}/12,\,5\,\sqrt{3}/24$ as a function
of $\tau$ for $n=10$ are presented in Table
\ref{tab:CS-asy-emp-val-A}. The empirical significance level closest
to $.05$ occurs at $\tau=.6$, (much smaller for other $\tau$ values)
which have the empirical power
%$\widehat{\beta}_{10}(\sqrt{3}/21) =.0767$,
$\widehat{\beta}_{10}(\sqrt{3}/12) = .1181$, and
$\widehat{\beta}_{10}(5\,\sqrt{3}/24) = .1187$. However, the
empirical significance levels imply that $n=10$ is not large enough
for normal approximation. With $n=20$, the empirical significance
levels gets closer to $.05$ for
$\tau=.3,\,.4,\,.5,\,.7,\,.8,\,.9,\,1.0$, with closest at $\tau=.4$
which has the empirical power $.1497$. With $n=100$, the empirical
significance levels are $\approx .05$ for $\tau \ge .3$ and the
highest empirical power is $.997$ at $\tau=1.0$.  Note that as $n$
increases, the empirical power estimates increase for $\tau \ge .2$
and the empirical significance levels get closer to $.05$ for $\tau
\ge .5$. This analysis indicate that in the one triangle case, the
sample size should be really large ($n \geq 100$) for the normal
approximation to be appropriate. Moreover, the smaller the $\tau$
value, the larger the sample needed for the normal approximation to
be appropriate. Therefore, we recommend the use of large $\tau$
values ($\tau \lesssim 1.0$) for association.

\begin{table}[]
\centering
\begin{tabular}{|c|c|c|c|c|c|c|c|c|c|c|}
\hline
$\tau$  & .1 &.2 & .3 & .4 & .5 & .6 & .7 & .8 & .9 & 1.0 \\
\hline
\multicolumn{11}{c}{$n=10$, $N=10000$} \\
\hline
$\widehat{\alpha}_A(n)$ & 0 & 0 & 0 & 0 & 0 & .0465 & .0164 & .0223 & .0209 & .0339\\
\hline
%$\widehat{\beta}^A_{n}(\tau,\sqrt{3}/21)$ & 0 & 0 & 0 & 0 & 0 & .0767 & .0286 & .0386 & .0419 & .0700\\
%\hline
$\widehat{\beta}^A_{n}(\tau,\sqrt{3}/12)$ & 0 & 0 & 0 & 0 & 0 & .1181 & .0569 & .0831 & .0882 & .1490 \\
\hline
$\widehat{\beta}^A_{n}(\tau,5\,\sqrt{3}/24)$ & 0 & 0 & 0 & 0 & 0 & .1187 & .0581 & .0863 & .0985 & .1771 \\
\hline
\multicolumn{11}{c}{$n=20$, $N=10000$} \\
\hline
$\widehat{\alpha}_A(n)$ &  .6603 & .2203 & .1069 & .0496 & .0338 & .0301 & .0290 & .0267 & .0333 & .0372\\
\hline
$\widehat{\beta}^A_{n}(\tau,\sqrt{3}/12)$ & .7398 & .3326 & .2154 & .1497 & .1442 & .1608 & .1818 & .2084 & .2663 & .3167 \\
\hline
\multicolumn{11}{c}{$n=100$, $N=1000$} \\
\hline
$\widehat{\alpha}_A(n)$ & .169 & .075 & .053 & .047 & .049 & .044 & .040 & .044 & .049 & .049\\
\hline
$\widehat{\beta}^A_{n}(\tau,\sqrt{3}/12)$ & .433 & .399 & .460 & .559 & .687 & .789 & .887 & .938 & .977 & .997 \\
\hline
\end{tabular}
 \caption{ \label{tab:CS-asy-emp-val-S} \label{tab:CS-asy-emp-val-A}
The empirical significance level and empirical power values under
$H^A_{\ve}$ for $\ve=5\,\sqrt{3}/24$, $\sqrt{3}/12$, $\sqrt{3}/21$
with $N=10000$, and $n=10$ at $\alpha=.05$.}
\end{table}

\subsubsection{Pitman Asymptotic Efficiency Under the Alternatives}
\label{sec:Pitman}
 Pitman asymptotic efficiency (PAE) provides for
an investigation of ``local asymptotic power''
--- local around $H_o$. This involves the limit as $n \rightarrow \infty$,
as well as the limit as $\ve \rightarrow 0$. See proof of Theorem 3
for the ranges of $\tau$ and $\ve$ for which relative density is
continuous as $n$ goes to $\infty$. A detailed discussion of PAE can
be found in (\cite{kendall:1979,eeden:1963}. For segregation or
association alternatives the PAE is given by $\PAE(\rho_n(\tau)) =
\frac{\left( \mu^{(k)}(\tau,\ve=0) \right)^2}{\nu(\tau)}$ where $k$
is the minimum order of the derivative with respect to $\ve$ for
which $\mu^{(k)}(\tau,\ve=0) \not= 0$.  That is, $\mu^{(k)}(r,\ve=0)
\not=0$ but $\mu^{(l)}(\tau,\ve=0)=0$ for $l=1,2,\ldots,k-1$.  Then
under segregation alternative $H^S_{\ve}$ and association
alternative $H^A_{\ve}$, the PAE of $\rho_n(\tau)$ is given by
$$
\PAE^S(\tau) =
   \frac{\left( \mu^{\prime\prime}_S(\tau,\ve=0) \right)^2}{\nu(\tau)} \mbox{ and }
\PAE^A(\tau) =
   \frac{\left( \mu^{\prime\prime}_A(\tau,\ve=0) \right)^2}{\nu(\tau)},
$$
respectively, since
$\mu_S^{\prime}(\tau,\ve=0)=\mu_A^{\prime}(\tau,\ve=0) = 0$.
Equation (\ref{eq:CSAsyvar}) provides the denominator; the numerator
requires $\mu_S(\tau,\ve)$ and $\mu_A(\tau,\ve)$ which are provided
in the Appendix,
% under both segregation and association alternatives,
where we only use the intervals of $\tau$ that do not
vanish as $\ve \rightarrow 0$.

\begin{figure}[ht]
\centering
 \psfrag{t}{ {\large $\tau$}}
 \psfrag{paeS}{ \large{$\PAE^S(\tau)$}}
 \epsfig{figure=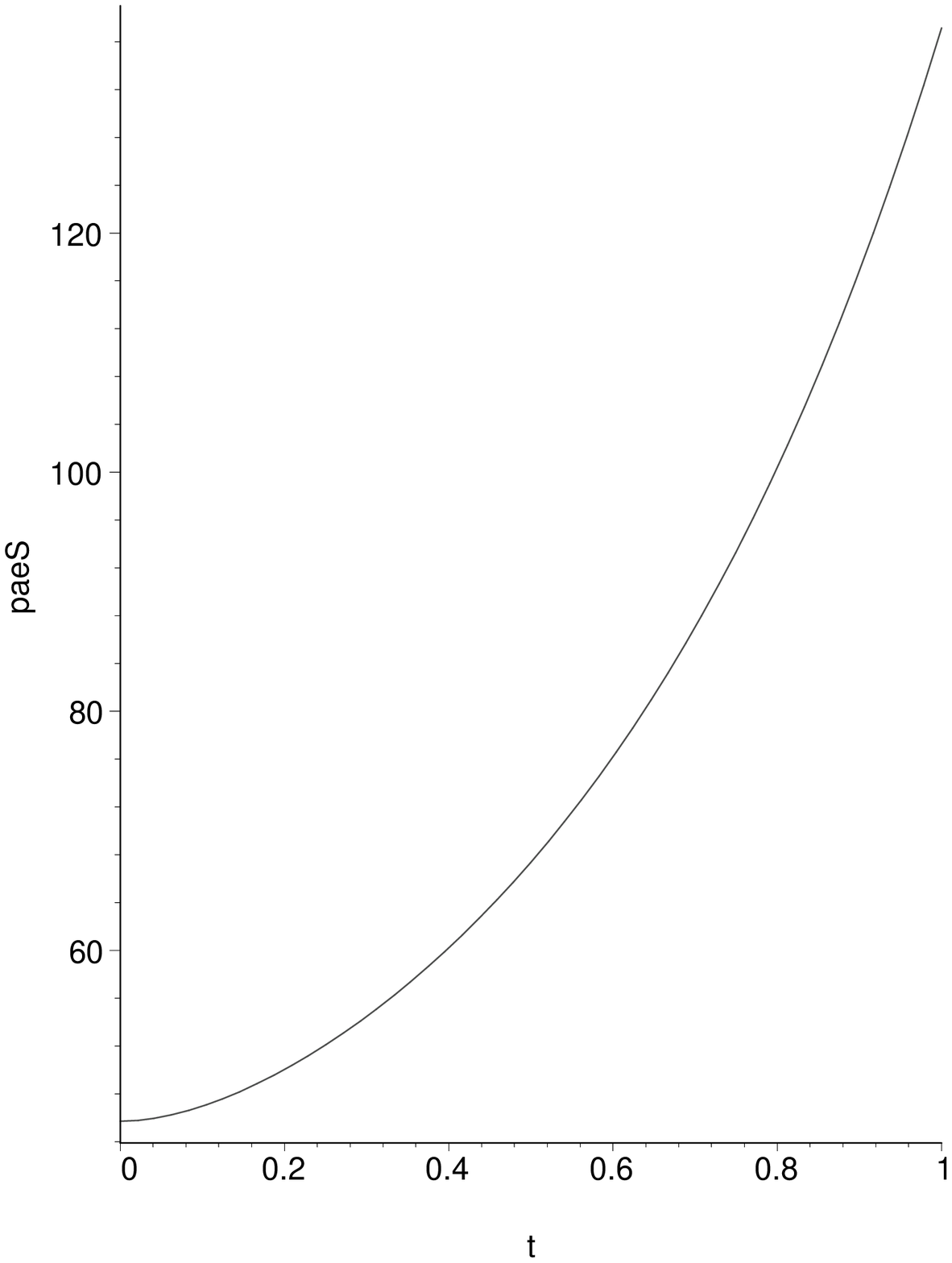, height=150pt,
width=200pt}
% \rotatebox{-0}{ \resizebox{2.25 in}{!}{\includegraphics{CSpae_segreg.eps} } }
 \psfrag{paeA}{ \large{$\PAE^A(\tau)$}}
 \epsfig{figure=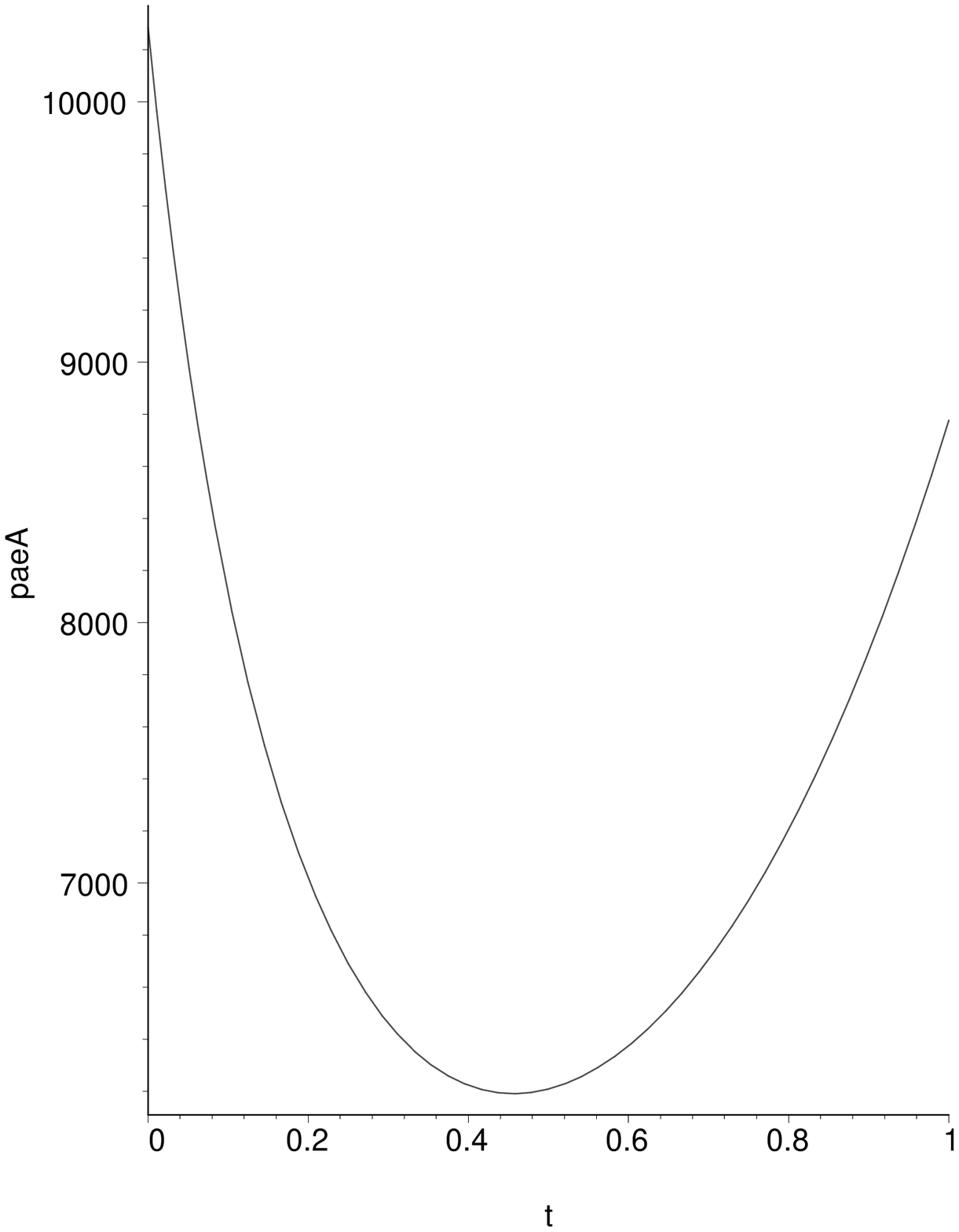, height=150pt, width=200pt}
%\rotatebox{-0}{ \resizebox{2.25 in}{!}{\includegraphics{CSpae_aggreg.eps} } }
 \caption{
 \label{fig:CS-PAE-Curves} Pitman asymptotic efficiency against
segregation (left) and association (right) as a function of $\tau$.
}
\end{figure}

In Figure \ref{fig:CS-PAE-Curves}, we
present the PAE as a function of $\tau$ for both segregation and association.

Notice that $\lim_{\tau \rightarrow 0}\PAE^S(\tau) = 320/7 \approx
45.7143$, $\argsup_{\tau \in (0,1]}\PAE^S(\tau)=1.0$, and
$\PAE^S(\tau=1) = 960/7 \approx 137.1429$. \emph{Based on the PAE
analysis, we suggest, for large $n$ and small $\ve$, choosing $\tau$
large (i.e., $\tau=1$) for testing against segregation.}

Notice that $\lim_{\tau \rightarrow 0} \PAE^A(\tau) = 72000/7
\approx 10285.7143$, $\PAE^A(\tau=1) = 61440/7 \approx 8777.1429$,
$\arginf_{\tau \in (0,1]} \PAE^A(\tau) \approx .4566$ with
$\PAE^A(\tau \approx .4566) \approx 6191.0939$. Based on the
asymptotic efficiency analysis, we suggest, for large $n$ and small
$\ve$, choosing $\tau$ small for testing against association.
However, for small and moderate values of $n$ the normal
approximation is not appropriate due to the skewness in the density
of $\rho_n(\tau)$. Therefore, \emph{for small and moderate $n$, we
suggest large $\tau$ values ($\tau \lesssim 1$).}

%We can also investigate Hodges-Lehmann asymptotic efficiency (HLAE) of $\rho_n(\tau)$. However, unlike PAE, HLAE does only involve $n \rightarrow \infty$  at a fixed $\ve > 0$.
%Hence HLAE requires the mean and, especially,
%the asymptotic variance of $\rho_n(\tau)$ {\it under a fixed alternative}.

\subsection{The Case with Multiple Delaunay Triangles}
\label{sec:mult-tri}
 Suppose $\Y$ is a finite collection of points
in $\R^2$ with $|\Y| \ge 3$. Consider the Delaunay triangulation
(assumed to exist) of $\Y$, where $T_j$ denotes the $j^{th}$
Delaunay triangle, $J$ denotes the number of triangles, and
$C_H(\Y)$ denotes the convex hull of $\Y$. We wish to investigate
$H_o: X_i \stackrel{iid}{\sim} \mathcal{U}(C_H(\Y))$ against
segregation and association alternatives.

Figure \ref{fig:deldata} is the graph of
realizations of $n=1000$ observations which are
independent and identically distributed
according to $\U(C_H(\Y))$
for $|\Y|=10$ and $J=13$ and under segregation and association
for the same $\Y$.

%\begin{figure}[ht]
%\centering
% \rotatebox{-0}{ \resizebox{1.8 in}{!}{
%\includegraphics{deldata_seg_eps1_n1000.eps} } }
% \rotatebox{-0}{ \resizebox{1.8 in}{!}{ \includegraphics{deldata_csr_n1000.eps} } }
% \rotatebox{-0}{ \resizebox{1.8 in}{!}{ \includegraphics{deldata_agg_eps2_n1000.eps} } }
% \caption{
%\label{fig:deldata} Realizations of segregation (left), $H_o$
%(middle), and association (right) for $|\Y|=10$. }
%\end{figure}

The digraph $D$ is constructed using
$\NCSt(j,\cdot)=N_{\Y_j}^{\tau}(\cdot)$ as described above, where
for $X_i \in T_j$ the three points in $\Y$ defining the Delaunay
triangle $T_j$ are used as $\Y_j$. Letting $w_j = A(T_j) /
A(C_H(\Y))$ with $A(\cdot)$ being the area functional, we obtain the
following as a corollary to Theorem 2.

{\bf Corollary 1:}
The asymptotic null distribution for $\rho_n(\tau,J)$ conditional on $\mathcal W=\{w_1,\ldots,w_J\}$
for $\tau \in (0,1]$
is given by $\N(\mu(\tau,J),\nu(\tau,J)/n)$ provided that $\nu(\tau,J)>0$ with
\begin{eqnarray}
\mu(\tau,J):=\mu(\tau) \,\sum_{j=1}^{J}w_j^2\;\; \mbox{ and }\;\;
\nu(\tau,J):= \nu(\tau) \,\sum_{j=1}^{J}w_j^3
+4\,\mu(\tau)^2\left[\sum_{j=1}^{J}w_j^3-\left(\sum_{j=1}^{J}w_j^2
\right)^2\right],
\end{eqnarray}
where $\mu(\tau)$ and $\nu(\tau)$ are given by Equations
(\ref{eq:CSAsymean}) and (\ref{eq:CSAsyvar}), respectively.

By an appropriate application of Jensen's inequality, we see that
$\sum_{j=1}^{J}w_j^3 \ge \left(\sum_{j=1}^{J}w_j^2 \right)^2.$
Therefore, the covariance  $\nu(\tau,J)=0$ iff both $\nu(\tau)=0$
and $\sum_{j=1}^{J}w_j^3=\left(\sum_{j=1}^{J}w_j^2 \right)^2$ hold,
so asymptotic normality may hold even when $\nu(\tau)=0$ (provided
that $\mu(\tau)>0$).

Similarly, for the segregation (association) alternatives where
$4\,\ve^2/3 \cdot 100 \%$ of the area around the vertices of each
triangle is forbidden (allowed), we obtain the above asymptotic
distribution of $\rho_n(\tau,J)$ with $\mu(\tau,J)$ being replaced
by $\mu_S(\tau,J,\ve)$, $\nu(\tau,J)$ by $\nu_S(\tau,J,\ve)$,
$\mu(\tau)$ by $\mu_S(\tau,\ve)$, and $\nu(\tau)$ by
$\nu_S(\tau,\ve)$. Likewise for association.

Thus in the case of $J>1$, we have a (conditional) test of $H_o: X_i
\stackrel{iid}{\sim} \U(C_H(\Y))$ which once again rejects against
segregation for large values of $\rho_n(\tau,J)$ and rejects against
association for small values of $\rho_n(\tau,J)$.

The segregation (with $\delta=1/16$, i.e., $\ve=\sqrt{3}/8$), null,
and association (with $\delta=1/4$, i.e., $\ve=\sqrt{3}/12$)
realizations (from left to right) are depicted in Figure
\ref{fig:deldata} with $n=1000$. For the null realization, the
p-value $p \ge .34$ for all $\tau$ values relative to the
segregation alternative, also  $p \ge .32$ for all $\tau$ values
relative to the association alternative. For the segregation
realization, we obtain $p \le .021$ for all $\tau \ge .2$. For the
association realization, we obtain $p \le .02$ for all $\tau \ge .2$
and $p=.07$ at $\tau=.1$. Note that this is only for one realization
of $\X_n$.

%\begin{figure}[ht]
%\centering
%\psfrag{power}{ \Huge{\bf{power}}}
%\rotatebox{-90}{ \resizebox{2.1 in}{!}{ \includegraphics{MTCSSegtilde1n100.ps}}}
%\rotatebox{-90}{ \resizebox{2.1 in}{!}{ \includegraphics{MTCSAggtilde2n100.ps}}}
%\caption{
% \label{fig:MTCS-tilde-n100}
%Monte Carlo power using the asymptotic critical value against $H^S_{\sqrt{3}/8}$ (left), $H^A_{\sqrt{3}/12}$ (right)
%as a function of $\tau$, for $n=100$, conditional on the realization of $\Y$ in Figure \ref{fig:deldata}.  The circles represent the empirical significance levels while triangles represent the empirical power values.}
%\end{figure}

%\begin{figure}[ht]
%\centering
%\rotatebox{-90}{ \resizebox{2.1 in}{!}{ \includegraphics{MTCSSegtilde1n500.ps}}}
%\rotatebox{-90}{ \resizebox{2.1 in}{!}{ \includegraphics{MTCSAggtilde2n500.ps}}}
%\caption{
% \label{fig:MTCS-tilde-n500}
%Monte Carlo power using the asymptotic critical value against $H^S_{\sqrt{3}/8}$ (left), $H^A_{\sqrt{3}/12}$ (right)
%as a function of $\tau$, for $n=500$, conditional on the realization of $\Y$ in Figure \ref{fig:deldata}.  The circles represent the empirical significance levels while triangles represent the empirical power values.}
%\end{figure}

We repeat the null and alternative realizations $1000$ times with
$n=100$ and $n=500$ and estimate the significance levels and
empirical power. The estimated values are presented in Table
\ref{tab:CS-MT-asy-emp-val}. With $n=100$, the empirical
significance levels are all greater than .05 and less than .10 for
$\tau \ge .6$ against both alternatives, much larger for other
values. This analysis suggests that $n=100$ is not large enough for
normal approximation. With $n=500$, the empirical significance
levels are around .1 for $.3 \le \tau < .5$ for segregation, and
around ---but slightly larger than--- .05 for $\tau \ge .5$.  Based
on this analysis, we see that, against segregation, our test is
liberal ---less liberal for larger $\tau$--- in rejecting $H_o$ for
small and moderate $n$, against association it is slightly liberal
for small and moderate $n$, and large $\tau$ values.  \emph{For both
alternatives, we suggest the use of large $\tau$ values.} Observe
that the poor performance of relative density in one-triangle case
for association does not persist in multiple triangle case. In fact,
for the multiple triangle case, $R(\tau)$ gets to be more
appropriate for testing against association compared to testing
against segregation.

\begin{table}[]
\centering
\begin{tabular}{|c|c|c|c|c|c|c|c|c|c|c|}
\hline
$\tau$  & .1 & .2 & .3 & .4 & .5 & .6 & .7 & .8 & .9 & 1.0\\
\hline
\multicolumn{11}{c}{$n=100$, $N=1000$, $J=13$} \\
\hline
$\widehat{\alpha}_S(n,J)$ & .496 & .366 & .302 & .242 & .190 & .103 & .102 & .092 & .095 & .091 \\
\hline
$\widehat{\beta}^S_{n}(\tau,\sqrt{3}/8,J)$ & .393 & .429 & .464 & .512 & .551 & .578 & .608 & .613 & .611 & .604 \\
\hline
\hline
$\widehat{\alpha}_A(n,J)$ & .726 & .452 & .322 & .310 & .194 & .097 & .081 & .072 & .063 & .067  \\
\hline
$\widehat{\beta}^A_{n}(r,\sqrt{3}/12,J)$ & .452 & .426 & .443 & .555 & .567 & .667 & .721 & .809 & .857 & .906 \\
\hline
\multicolumn{11}{c}{$n=500$, $N=1000$, $J=13$} \\
\hline
$\widehat{\alpha}_S(n,J)$ & 0.246 & 0.162 & 0.114 & 0.103 & 0.097 & 0.092 & 0.095 & 0.093 & 0.095 & 0.090 \\
\hline
$\widehat{\beta}^S_{n}(r,\sqrt{3}/8,J)$ & 0.829 & 0.947 & 0.982 & 0.988 & 0.995 & 0.995 & 0.997 & 0.998 & 0.997 & 0.997 \\
\hline
\hline
$\widehat{\alpha}_A(n,J)$ & 0.255 & 0.117 & 0.077 & 0.067 & 0.052 & 0.059 & 0.061 & 0.054 & 0.056 & 0.058 \\
\hline
$\widehat{\beta}^A_{n}(\tau,\sqrt{3}/12,J)$ & 0.684 & 0.872 & 0.953 & 0.991 & 0.999 & 1.000 & 1.000 & 1.000 & 1.000 & 1.000 \\
\hline
\end{tabular}\caption{
\label{tab:CS-MT-asy-emp-val}
The empirical significance level and empirical power values under $H^S_{\sqrt{3}/8}$ and $H^A_{\sqrt{3}/12}$, $N=1000$, $n=100$, and $J=13$, at $\alpha=.05$ for the realization of $\Y$ in Figure \ref{fig:deldata}.}
\end{table}

The conditional test presented here is appropriate when $w_j \in
\mathcal W$ are fixed, not random. An unconditional version requires
the joint distribution of the number and relative size of Delaunay
triangles when $\Y$ is, for instance, a Poisson point pattern. Alas,
this joint distribution is not available (\cite{okabe:2000}).

%\begin{remark}
%We can derive related test statistics in multiple triangle case similar to the ones in Section \ref{sec:rel-test-stat-MT}, by replacing $r$ with $\tau$.
%\end{remark}

\subsubsection{Pitman Asymptotic Efficiency Analysis for Multiple Triangle Case}
The PAE analysis is given for $J=1$. For $J>1$, the analysis will
depend on both the number of triangles as well as the sizes of the
triangles.  So the optimal $\tau$ values with respect to these
efficiency criteria for $J=1$ are not necessarily optimal for $J>1$,
so the analyses need to be updated, conditional on the values of $J$
and $\mathcal W$.

Under the segregation alternative $H^S_{\ve}$, the PAE of
$\rho_n(\tau)$ is given by
$$
\PAE_J^S(\tau) = \frac{(
\mu_S^{\prime\prime}(\tau,J,\ve=0))^2}{\nu(\tau,J)}=
   \frac{\left( \mu_S^{\prime\prime}(\tau,\ve=0)\,\sum_{j=1}^{J}w_j^2 \right)^2}{\nu(\tau) \,\sum_{j=1}^{J}w_j^3 +4\,\mu_S(\tau)^2\Bigl(\sum_{j=1}^{J}w_j^3-\bigl(\sum_{j=1}^{J}w_j^2 \bigr)^2\Bigr)}.
$$
Under association alternative $H^A_{\ve}$ the PAE of $\rho_n(\tau)$
is similar.

%\begin{figure}[ht]
%\centering
%\psfrag{tau}{{\Huge $\tau$}}
%\rotatebox{-0}{ \resizebox{2.25 in}{!}{ \includegraphics{CS_MT_pae_segreg.eps} } }
%\rotatebox{-0}{ \resizebox{2.25 in}{!}{ \includegraphics{CS_MT_pae_aggreg.eps} } }
%\caption{
%\label{fig:CS-MT-PAE-Curves}
%Pitman asymptotic efficiency against segregation (left) and association (right)
%as a function of $\tau$ with $J=13$.
%Notice that vertical axes are differently scaled.
%}
%\end{figure}

%In Figure \ref{fig:CS-MT-PAE-Curves}, we present the PAE as a
%function of $\tau$ for both segregation and association conditional
%on the realization of $\Y$ in Figure \ref{fig:deldata}. Notice that
The PAE curves for $J=13$ (as in Figure \ref{fig:deldata}) are
similar to the ones for the $J=1$ case (See Figures
\ref{fig:CS-PAE-Curves})
% and \ref{fig:CS-MT-PAE-Curves})
hence are omitted. Some values of note are $\lim_{\tau \rightarrow
0}\PAE_J^S(\tau) \approx 38.1954$, $\argsup_{\tau \in
(0,1]}\PAE_J^S(\tau) =1$ with $\PAE_J^S(\tau=1) \approx 100.7740$.
As for association, $\lim_{\tau \rightarrow 0} \PAE_J^A(\tau)
\approx 8593.9734$, $ \PAE_J^A(\tau=1) \approx 6449.5356$,
$\arginf_{\tau \in (0,1]} \PAE_J^A(\tau) \approx .4948$ with
$\PAE_J^A(\tau \approx .4948) \approx 5024.2236$. Based on the
Pitman asymptotic efficiency analysis, we suggest, \emph{for large
$n$ and small $\ve$, choosing large $\tau$ for testing against
segregation and small $\tau$ against association}.  However,
\emph{for moderate and small $n$, we suggest large $\tau$ values for
association} due to the skewness of the density of $\rho_n(\tau)$.

%Under segregation, the HLAE is given by
%$$
%\HLAE_J^S(\tau,\ve):=\frac{(\pi^J_a(\tau,\ve)-\pi^J_a(\NCSt))^2}{\nu_J(\tau,\ve)}=\frac{\Bigl(\mu(\tau,\ve)\,\bigl(\sum_{j=1}^{J}w_j^2 \bigr)-\mu(\tau)\,\bigl(\sum_{j=1}^{J}w_j^2 \bigr)\Bigr)^2}{\nu(\tau,\ve) \,\sum_{j=1}^{J}w_j^3 +4\,\mu(\tau,\ve)^2\Bigl(\sum_{j=1}^{J}w_j^3-\bigl(\sum_{j=1}^{J}w_j^2 \bigr)^2\Bigr)}.
%$$
% Notice that $\HLAE_J^S(\tau,\ve=0)=0$.

%The functional form of $\HLAE_J^A(\tau,\ve)$ is similar which implies $\HLAE_J^A(\tau,\ve=0)=0$.

\subsection{Extension to Higher Dimensions}
\label{sec:NCS-higher-D}

The extension of $\NCSt$ to $\R^d$ for $d > 2$ is straightforward.
Let $\Y = \{\y_1,\y_2,\cdots,\y_{d+1}\}$ be $d+1$ points in general
position. Denote the simplex formed by these $d+1$ points as
$\mathcal S (\Y)$. (A simplex is the simplest polytope in $\R^d$
having $d+1$ vertices, $d\,(d+1)/2$ edges and $d+1$ faces of
dimension $(d-1)$.) For $\tau \in [0,1]$, define the $\tau$-factor
central similarity proximity map as follows. Let $\varphi_j$ be the
face opposite vertex $\y_j$ for $j=1,2,\ldots,d+1$, and ``face
regions'' $R(\varphi_1),\ldots,R(\varphi_{d+1})$ partition $\mathcal
S(\Y)$ into $d+1$ regions, namely the $d+1$ polytopes with vertices
being the center of mass together with $d$ vertices chosen from
$d+1$ vertices. For $x \in \mathcal S(\Y) \setminus \Y$, let
$\varphi(x)$ be the face in whose region $x$ falls; $x \in
R(\varphi(x))$. (If $x$ falls on the boundary of two face regions,
 we assign $\varphi(x)$ arbitrarily.)
For ${\tau} \in (0,1]$, the {\em $\tau$-factor} central similarity
proximity region $\NCSt(x)=\NY^{\tau}(x)$ is defined to be the
simplex $\mathcal S_{\tau}(x)$ with the following properties:
\begin{itemize}
\item[(i)] $\mathcal S_{\tau}(x)$ has a face $\varphi_{\tau}(x)$ parallel to $\varphi(x)$ such that $\tau\,
d(x,\varphi(x))=d(\varphi_{\tau}(x),x)$ where $d(x,\varphi(x))$ is the Euclidean (perpendicular) distance from $x$ to $\varphi(x)$,
\item[(ii)] $\mathcal S_{\tau}(x)$ has the same orientation as and is similar to $\mathcal S(\Y)$,
\item[(iii)] $x$ is at the center of mass of $\mathcal S_{\tau}(x)$.  Note that $\tau>1$ implies that $x \in \NCSt(x)$.
\end{itemize}
For $\tau=0$, define $\NCSt(x)=\{x\}$ for all $x \in \mathcal
S(\Y)$.

Theorem 1 generalizes, so that any simplex $\mathcal S$ in $\R^d$
can be transformed into a regular polytope (with edges being equal
in length and faces being equal in area) preserving uniformity.
Delaunay triangulation becomes Delaunay tesselation in $\R^d$,
provided no more than $d+1$ points being cospherical (lying on the
boundary of the same sphere). In particular, with $d=3$, the general
simplex is a tetrahedron (4 vertices, 4 triangular faces and 6
edges), which can be mapped into a regular tetrahedron (4 faces are
equilateral triangles) with vertices
$(0,0,0)\,(1,0,0)\,(1/2,\sqrt{3}/2,0),\,(1/2,\sqrt{3}/6,\sqrt{6}/3)$.

Asymptotic normality of the $U$-statistic and consistency of the tests hold for $d>2$.

\section{Discussion and Conclusions}
\label{sec:CS-discussion}
 In this article, we investigate the
mathematical and statistical properties of a new proximity catch
digraph (PCD) and its use in the analysis of spatial point patterns.
The mathematical results are the detailed computations of means and
variances of the $U$-statistics under the null and alternative
hypotheses. These statistics require keeping good track of the
geometry of the relevant neighborhoods, and the complicated
computations of integrals are done in the symbolic computation
package, MAPLE. The methodology is similar to the one given by
\cite{ceyhan:NPE-rel-dens}. However, the results are simplified by
deliberate choices we make. For example, among many possibilities,
the proximity map is defined in such a way that the distribution of
the domination number and relative density is geometry invariant for
uniform data in triangles, which allows the calculations on the
standard equilateral triangle rather than for each triangle
separately.
%Moreover, $\tau$-factor central similarity PCD has
%simpler moments and has better asymptotic performance (has higher
%asymptotic efficiency) for association alternatives when compared to
%$r$-factor proportional-edge PCD (\cite{ceyhan:NPE-rel-dens}). We should
%also point out that, PAE measures are bounded for $\tau$-factor
%central similarity PCD.

In various fields, there are many tests available for spatial point
patterns. An extensive survey is provided by Kulldorff who
enumerates more than 100 such tests, most of which need adjustment
for some sort of inhomogeneity (\cite{kulldorff:2006}). He also
provides a general framework to classify these tests. The most
widely used tests include Pielou's test of segregation for two
classes (\cite{pielou:1961}) due to its ease of computation and
interpretation and Ripley's $K(t)$ and $L(t)$ functions
(\cite{ripley:1981}).

The first proximity map similar to the $\tau$-factor proximity map
$\NCSt$ in literature is the spherical proximity map
$N_S(x):=B(x,r(x))$; see, e.g., \cite{priebe:2001}. A slight
variation of $N_S$ is the arc-slice proximity map
$N_{AS}(x):=B(x,r(x)) \cap T(x)$ where $T(x)$ is the Delaunay cell
that contains $x$ (see (\cite{ceyhan:CS-JSM-2003})). Furthermore, Ceyhan
and Priebe introduced the (unparametrized) central similarity
proximity map $N_{CS}$ in (\cite{ceyhan:CS-JSM-2003}) and another family
of PCDs in (\cite{ceyhan:dom-num-NPE}).

The spherical proximity map $N_S$
%(the proximity map associated with CCCD)
is used in classification in the literature, but not for
testing spatial patterns between two or more classes. We develop a
technique to test the patterns of segregation or association. There
are many tests available for segregation and association in ecology
literature. See (\cite{dixon:1994}) for a survey on these tests and
relevant references. Two of the most commonly used tests are
Pielou's $\chi^2$ test of independence and Ripley's test based on
$K(t)$ and $L(t)$ functions. However, the test we introduce here is
not comparable to either of them. Our test is a conditional test ---
conditional on a realization of $J$ (number of Delaunay triangles)
and $\mathcal W$ (the set of relative areas of the Delaunay
triangles) and we require the number of triangles $J$ is fixed and
relatively small compared to $n=|\X_n|$. Furthermore, our method
deals with a slightly different type of data than most methods to
examine spatial patterns. The sample size for one type of point
(type $\X$ points) is much larger compared to the the other (type
$\Y$ points). This implies that in practice, $\Y$ could be
stationary or have much longer life span than members of $\X$. For
example, a special type of fungi might constitute $\X$ points, while
the tree species around which the fungi grow might be viewed as the
$\Y$ points.

The sampling structure for our asymptotic analysis is infill
asymptotics (\cite{cressie:1991}). Moreover, our statistic that can
be written as a $U$-statistic based on the locations of type $X$
points with respect to type $Y$ points. This is one advantage of the
proposed method: most statistics for spatial patterns can not be
written as $U$-statistics. The $U$-statistic form avails us the
asymptotic normality, once the mean and variance is obtained by
detailed geometric calculations.

The null hypothesis we consider is considerably more restrictive
than current approaches, which can be used much more generally. In
particular, we consider the completely spatial randomness pattern on
the convex hull of $\Y$ points.

Based on the asymptotic analysis and finite sample performance of
relative density of $\tau$-factor central similarity PCD, we
recommend large values of $\tau$ ($\tau \lesssim 1$) should be used,
regardless of the sample size for segregation. For association, we
recommend large values of $\tau$ ($\tau \lesssim 1$) for small to
moderate sample sizes, and small values of $\tau$ ($\tau \gtrsim
1$). However, in a practical situation, we will not know the pattern
in advance. So as an automatic data-based selection of $\tau$ to
test CSR against segregation or association, one can start with
$\tau=1$, and if the relative density is found to be smaller than
that under CSR (which is suggestive of association), use any $\tau
\in [.8,1.0]$ for small to moderate sample sizes ($n \lesssim 200$),
and use $\tau \gtrsim 0$ (say $\tau=.1$) for large sample sizes
$n>200$. If the relative density is found to be larger than that
under CSR (which is suggestive of segregation), then use large
$\tau$ (any $\tau \in [.8,1.0]$) regardless of the sample size.
However, for large $\tau$ (say, $\tau \in [.8,1.0]$)), $\tau=1$ has
more geometric appeal than the rest, so it can be used when large
$\tau$ is recommended.

Although the statistical analysis and the mathematical properties
related to the $\tau$-factor central similarity proximity catch
digraph are done in $\R^2$, the extension to $\R^d$ with $d > 2$ is
straightforward. Moreover, the geometry invariance, asymptotic
normality of the $U$-statistic and consistency of the tests hold for
$d>2$.

Throughout the article, we avoid to provide a real life example,
because the procedure in its current form ignores the $X$ points
outside the convex hull of $Y$ points (which is referred as the
\emph{boundary influence} or \emph{edge effect} in ecology
literature). Furthermore, the spatial patterns of segregation and
association are closely related to the pattern classification
problem. These aspects are topics of ongoing research.

\section*{Acknowledgements}
This work partially supported by Office of Naval Research Grant
N00014-01-1-0011 and by  Defense Advanced Research Projects Agency Grant F49620-01-1-0395.

%\bibliography{References}

\begin{thebibliography}{}

\bibitem[Ceyhan and Priebe, 2003]{ceyhan:CS-JSM-2003}
Ceyhan, E. and Priebe, C. (2003).
\newblock Central similarity proximity maps in {Delaunay} tessellations.
\newblock In {\em Proceedings of the Joint Statistical Meeting, Statistical
  Computing Section, American Statistical Association}.

\bibitem[Ceyhan et~al., 2004]{ceyhan:TR-CS-rel-dens}
Ceyhan, E., Priebe, C., and Marchette, D. (2004).
\newblock Relative density of random $\tau$-factor proximity catch digraph for
  testing spatial patterns of segregation and association.
\newblock Technical Report 645, Department of Applied Mathematics and
  Statistics, The Johns Hopkins University, Baltimore, MD, 21218.

\bibitem[Ceyhan and Priebe, 2005]{ceyhan:dom-num-NPE}
Ceyhan, E. and Priebe, C.~E. (2005).
\newblock The use of domination number of a random proximity catch digraph for
  testing spatial patterns of segregation and association.
\newblock {\em Statistics and Probability Letters}, 73:37--50.

\bibitem[Ceyhan et~al., 2006]{ceyhan:NPE-rel-dens}
Ceyhan, E., Priebe, C.~E., and Wierman, J.~C. (2006).
\newblock Relative density of the random $r$-factor proximity catch digraphs
  for testing spatial patterns of segregation and association.
\newblock {\em Computational Statistics \& Data Analysis}, 50(8):1925--1964.

\bibitem[Coomes et~al., 1999]{coomes:1999}
Coomes, D.~A., Rees, M., and Turnbull, L. (1999).
\newblock Identifying aggregation and association in fully mapped spatial data.
\newblock {\em Ecology}, 80(2):554--565.

\bibitem[Cressie, 1991]{cressie:1991}
Cressie, N. A.~C. (1991).
\newblock {\em Statistics for Spatial Data}.
\newblock Wiley, New York.

\bibitem[DeVinney et~al., 2002]{devinney:2002a}
DeVinney, J., Priebe, C.~E., Marchette, D.~J., and Socolinsky, D. (2002).
\newblock Random walks and catch digraphs in classification.
\newblock
  \url{http://www.galaxy.gmu.edu/interface/I02/I2002Proceedings/DeVinneyJason/%
DeVinneyJason.paper.pdf}.
\newblock Proceedings of the $34^{\text{th}}$ Symposium on the Interface:
  Computing Science and Statistics, Vol. 34.

\bibitem[Diggle, 2003]{diggle:2003}
Diggle, P.~J. (2003).
\newblock {\em Statistical Analysis of Spatial Point Patterns}.
\newblock Arnold Publishers, London.

\bibitem[Dixon, 1994]{dixon:1994}
Dixon, P.~M. (1994).
\newblock Testing spatial segregation using a nearest-neighbor contingency
  table.
\newblock {\em Ecology}, 75(7):1940--1948.

\bibitem[Dixon, 2002]{dixon:2002b}
Dixon, P.~M. (2002).
\newblock Nearest-neighbor contingency table analysis of spatial segregation
  for several species.
\newblock {\em Ecoscience}, 9(2):142--151.

\bibitem[Eeden, 1963]{eeden:1963}
Eeden, C.~V. (1963).
\newblock The relation between {Pitman's} asymptotic relative efficiency of two
  tests and the correlation coefficient between their test statistics.
\newblock {\em The Annals of Mathematical Statistics}, 34(4):1442--1451.

\bibitem[Gotelli and Graves, 1996]{gotelli:1996}
Gotelli, N.~J. and Graves, G.~R. (1996).
\newblock {\em Null Models in Ecology}.
\newblock Smithsonian Institution Press.

\bibitem[Hamill and Wright, 1986]{hamill:1986}
Hamill, D.~M. and Wright, S.~J. (1986).
\newblock Testing the dispersion of juveniles relative to adults: A new
  analytical method.
\newblock {\em Ecology}, 67(2):952--957.

\bibitem[Janson et~al., 2000]{janson:2000}
Janson, S., {\L}uczak, T., and Rucin\'{n}ski, A. (2000).
\newblock {\em Random Graphs}.
\newblock Wiley-Interscience Series in Discrete Mathematics and Optimization,
  John Wiley \& Sons, Inc., New York.

\bibitem[Jaromczyk and Toussaint, 1992]{jaromczyk:1992}
Jaromczyk, J.~W. and Toussaint, G.~T. (1992).
\newblock Relative neighborhood graphs and their relatives.
\newblock {\em Proceedings of IEEE}, 80:1502--1517.

\bibitem[Kendall and Stuart, 1979]{kendall:1979}
Kendall, M. and Stuart, A. (1979).
\newblock {\em The Advanced Theory of Statistics, Volume 2., 4th edition}.
\newblock Griffin, London.

\bibitem[Kulldorff, 2006]{kulldorff:2006}
Kulldorff, M. (2006).
\newblock Tests for spatial randomness adjusted for an inhomogeneity: A general
  framework.
\newblock {\em Journal of the American Statistical Association},
  101(475):1289--1305(17).

\bibitem[Lahiri, 1996]{lahiri:1996}
Lahiri, S.~N. (1996).
\newblock On consistency of estimators based on spatial data under infill
  asymptotics.
\newblock {\em Sankhya: The Indian Journal of Statistics, Series A},
  58(3):403--417.

\bibitem[Lehmann, 1988]{lehmann:1988}
Lehmann, E.~L. (1988).
\newblock {\em Nonparametrics: Statistical Methods Based on Ranks}.
\newblock Prentice-Hall, Upper Saddle River, NJ.

\bibitem[Marchette and Priebe, 2003]{marchette:2003}
Marchette, D.~J. and Priebe, C.~E. (2003).
\newblock Characterizing the scale dimension of a high dimensional
  classification problem.
\newblock {\em Pattern Recognition}, 36(1):45--60.

\bibitem[Nanami et~al., 1999]{nanami:1999}
Nanami, S.~H., Kawaguchi, H., and Yamakura, T. (1999).
\newblock Dioecy-induced spatial patterns of two codominant tree species,
  \emph{{P}odocarpus nagi} and \emph{{N}eolitsea aciculata}.
\newblock {\em Journal of Ecology}, 87(4):678--687.

\bibitem[Okabe et~al., 2000]{okabe:2000}
Okabe, A., Boots, B., and Sugihara, K. (2000).
\newblock {\em Spatial Tessellations: Concepts and Applications of Voronoi
  Diagrams}.
\newblock Wiley.

\bibitem[Pielou, 1961]{pielou:1961}
Pielou, E.~C. (1961).
\newblock Segregation and symmetry in two-species populations as studied by
  nearest-neighbor relationships.
\newblock {\em Journal of Ecology}, 49(2):255--269.

\bibitem[Priebe et~al., 2001]{priebe:2001}
Priebe, C.~E., DeVinney, J.~G., and Marchette, D.~J. (2001).
\newblock On the distribution of the domination number of random class catch
  cover digraphs.
\newblock {\em Statistics and Probability Letters}, 55:239--246.

\bibitem[Priebe et~al., 2003a]{priebe:2003b}
Priebe, C.~E., Marchette, D.~J., DeVinney, J., and Socolinsky, D. (2003a).
\newblock Classification using class cover catch digraphs.
\newblock {\em Journal of Classification}, 20(1):3--23.

\bibitem[Priebe et~al., 2003b]{priebe:2003a}
Priebe, C.~E., Solka, J.~L., Marchette, D.~J., and Clark, B.~T. (2003b).
\newblock Class cover catch digraphs for latent class discovery in gene
  expression monitoring by {DNA} microarrays.
\newblock {\em Computational Statistics and Data Analysis on Visualization},
  43-4:621--632.

\bibitem[Ripley, 1981]{ripley:1981}
Ripley, B.~D. (1981).
\newblock {\em Spatial Statistics}.
\newblock Wiley, New York.

\bibitem[Toussaint, 1980]{toussaint:1980}
Toussaint, G.~T. (1980).
\newblock The relative neighborhood graph of a finite planar set.
\newblock {\em Pattern Recognition}, 12(4):261--268.

\end{thebibliography}
%\bibliographystyle{apalike}
%\bibliographystyle{chicago}
%\bibliographystyle{amsalpha}
%\bibliographystyle{plainnat}

\newpage
\section*{APPENDIX}
\subsection*{Proof of Theorem 1}
A composition of translation, rotation, reflections, and scaling
will take any given triangle $T_o = T(\y_1,\y_2,\y_3)$ to the
``basic'' triangle $T_b = T\bigl((0,0),(1,0),(c_1,c_2)\bigr)$ with
$0 < c_1 \le 1/2$, $c_2 > 0$ and $(1-c_1)^2+c_2^2 \le 1$, preserving
uniformity. The transformation $\phi_e: \R^2 \rightarrow \R^2$ given
by $\phi_e(u,v) =
\left(u+\frac{1-2\,c_1}{\sqrt{3}}\,v,\frac{\sqrt{3}}{2\,c_2}\,v
\right)$ takes $T_b$ to the equilateral triangle $T_e =
T\bigl((0,0),(1,0),\bigl(1/2,\sqrt{3}/2\bigr)\bigr)$. Investigation
of the Jacobian shows that $\phi_e$ also preserves uniformity.
Furthermore, the composition of $\phi_e$ with the rigid motion
transformations maps
     the boundary of the original triangle $T_o$
  to the boundary of the equilateral triangle $T_e$,
     the median lines of $T_o$
  to the median lines of $T_e$,
and  lines parallel to the edges of $T_o$
  to lines parallel to the edges of $T_e$
and straight lines that cross $T_o$ to the straight lines that cross
$T_e$. Since the joint distribution of any collection of the
$h_{ij}$ involves only probability content of unions and
intersections of regions bounded by precisely such lines, and the
probability content of such regions is preserved since uniformity is
preserved, the desired result follows. $\blacksquare$

\subsection*{Derivation of $\mu(\tau)$ and $\nu(\tau)$}
 Let $M_j$
be the midpoint of edge $e_j$ for $j=1,2,3$, $M_C$ be the center of
mass, and $T_s:=T(\y_1,M_3,M_C)$. By symmetry $\mu(\tau)=P\bigl(X_2
\in \NCSt(X_1)\bigr)=6\,P\bigl(X_2 \in \NCSt(X_1),\, X_1 \in
T_s\bigr)$. Then
\begin{eqnarray*}
P\bigl(X_2 \in \NCSt(X_1),\, X_1 \in T_s\bigr) &= & \int_0^{1/2}\int_0^{\ell_{am}(x)} \frac{A(\NCSt(x_1))}{A(\TY)^2}\,dydx\\
& = & \tau^2/36
\end{eqnarray*}
where $A\bigl(\NCSt(x_1)\bigr)=3\,\sqrt{3}\,\tau^2\,y^2$,
$A(\TY)=\sqrt{3}/4$, and $\ell_{am}(x)=x/\sqrt{3}$. Hence
$\mu(\tau)=\tau^2/6$.

Next, we find the asymptotic variance term.
Let
\begin{eqnarray*}
P^{\tau}_{2N}:=P\bigl(\{X_2,X_3\} \subset \NCSt(X_1)\bigr),\;\;\;P^{\tau}_{2G}:=P\bigl(\{X_2,X_3\} \subset \G_1(X_1,\NCSt)\bigr)\; \; \mbox{ and } \;\;\\
P^{\tau}_M:=P\bigl(X_2 \in \NCSt(X_1), X_3 \in \G_1(X_1,\NCSt\bigr).
\end{eqnarray*}
where $\G_1(x,\NCSt)$ is the \emph{$\G_1$-region} of $x$ based on
$\NCSt$ and defined as $\G_1(x,\NCSt):=\{y \in \TY:\;x \subset
\NCSt(y)\}$. See (\cite{ceyhan:dom-num-NPE}) for more detail.

 Then $\Cov[h_{12},h_{13}]=\E[h_{12}\,h_{13}]-\E[h_{12}]\E[h_{13}]$ where
\begin{eqnarray*}
\E[h_{12}\,h_{13}] &=&P\bigl(\{X_2,X_3\} \subset \NCSt(X_1)\bigr)+2\,P\bigl(X_2 \in \NCSt(X_1), X_3 \in \G_1(X_1,\NCSt)\bigr)\\
& & +P\bigl(\{X_2,X_3\} \subset \G_1(X_1,\NCSt)\bigr) =
P^{\tau}_{2N}+2\,P^{\tau}_M+P^{\tau}_{2G}.
\end{eqnarray*}
Hence $\nu(\tau)=\Cov[h_{12},h_{13}]  =
\bigl(P^{\tau}_{2N}+2\,P^{\tau}_M+P^{\tau}_{2G}\bigr)-[2\,\mu(\tau)]^2.$

\begin{figure} [ht]
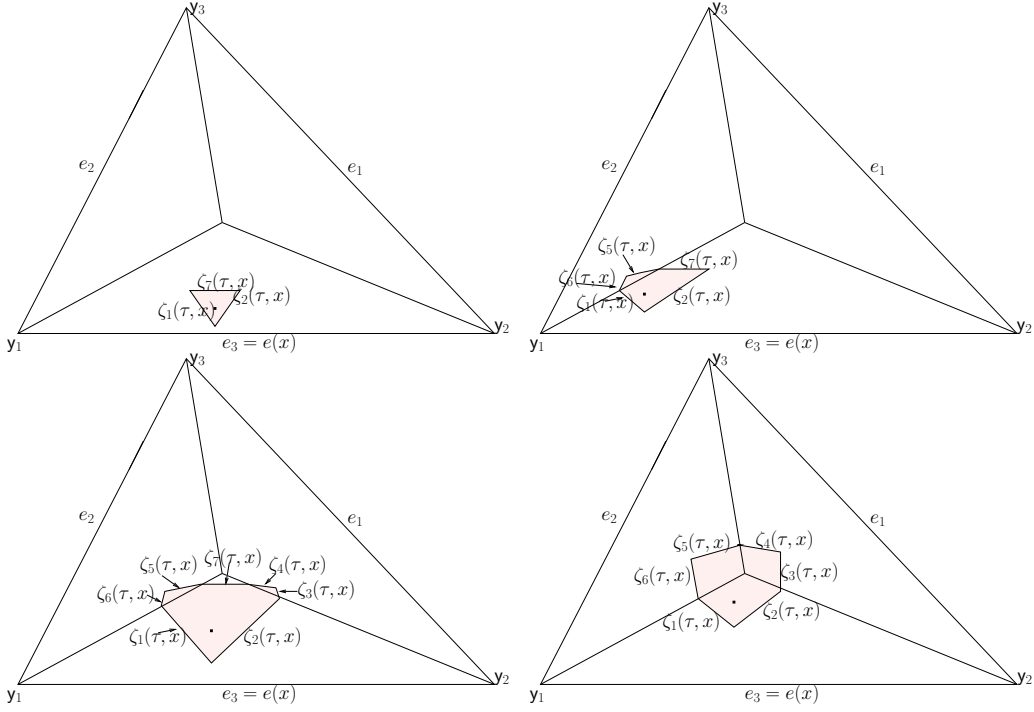

   \centering
   \scalebox{.3}{\input{N_CSGam1.pstex_t}}
   \scalebox{.3}{\input{N_CSGam2.pstex_t}}
   \scalebox{.3}{\input{N_CSGam3.pstex_t}}
   \scalebox{.3}{\input{N_CSGam4.pstex_t}}
   \caption{The prototypes of the four cases of $\G_1(x_1,\NCSt)$ for $x_1 \in T(\y_1,M_3,M_{C})$ with $\tau=1/2$.}
\label{fig:G1-NCS-Cases-1}
\end{figure}

\begin{figure} [ht]
    \centering
   \scalebox{.4}{\input{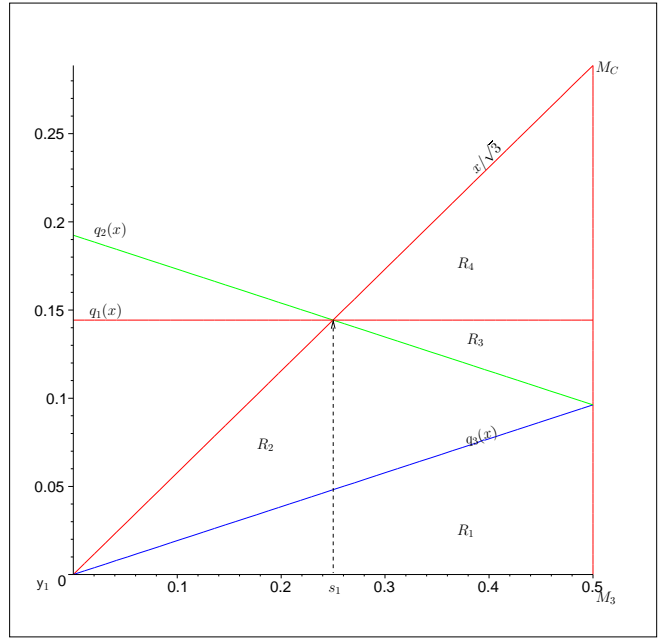}}
    \caption{The regions corresponding to the prototypes of the four cases with $\tau=1/2$.}
    \label{regions-for-NCS}
\end{figure}

To find the covariance, we need to find the possible types of
$\G_1(x_1,\NCSt)$ and $\NCSt(x_1)$ for $\tau \in (0,1]$. There are
four cases regarding $\G_1(x_1,\NCSt)$ and one case for
$\NCSt(x_1)$. See Figure \ref{fig:G1-NCS-Cases-1} for the prototypes
of these four cases of $\G_1(x_1,\NCSt)$ where, for $(x_1,y_1) \in
\TY$, the explicit forms of $\zeta_j(\tau,x)$ are
\begin{eqnarray*}
\zeta_1(\tau,x)&=&\frac {(\sqrt{3}y_1+3\,x_1-3\,x)}{\sqrt{3}\,(1+2\,\tau)},\\
\zeta_2(\tau,x)&=&-\frac {(-\sqrt{3}y_1+3\,x_1-3\,x)}{\sqrt{3}\,(1+2\,\tau)},\\
\zeta_3(\tau,x)&=&\frac {(3\,x_1+3\,\tau-3\,\tau\,x-3\,x-\sqrt{3}y_1)}{\sqrt{3}\,(-1+\tau)},\\
\zeta_4(\tau,x)&=&-\frac {-\tau\,\sqrt{3}+\tau\,\sqrt{3}x-2\,y_1}{2+\tau},\\
\zeta_5(\tau,x)&=&\frac {\tau\,\sqrt{3}x+2\,y_1}{2+\tau},\\
\zeta_6(\tau,x)&=&\frac {(-3\,x-3\,\tau\,x+3\,x_1+\sqrt{3}y_1)}{\sqrt{3}\,(1-\tau)},\\
\zeta_7(\tau,x)&=&\frac {y_1}{1-\tau}.
\end{eqnarray*}
 Each case $j$ corresponds to the region $R_j$ in Figure \ref{regions-for-NCS}, where
$$q_1(x)=\frac{1-\tau}{2\,\sqrt{3}},\;\;\; q_2(x)=\frac {(x-1)(\tau-1)}{\sqrt{3}\,(1+\tau)},\;\;\;q_3(x)=\frac {(1-\tau)x}{\sqrt{3}\,(1+\tau)},\;\;\;\mbox{ and }s_1=(1-\tau)/2.$$

The explicit forms of $R_j$, $j=1,\ldots,4$ are as follows:
\begin{eqnarray*}
R_1& = &\{(x,y)\in  [0,1/2]\times [0,q_3(x)]\},\\
R_2& = &\{(x,y)\in [0,s_1] \times [q_3(x),\ell_{am}(x)]\cup [s_1,1/2] \times [q_3(x),q_2(x)]\},\\
R_3& = &\{(x,y)\in [s_1,1/2]\times [q_2(x),q_1(x)]\},\\
R_4& = &\{(x,y)\in [s_1,1/2] \times [q_1(x),\ell_{am}(x)]\}.
\end{eqnarray*}

By symmetry,
 $$P\bigl(\{X_2,X_3\} \subset \NCSt(X_1)\bigr)=6\,P\bigl(\{X_2,X_3\} \subset \NCSt(X_1),\; X_1 \in T_s\bigr),$$
and
$$ P\bigl(\{X_2,X_3\} \subset \NCSt(X_1),\; X_1 \in T_s\bigr) = \int_0^{1/2}\int_0^{\ell_{am}(x)} \frac{A(\NCSt(x_1))^2}{A(\TY)^3}dydx=\tau^4/90, $$
where $A(\NCSt(x_1))=3\,\sqrt{3}\,\tau^2\,y^2 $.  Hence,
$$P\bigl(\{X_2,X_3\} \subset \NCSt(X_1)\bigr)=\tau^4/15.$$

Next, by symmetry,
 $$P\bigl(\{X_2,X_3\} \subset \G_1(X_1,\NCSt)\bigr)=6\,P\bigl(\{X_2,X_3\} \subset \G_1(X_1,\NCSt),\; X_1 \in T_s\bigr),$$
and
 $$P\bigl(\{X_2,X_3\} \subset \G_1(X_1,\NCSt),\; X_1 \in T_s\bigr)=\sum_{j=1}^4 P\bigl(\{X_2,X_3\} \subset \G_1(X_1,\NCSt),\; X_1 \in R_j\bigr).$$
For $x_1 \in R_1$,
\begin{eqnarray*}
P\bigl(\{X_2,X_3\} \subset \G_1(X_1,\NCSt),\; X_1 \in R_1\bigr)&=&\int_0^{1/2}\int_0^{q_3(x)} \frac{A(\G_1(x_1,\NCSt))^2}{A(\TY)^3}dydx\\
& = & \frac {{\tau}^{4}(1-\tau)}{90\,(1+2\,\tau)^{2}(1+\tau)^{5}},
\end{eqnarray*}
where $A\bigl(\G_1(x_1,\NCSt)\bigr)=3\,{\frac
{{\tau}^{2}\sqrt{3}y^{2}}{(\tau-1)^{2}(2\,\tau+1)}}$.

For $x_1 \in R_2$,
\begin{eqnarray*}
\lefteqn{P\bigl(\{X_2,X_3\} \subset \G_1(X_1,\NCSt),\; X_1 \in R_2\bigr) = \int_{0}^{s_1}\int_{q_3(x)}^{\ell_{am}(x)} \frac{A(\G_1(x_1,\NCSt))^2}{A(\TY)^3}dydx}\\
& &+\int_{s_1}^{1/2}\int_{q_3(x)}^{q_2(x)} \frac{A(\G_1(x_1,\NCSt))^2}{A(\TY)^3}dydx\\
& = &\frac {{\tau}^{5}(4\,{\tau}^{6}+6\,{\tau}^{5}-12\,{\tau}^{4}-21\,{\tau}^{3}+14\,{\tau}^{2}+40\,\tau+20)(1-\tau)}{45\,(2\,\tau+1)^{2}(\tau+2)^{2}(\tau+1)^{5}}.
\end{eqnarray*}
where $A\bigl(\G_1(x_1,\NCSt)\bigr)=\frac
{3\,\sqrt{3}(x^{2}\tau+2\,\sqrt{3}x\,y\,\tau-y^{2}\tau-x^{2}+2\,\sqrt{3}x\,y-3\,y^{2})\tau}{4\,(1-\tau)(2\,\tau+1)(\tau+2)}$.

For $x_1 \in R_3$,
\begin{eqnarray*}
P\bigl(\{X_2,X_3\} \subset \G_1(X_1,\NCSt),\; X_1 \in R_3\bigr)& = &\int_{s_1}^{1/2}\int_{q_2(x)}^{q_1(x)} \frac{A(\G_1(x_1,\NCSt))^2}{A(\TY)^3}dydx\\
& = &\frac {{\tau}^{6}(1-\tau)(6\,{\tau}^{6}-35\,{\tau}^{4}+130\,{\tau}^{2}+160\,\tau+60)}{90\,(2\,\tau+1)^{2}(\tau+2)^{2}(\tau+1)^{5}}.
\end{eqnarray*}
where\\
 $A\bigl(\G_1(x_1,\NCSt)\bigr)=-\frac {3\,\sqrt{3}(2\,x^{2}{\tau}^{2}+2\,y^{2}{\tau}^{2}-4\,x^{2}\tau-2\,x\,{\tau}^{2}+4\,y^{2}\tau+2\,\sqrt{3}y\,{\tau}^{2}+2\,x^{2}+4\,x\,\tau+6\,y^{2}+{\tau}^{2}-2\,x-2\,\sqrt{3}y-2\,\tau+1)\tau}{4\,(2\,\tau+1)(\tau-1)^{2}(\tau+2)}$.

For $x_1 \in R_4$,
\begin{eqnarray*}
\lefteqn{P\bigl(\{X_2,X_3\} \subset \G_1(X_1,\NCSt),\; X_1 \in R_4\bigr)=\int_{s_1}^{1/2}\int_{q_1(x)}^{\ell_{am}(x)} \frac{A(\G_1(x_1,\NCSt))^2}{A(\TY)^3}dydx}\\
& &+\int_{s_4}^{s_5}\int_{q_3(x)}^{\ell_{am}(x)} \frac{A(\G_1(x_1,\NCSt))^2}{A(\TY)^3}dydx+\int_{s_5}^{1/2}\int_{q_3(x)}^{q_{12}(x)} \frac{A(\G_1(x_1,\NCSt))^2}{A(\TY)^3}dydx\\
& = &\frac {{\tau}^{6}({\tau}^{2}-5\,\tau+10)}{15\,(2\,\tau+1)^{2}(\tau+2)^{2}}.
\end{eqnarray*}
where $A\bigl(\G_1(x_1,\NCSt)\bigr)=-\frac
{\sqrt{3}(3\,x^{2}+3\,y^{2}-3\,x-\sqrt{3}y-\tau+1)\tau}{2\,(2\,\tau+1)(\tau+2)}$.

So
$$ P\bigl(\{X_2,X_3\} \subset \G_1(X_1,\NCSt)\bigr)=6\,\left(-\frac {({\tau}^{2}-7\,\tau-2){\tau}^{4}}{90\,(\tau+1)(2\,\tau+1)(\tau+2)}\right)=-\frac {({\tau}^{2}-7\,\tau-2){\tau}^{4}}{15\,(\tau+1)(2\,\tau+1)(\tau+2)}.$$

Furthermore, by symmetry,
\begin{eqnarray*}
P\bigl(X_2 \in \NCSt(X_1),\;X_3\in \G_1(X_1,\NCSt)\bigr)=\\
6\,P\bigl(X_2 \in \NCSt(X_1),\;X_3\in \G_1(X_1,\NCSt),\; X_1 \in
T_s\bigr),
\end{eqnarray*}
and
\begin{eqnarray*}
P\bigl(X_2 \in \NCSt(X_1),\;X_3\in \G_1(X_1,\NCSt),\; X_1 \in T_s\bigr)\\
=\sum_{j=1}^4 P\bigl(X_2 \in \NCSt(X_1),\;X_3\in \G_1(X_1,\NCSt),\;
X_1 \in R_j \bigr).
\end{eqnarray*}

where $ P\bigl(X_2 \in \NCSt(X_1),\;X_3\in \G_1(X_1,\NCSt),\; X_1
\in R_j\bigr)$ can be calculated with the same region of integration
with integrand being replaced by
$\frac{A(\NCSt(x_1))\,A(\G_1(x_1,\NCSt))}{A(\TY)^3}$.

Then\\
$P\bigl(X_2 \in \NCSt(X_1),\;X_3\in
\G_1(X_1,\NCSt)\bigr)=6\,\left(\frac
{(2\,{\tau}^{4}-3\,{\tau}^{3}-4\,{\tau}^{2}+10\,\tau+4){\tau}^{4}}{180\,(2\,\tau+1)(\tau+2)}\right)=\frac
{(2\,{\tau}^{4}-3\,{\tau}^{3}-4\,{\tau}^{2}+10\,\tau+4){\tau}^{4}}{30\,(2\,\tau+1)(\tau+2)}.$

Hence $$\E[h_{12}\,h_{13}]=\frac {{\tau}^{4}(2\,{\tau}^{5}-{\tau}^{4}-5\,{\tau}^{3}+12\,{\tau}^{2}+28\,\tau+8)}{15\,(\tau+1)(2\,\tau+1)(\tau+2)}.$$

Therefore, $$\nu(\tau)=\frac
{{\tau}^{4}(6\,{\tau}^{5}-3\,{\tau}^{4}-25\,{\tau}^{3}+{\tau}^{2}+49\,\tau+14)}{45\,(\tau+1)(2\,\tau+1)(\tau+2)}.$$

For $\tau=0 $, it is trivial to see that $\nu(\tau)=0$. %$\blacksquare$

\section*{Sketch of Proof of Theorem 3}
 Under the alternatives, i.e. $\ve>0$ ,
$\rho_n(\tau)$ is a $U$-statistic with the same symmetric kernel
$h_{ij}$ as in the null case. The mean
$\mu_S(\tau,\ve)=\E_{\ve}[\rho_n(\tau)] = \E_{\ve}[h_{12}]/2$ (and
$\mu_A(\tau,\ve)$), now a function of both $\tau$ and $\ve$, is
again in $[0,1]$. $\nu_S(\tau,\ve)=\Cov_{\ve}[h_{12},h_{13}]$ (and
$\nu_A(\tau,\ve)$), also a function of both $\tau$ and $\ve$, is
bounded above by $1/4$, as before. Thus asymptotic normality obtains
provided that $\nu_S(\tau,\ve) > 0$ ($\nu_A(\tau,\ve) > 0$);
otherwise $\rho_n(\tau)$ is degenerate.  The explicit forms of
$\mu_S(\tau,\ve)$ and $\mu_A(\tau,\ve)$ are given, defined
piecewise, in the Appendix. Note that under $H^S_{\ve}$,
$$\nu_S(\tau,\ve)>0 \mbox{  for }(\tau,\ve) \in
\left( \bigl( 0,1 \bigr] \times \Bigl( 0,3\,\sqrt{3}/10
\Bigr]\right) \bigcup
\left(\Biggl(\frac{2\,(\sqrt{3}-3\,\ve)}{4\,\ve-\sqrt{3}},1\Biggr]
\times \Bigl( 3\,\sqrt{3}/10,\sqrt{3}/3 \Bigr)\right),$$ and under
$H^A_{\ve}$,
$$\nu_A(\tau,\ve)>0 \mbox{ for }(\tau,\ve) \in \bigl(0,1\bigr] \times \Bigl(0,\sqrt{3}/3 \Bigr). \blacksquare $$

\section*{Sketch of Proof of Theorem 4}
 Since the variance of the asymptotically normal
test statistic, under both the null and the alternatives, converges
to 0 as $n \rightarrow \infty$ (or is degenerate), it remains to
show that the mean under the null, $\mu(\tau)=\E[\rho_n(\tau)]$, is
less than (greater than) the mean under the alternative,
$\mu_S(\tau,\ve)=\E_{\ve}[\rho_n(\tau)]$ ($\mu_A(\tau,\ve)$) against
segregation (association) for $\ve > 0$. Whence it will follow that
power converges to 1 as $n \rightarrow \infty$.

It is possible, albeit tedious, to compute $\mu_S(\tau,\ve)$ and
$\mu_A(\tau,\ve)$ under the two alternatives. The calculations are
deferred to the technical report by \cite {ceyhan:TR-CS-rel-dens} due to its
extreme length and technicality, but the resulting explicit forms
are provided in the Appendix. Detailed analysis of $\mu_S(\tau,\ve)$
and $\mu_A(\tau,\ve)$ indicates that under segregation
$\mu_S(\tau,\ve)>\mu(\tau)$ for all $\ve >0$ and $\tau \in (0,1]$.
Likewise, detailed analysis of $\mu_A(\tau,\ve)$ indicates that
under association $\mu_A(\tau,\ve)<\mu(\tau)$ for all $\ve >0$ and
$\tau \in (0,1]$.
 We direct the reader to the technical report for the details of the calculations.
Hence the desired result follows for both alternatives.
$\blacksquare$

\section*{Proof of Corollary 1}
In the multiple triangle case,
\begin{eqnarray*}
\mu(\tau,J)=\E[\rho_n(\tau)]=\frac{1}{n\,(n-1)}\sum\hspace*{-0.1 in}\sum_{i < j \hspace*{0.25 in}}   \hspace*{-0.1 in} \,\E[h_{ij}]=\\
\frac{1}{2}\E[h_{12}]=\E[I(A_{12})] =P\bigl(A_{12}\bigr)=P\bigl(X_2
\in \NCSt(X_1)\bigr).
\end{eqnarray*}
But, by definition of $\NCSt(\cdot)$, $X_2 \not\in \NCSt(X_1)$ a.s.
if $X_1$ and $X_2$ are in different triangles. So by the law of
total probability
\begin{eqnarray*}
\mu(\tau,J)&:=&P\bigl(X_2 \in \NCSt(X_1)\bigr)= \sum_{j=1}^{J}P\bigl(X_2 \in \NCSt(X_1)\,|\,\{X_1,X_2\} \subset T_j \bigr)\,P\bigl(\{X_1,X_2\} \subset T_j \bigr)\\
&=& \sum_{j=1}^{J}\mu(\tau)\,P\bigl( \{X_1,X_2\} \subset T_j \bigr) \;\;\;\mbox{ (since $P\bigl(X_2 \in \NCSt(X_1)\,|\,\{X_1,X_2\} \subset T_j\bigr)=\mu(\tau)$)}\\
&=& \mu(\tau) \, \sum_{j=1}^{J}\bigl(A(T_j) /
A(C_H(\Y))\bigr)^2\;\;\; \mbox{ (since $P\bigl(\{X_1,X_2\} \subset
T_j\bigr)=\bigl(A(T_j) / A(C_H(\Y))\bigr)^2$)}
\end{eqnarray*}
Letting $w_j:=A(T_j) / A(C_H(\Y))$, we get
$\mu(\tau,J)=\mu(\tau)\cdot(\sum_{j=1}^{J}w_j^2)$ where $\mu(\tau)$
is given by Equation (\ref{eq:CSAsymean}).

Furthermore, the asymptotic variance is
\begin{eqnarray*}
\nu(\tau,J)&=&\E[h_{12}\,h_{13}]-\E[h_{12}]\E[h_{13}]\\
& = & P\bigl(\{X_2,X_3\} \subset \NCSt(X_1)\bigr)+2\,P\bigl(X_2 \in \NCSt(X_1), X_3 \in \G_1(X_1,\NCSt)\bigr)\\
& &+P\bigl(\{X_2,X_3\} \subset
\G_1(X_1,\NCSt)\bigr)-4\,(\mu(\tau,J))^2.
\end{eqnarray*}
Then for $J>1$, we have
\begin{eqnarray*}
P\bigl(\{X_2,X_3\} \subset \NCSt(X_1)\bigr)&=&\sum_{j=1}^{J}P\bigl(\{X_2,X_3\} \subset \NCSt(X_1)\,|\, \{X_1,X_2,X_3\} \subset T_j\bigr)\, P\bigl(\{X_1,X_2,X_3\} \subset T_j\bigr)\\
& = &\sum_{j=1}^{J}P^{\tau}_{2N}\, \bigl(A(T_j) / A(C_H(\Y))\bigr)^3 =P^{\tau}_{2N}\, \left(\sum_{j=1}^{J}w_j^3 \right).
\end{eqnarray*}
Similarly, $P\bigl(X_2 \in \NCSt(X_1), X_3 \in \G_1(X_1,\NCSt)\bigr)=P^{\tau}_{M}\,\left(\sum_{j=1}^{J}w_j^3 \right) \mbox{  and  }P\bigl(\{X_2,X_3\} \subset \G_1(X_1,\NCSt)\bigr)=P^{\tau}_{2G}\,\left(\sum_{j=1}^{J}w_j^3 \right)$, hence, $\nu(\tau,J)=\bigl(P^{\tau}_{2N}+2\,P^{\tau}_M+P^{\tau}_{2G}\bigr)\,\left(\sum_{j=1}^{J}w_j^3 \right)-4\, \mu(\tau,J) ^2=\nu(\tau)\,\left(\sum_{j=1}^{J}w_j^3 \right)+4\,\mu(\tau)^2\,\left(\sum_{j=1}^{J}w_j^3-\left(\sum_{j=1}^{J}w_j^2 \right)^2\right),$ so conditional on $\mathcal W$, if $\nu(\tau,J)>0$ then $\sqrt{n}\,\bigl(\rho_n(\tau)-\tilde \mu(\tau)\bigr) \stackrel {\mathcal L}{\longrightarrow} \N(0,\nu(\tau,J))$. %$\blacksquare$

\end{document}